\documentclass[review]{elsarticle}

\pdfoutput = 1
\usepackage{lineno,hyperref}
\modulolinenumbers[5]

\usepackage{caption}
\usepackage{subcaption}
\usepackage{graphicx}
\usepackage{amsmath}
\usepackage{algorithm}
\usepackage{algorithmic}
\usepackage{rotating}
\usepackage{adjustbox}
\usepackage{arydshln}
\usepackage{multirow}

\makeatletter
\newcommand*{\rom}[1]{\expandafter\@slowromancap\romannumeral #1@}
\makeatother

\journal{arXiv}

\begin{document}

\begin{frontmatter}

\title{Predicting Berth Stay for Tanker Terminals: A Systematic and Dynamic Approach}

\author[a,b]{Deqing Zhai\corref{mycorrespondingauthor}}
\ead{dzhai001@e.ntu.edu.sg}

\author[a]{Xiuju Fu\corref{mycorrespondingauthor}}
\ead{fuxj@ihpc.a-star.edu.sg}

\author[a]{Xiao Feng Yin}
\author[a]{Haiyan Xu}
\author[a]{Wanbing Zhang}

\cortext[mycorrespondingauthor]{Corresponding author}
\address[a]{1 Fusionopolis Way, Institute of High Performance Computing,\\ Agency for Science, Technology and Research, Singapore 138632}
\address[b]{50 Nanyang Avenue, School of Electrical and Electronic Engineering, \\ Nanyang Technological University, Singapore 639798}

\begin{abstract}
	Given the trend of digitization and increasing number of maritime transport, prediction of vessel berth stay has been triggered for requirements of operation research and scheduling optimization problem in the era of maritime big data, which takes a significant part in port efficiency and maritime logistics enhancement. This study proposes a systematic and dynamic approach of predicting berth stay for tanker terminals. The approach covers three innovative aspects: 1) Data source employed is multi-faceted, including cargo operation data from tanker terminals, time-series data from automatic identification system (AIS), etc. 2) The process of berth stay is decomposed into multiple blocks according to data analysis and information extraction innovatively, and practical operation scenarios are also developed accordingly. 3) The predictive models of berth stay are developed on the basis of prior data analysis and information extraction under two methods, including regression and decomposed distribution. The models are evaluated under four dynamic scenarios with certain designated cargoes among two different terminals. The evaluation results show that the proposed approach can predict berth stay with the accuracy up to 98.81\% validated by historical baselines, and also demonstrate the proposed approach has dynamic capability of predicting berth stay among the scenarios. The model may be potentially applied for short-term pilot-booking or scheduling optimizations within a reasonable time frame for advancement of port intelligence and logistics efficiency.
\end{abstract}


\end{frontmatter}


\section{Introduction}
\begin{figure}[htbp]
		\centering
		\includegraphics[width=\textwidth]{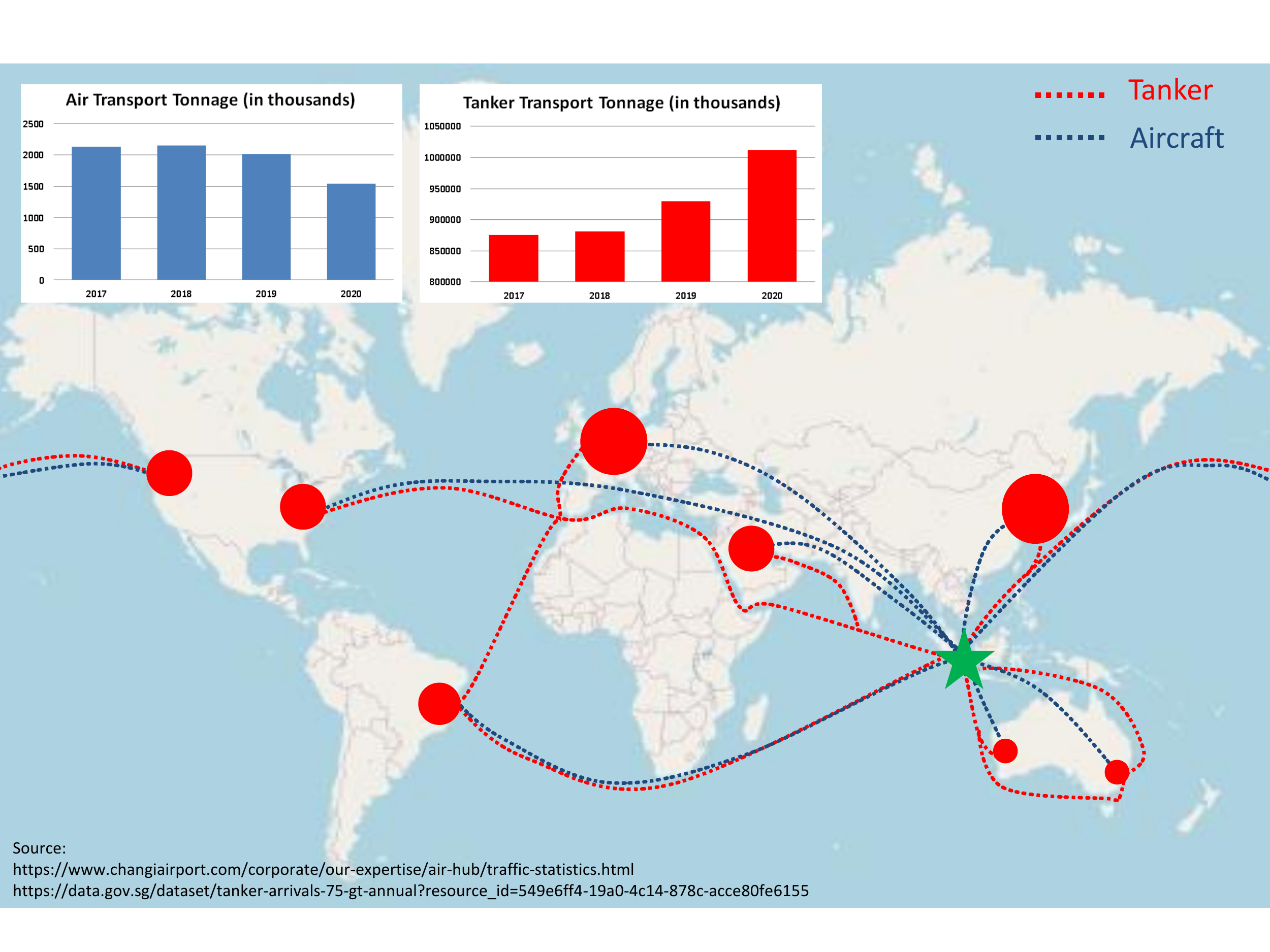} 
		\caption{Global Logistics to Singapore (Tanker vs. Air)}
		\label{Fig:Transport_Chain}
	\end{figure}

According to the statistics of International Maritime Organization (IMO) in 2016, about 80\% - 90\% worldwide trades were carried out by sea \cite{Goodwin2016}. Tanker shipping operation is a significant portion of maritime shipping. In 2020 the global oil tanker fleet had a capacity of around 601 million deadweight tonnage. In terms of tonnage, oil tankers account for around 29\% of global seaborne trade \cite{Goodwin2016}. To  be more specific to Singapore island, the global logistic networks mostly depend upon tanker and air shipping. Therefore, a brief transport description of global logistic networks between tanker and air shipping is described in Figure \ref{Fig:Transport_Chain}. It points out that tanker shipping is more heavily tilted and progressively increasing in the whole logistic picture of Singapore. Moreover, when transport mode transition from sea to land, tanker shipment also significantly propagates its impacts on corresponding hinterland transports with numerous port calls globally. Therefore, understanding and being able to forecast the upstream of transport patterns and behaviors would enhance the efficiency of logistic networks both locally and globally. It is important to the tasks, such as advanced pilot booking, advanced cargo operation scheduling, turnaround time saving, etc. As a result of those advancements, different stakeholders could gain benefits from both environmental contribution and economical growth.

\section{Challenge}
As progressively moving toward an era of digitization, data is the root of everything. As data comes from two sources in this study: one is operation data from terminal management and another is vessel's automatic identification system (AIS) data from apparatus onboard. One of most significant challenges encountered is the data quality and transparency in tanker shipping. Due to errors introduced by operators and AIS device in the process of recording, mis-recording is inevitably avoided, which could affect data quality intensively. In order to draw accurate and reliable conclusions of analyses and modelling, the data has to be pre-processed and cleaned before feeding into relevant processing modules. In addition, data transparency is another challenging issue. Since operations of terminals have the ownership themselves, the operation data is supposed to be in nature of non-sharing and confidential. Therefore, making data transparent between multi-parties is one of most challenging barriers. 

Besides data source challenges, the nature of tanker terminal for cargo operations is also a series of complex and challenging processes. Based on domain knowledge, different tanker cargoes have different chemical characteristics, such as boiling point, freezing point, density, temperature, viscosity, etc. Therefore, different approaches are applied to different cargo operations like using pumps to discharge from tanker to shore or to load from shore to tanker. As a result, different cargo operations have different time frames of completing the operations. Hence, modelling tanker cargo operation is also one of most challenging parts.

There are also other uncertainties in the process of berth stay. For example, 1) Different modes of cargo operations: Multiple cargo operations are found either in concurrent or sequential, which means that cargoes are processed simultaneously or one-by-one. 2) Different handling procedures of cargo operations: The tanks have to be cleaned before or/and after cargo operations for certain groups of cargoes due to chemical characteristics and regulations. This specific example shows a typical different procedures between different cargo operations. Therefore, the uncertainties in the process vary accordingly, which also raise a challenging problem to decipher these uncertainties in this study.

\section{Innovation and Contribution}
To the best of our knowledge, there are quite few studies of berth stay prediction to marine vessel operations. In the limited studies, most are related to container vessels, due to their simple standard and quantifiable characteristics of container volume and number of crane operations \cite{Carlo2013}. However, tanker vessels have totally different patterns as described earlier.

\begin{figure}[htbp]
		\centering
		\includegraphics[width=\textwidth]{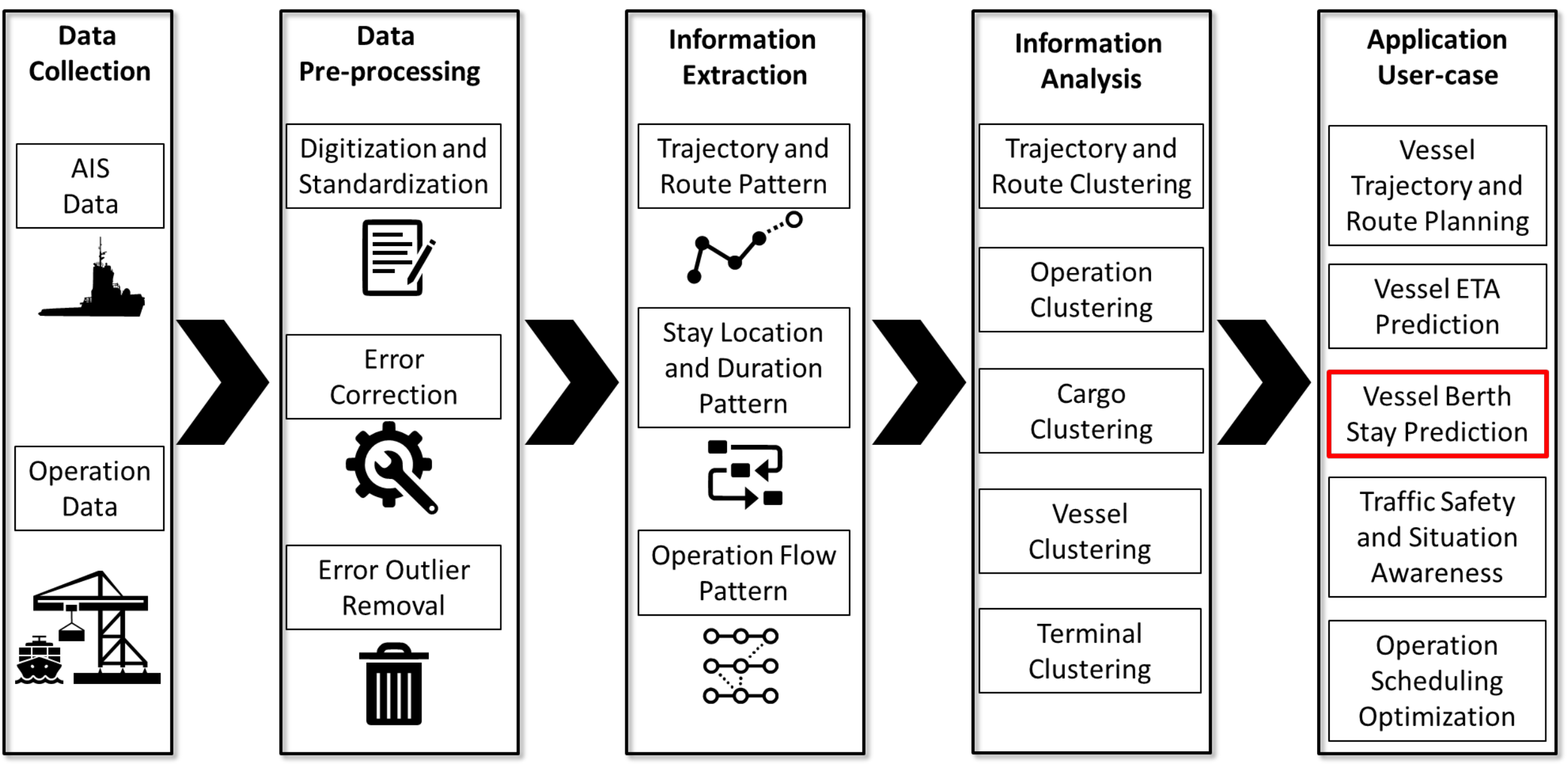} 
		\caption{Flowchart of Workflow}
		\label{Fig:Workflow}
	\end{figure}

To address all of the challenging problems mentioned above, a systematic and dynamic approach of predicting berth stay for tanker vessels is proposed. The approach refers to three innovative aspects: 1) Data source is multi-faceted, such as operation data from terminal management and time-series data from AIS. 2) The process of berth stay is innovatively decomposed into multiple blocks according to our data analysis and information extraction. 3) The predictive model of berth stay is decomposed and customized on the basis of data analysis and information extraction. In order to achieve a comprehensive evaluation of the proposed approach, single tanker terminal is insufficient. Therefore, two tanker terminals in Singapore are selected to cross-validate the proposed approach in this study, and a compact workflow to address the problems is presented in Figure \ref{Fig:Workflow}. This study focuses to address this berth stay prediction for tanker vessels.

The highlights of contribution are categorized into four parts, such as data source pre-processing, cargo operation modelling, berth stay decomposition and berth stay modelling. The detailed descriptions are shown as follows:

	\begin{itemize}
		\item[(1)] \textit{\textbf{Data Standardization and Cleaning}}: Given multiple types and formats of datasets, a data standardization of entries is proposed to facilitate operation research. Based on information extraction and analysis, port operation events and targeted cargoes are also standardized for tankers. Based on time-series data analysis, data abnormal detection and cleaning are performed.
		
		\item[(2)] \textit{\textbf{Predictive Modelling for Cargo Operation}}: With the standardized and cleaned datasets, predictive modelling for tanker operation is established by linear regression that is validated by data analysis.
		
		\item[(3)] \textit{\textbf{Tanker Berth Stay Workflow Decomposition}}: Given the detailed operation data, different cargoes have own properties and characteristics while being handled in terminals. Based on information extraction and data analyses, berth stay is decomposed into standard workflows with properly defined and designated processing blocks.
		
		\item[(4)] \textit{\textbf{Predictive Modelling for Tanker Berth Stay}}: Based on the information extraction and analysis of datasets, such as operation prediction and workflows, specific tanker berth stay prediction is performed corresponding to different applications and use-cases defined.
	\end{itemize}

With aforementioned, the paper is organized as follows: Section 2 reviews relevant literature regarding on scheduling optimization in maritime traffic. Section 3 formulates the problems that are to be tackled. The proposed methodology is followed up in Section 4. Section 5 presents the results and corresponding discussion on the basis of proposed methodology. Final conclusion, limitation and future direction are drawn in Section 6.

\section{Background on Tanker Vessel Operation and Prediction}
Back in 1999, Wang apparently showed that different activity-cargo types from different mixture of vessels in all directions are one of very challenging problems \cite{Wang1999}. There are four basic components of cargo operation, which are vessel operation, terminal operation, storage operation and receipt/delivery operation. Among these four key operations, two of them are related to berth stay, which are vessel operation and terminal operation. In this study, both vessel operation and terminal operation are considered. In the study of Stepec \textit{et al.}, predictive models of turnaround time were investigated, which is highly associated with berth stay duration. In the study, the importance of different features was examined, and it showed that features, such as cargo type, cargo size, shipment type, day of entry and berth location, were highly important toward accurate prediction \cite{Stepec2020}. Hence, this study adopts the relevant and available features, such as cargo type, cargo size and shipment type, for building predictive models of cargo operation and berth stay. Based on the master thesis of Premathilaka, a case study was investigated for turnaround time of container vessels in port of Colombo. Even though the case study was not related to tankers as this study covers, there are common factors that can be considered. In cargo operations, there are factors that influencing the operation time, such as bad weather, shift change, breakdowns, hatch cover handling, out-of-gauge handling and close bay handling, etc \cite{Premathilaka2018, Meng2017}. It is clear to note that weather, breakdown and shift are common factors to affect the operation time as well for tankers. In this study, the breakdowns and shift are considered, while weather factor actually is also indirectly embedded in the datasets represented by breakdowns duration and cargo operation speed. In terminal operations, Chapter 16 of International Safety Guide for Inland Navigation Tank-barges and Terminals provides information on relevant operations and procedures for tankers berthing terminals \cite{ISGChapter16}. This information facilitates to build up procedures of cargo operation and berth stay in this study. Besides the impact factors of on cargo operation and berth stay, Peter Olesen \textit{et al.} shows that information sharing could reduce a lot of time and resource for non-value adding activities in his study \cite{Olesen2013}. This is indirectly proofing that synthetic and aggregated data through data fusion is one of important factors for enhancing port efficiency. However, there are also challenges to be faced ahead with ``data''. Some of key challenges are cyber attack, mis-labeled data, mis-used data and shortage of skilled workforce \cite{Trelleborg2016}. These could eventually result in ``bad data''. If the ``bad data'' were not carefully identified or extracted, then the consequences would just be ``Garbage-In, Garbage-Out''. 

In the study of Ng \textit{et al.}, Port of Klang in Malaysia had been investigated on turnaround time for petroleum terminal. Various timing segments were defined in this study and a regression model was developed to enable terminal for optimal operations to improve turnaround time \cite{Ng2008}, which is equivalent to enhance berth stay prediction in this study. Based on the review of Macharis \textit{et al.}, minimal cost and time running for each shipment is sought taking into consideration, especially for cargo operation problems \cite{Macharis2004}. This is also one of key motivations in this study. As Herrero pointed out in her work, efficiency analyses of different approaches were conducted. For fleet operation, stochastic nature of operations lead to randomness highly suggested, so that stochastic approaches should be applied in such applications \cite{Herrero2005}. This is a guidance to our work in dealing with various stochastic processes in berth stay prediction too. According to the review paper of Carlo \textit{et al.}, there is an assumption that had been made by many researchers especially for scheduling and optimization, ``All operational time are deterministic.'' \cite{Carlo2013} However, to our best knowledge, there are few studies that really determine this deterministically operational time of maritime shipping vessels. Thus, this study aims to bridge the gaps in between. In the review paper of Karimi-Mamaghan \textit{et al.}, majority of papers used machine learning (ML) techniques to accomplish modeling problems effectively \cite{Mamaghan2021, Gambella2021}, such as fuzzy Bayesian \cite{Salleh2016}, CART \cite{Bishop2006}, nature-inspired \cite{Holland1992, Moscato1989, Storn1997, Kennedy2006} and non nature-inspired optimization \cite{Feo1995, Pelamatti2020, Boyd2004, Kochenberger2005} for efficient parameter tuning and accurate modelling.

\section{Description of Tanker Cargo Operation Problem}
In this section, tanker cargo operation at berth is elaborated in details. Before stating the problem, some definitions are addressed as follows: 

\noindent\textbf{Definition 1}: \textit{Cargo Operation} is a period of time that a vessel is under operation at berth, such as loading, discharging, etc. This is defined as a period dated from commence cargo operation to complete cargo operation.

\noindent\textbf{Definition 2}: \textit{Berth Stay} is a period of time that a vessel has been effectively berthed at jetty. It is defined as a period dated from all fast to all clear. All fast refers to a vessel that completes mooring, and all clear means that cargo arm or marpol prewash arm is disconnected between vessel and shore.

\noindent\textbf{Definition 3}: \textit{Prediction Error} ($\epsilon$) is defined as follows:

\begin{align}
	\epsilon = T_{predicted} - T_{actual}
\end{align}

\noindent, where $T_{predicted}$ and $T_{actual}$ are applicable to both \textit{Cargo Operation} and \textit{Berth Stay} for predictions and historical baselines, respectively.

In this study, \textit{Berth Stay} prediction is focused. This prediction means the minimum effective duration that a vessel requires at a berth to complete necessary operations. This information is critical for scheduling pilot booking and follow-ups. However, \textit{Cargo Operation} is one of key elements in \textit{Berth Stay}, and it also closely depends on cargo and terminal status. Hence, \textit{Cargo Operation} is emphatically studied together with \textit{Berth Stay}, and corresponding results of prediction errors based on previous definitions are presented in the following sections.

\section{Methodology}
In this section, methodology of this study is described in detail. It consists of data pre-processing, predictive modeling approaches for cargo operation and berth stay.

\subsection{Data Pre-processing}
Since current stage of maritime shipping operations is still not digitized yet, operational records are still manually input by operators or managers on-site among different terminals. Hence, each terminal has its own format of operational records and human errors are also introduced in the meantime. By understanding details in shipping operations among terminals, a standard data entry format has been established, and a cleaning process is followed up.

\subsubsection{Data Standardization}
\begin{figure}[htbp]
		\centering
		\includegraphics[width=\textwidth]{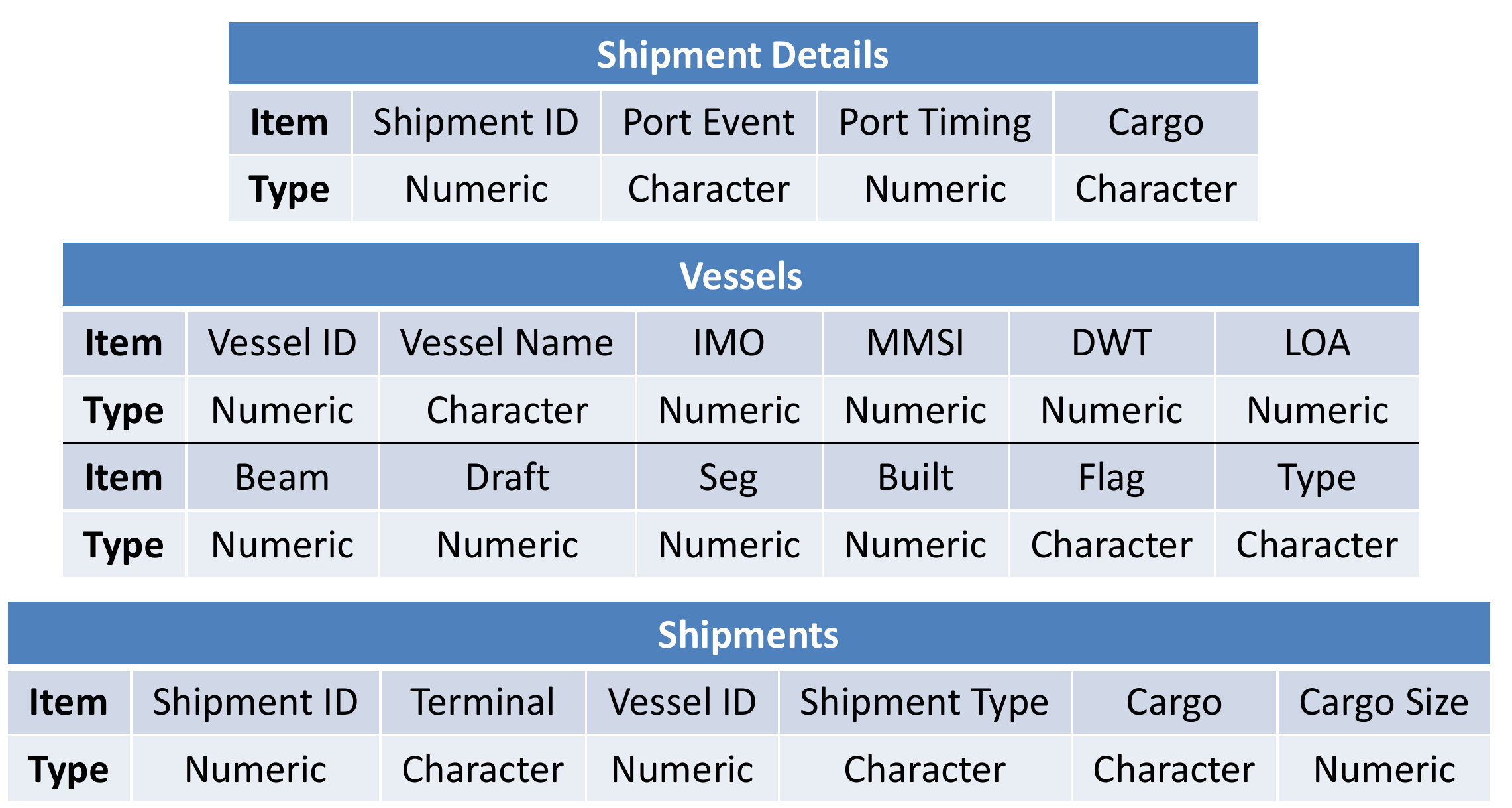} 
		\caption{Data Standardization}
		\label{Fig:Data_Standardization}
	\end{figure}

Data standardization is one of most necessary pre-processes for data cleaning and later predictive modeling. Without standardized datasets, the follow-ups are difficult to proceed. After raw operational datasets and domain knowledge inputs have been provided by industrial collaborators, a standardized operational data format has been formed up as shown in Figure \ref{Fig:Data_Standardization}. Due to different nomenclatures and terms during port event processes among different terminals, some key port events and key cargoes have also been standardized to avoid mis-understanding as shown in Table \ref{Table:Key_Port_Event} and Table \ref{Table:Key_Cargo}.

	\begin{table}[htbp]
		\caption{Key Operation Events at Berth}
		\label{Table:Key_Port_Event}
		\begin{center}
			\begin{adjustbox}{max width=0.6\linewidth}
				\begin{tabular}{ccc}
					\hline
					Event & Shipment Type & Event Type \\
					\hline
					All Fast & Loading/Discharging & General \\
					Surveyor On Board & Loading/Discharging & General \\
					Commence Safety Meeting & Loading/Discharging & General \\
					Complete Safety Meeting & Loading/Discharging & General \\
					Commence Ullage Before Operation & Loading/Discharging & General \\
					Complete Ullage Before Operation & Loading/Discharging & General \\
					Commence Sampling & Discharging & Cargo \\
					Sample Pass & Discharging & Cargo \\
					Cargo Arm Connected & Loading/Discharging & Cargo \\
					Commence Operation & Loading/Discharging & Cargo \\
					Complete Operation & Loading/Discharging & Cargo \\
					Commence Tank Inspection & Loading/Discharging & General \\
					Complete Tank Inspection & Loading/Discharging & General \\
					Cargo Arm Disconnected & Loading/Discharging & Cargo \\
					Commence Shifting & Loading/Discharging & General \\
					Complete Shifting & Loading/Discharging & General \\
					Marpol Prewash Arm Connected & Discharging & Cargo \\
					Commence Prewash & Discharging & Cargo \\
					Complete Prewash & Discharging & Cargo \\
					Commence Stripping to Shore & Discharging & Cargo \\
					Complete Stripping to Shore & Discharging & Cargo \\
					Marpol Prewash Arm Disconnected & Discharging & Cargo \\
					\hline        
				\end{tabular}
			\end{adjustbox}
		\end{center}
	\end{table}

	\begin{table}[htbp]
		\caption{Key Cargoes}
		\label{Table:Key_Cargo}
		\begin{center}
			\begin{adjustbox}{max width=0.7\linewidth}
				\begin{tabular}{cl}
					\hline
					Cargo Group & Cargo Name \\
					\hline
					G1 & BASE OIL (e.g. 70N, 100N, 150N, 400N, 500N, 600N, 2500N) \\
					G2 & EHC (e.g. EHC 50/110) \\
					\hline        
				\end{tabular}
			\end{adjustbox}
		\end{center}
	\end{table}

\subsubsection{Data Cleaning}
Due to manual inputs, human errors are inevitably involved in data recording. Most of error cases happen in cargo information, port event and port timing. 

\begin{itemize}
	\item[(a)] Cargo Information Error: Cargo names are not standardized and mis-leading. For instance, there are multiple cargoes in the group of base oil, such as 150N, 500N, 600N, etc. However, non-standard manual inputs could be 150, N150, 500, etc for similar cargoes. These typo errors propagate in the training processes of predictive models, so they would eventually affect the performance of predictive models.
	
	\item[(b)] Port Event Error: This type of error has significant impacts on predictive modeling development. For example, a port event, such as ``cargo arm disconnected'' for a certain type of cargo, is incorrectly recorded as other follow-up event, such as ``Marpol Prewash Arm Connected''. This leads to different timing of event taken and also affects whole chain of operations, which do not exactly reflect actual cases. This error also propagates to worsen the performance of predictive models.
	
	\item[(c)] Port Timing Error: Similar to the prior error, this error is mainly due to two categories. On one hand, the port timing is correctly associated with port event, but the timing is incorrectly recorded by human, such as recorded ``2018-01-10 10:00:00'' as ``2018-10-01 10:00:00''. On the other hand, the port timing is not the one associated with the true designated port event, such as a port timing ``2018-01-10 10:00:00'' associated to a false port event ``Commence Operation'' where the true port event is ``Commence Tank Inspection''.
\end{itemize}

For the above observed types of errors, most of errors can be corrected by investigating the whole chain's timing and events. Very few data with errors that could not be corrected is discarded in this case for the sake of good data quality and fidelity in predictive modeling.

\subsection{Predictive models for Cargo Operation}
As cargo operation is one of key operation events at berth and the most time consuming part in berth stay, predictive modeling for cargo operation is an essential step to move forward. In the following sections, inputs of predictive models and approaches are described in details.

\subsubsection{Input}
To build up predictive models for cargo operation, the datasets of cargo, terminal/berth and vessel are prepared. The designated period is from ``Commence Operation'' to ``Complete Operation''. The cargo information consists of cargo name, cargo size in terms of metric ton, shipment type (i.e. loading or discharging). Vessel information covers vessel identity, such as vessel name, IMO and MMSI, vessel physical parameters, like length, width, etc. and vessel type and flag. Some inputs may not be fully applied in this study, but their potential values could be explored and adopted for future research.

\subsubsection{Approach}
In this part, predictive modeling techniques for cargo operation are elaborated in details. Given the specific groups of cargoes as shown in Table \ref{Table:Key_Cargo} after sufficient data analyses, each correlation between cargo operation and particular cargo is well linearly associated. Therefore, cargo operation prediction is trained by linear regression, while berth stay prediction will be trained more sophisticatedly and demonstrated in later sections. 

For linear regression, one of most straight forward and effective methods is least squares method. Given inputs of cargo size under a specific type of cargo, $\boldsymbol{X_{0}}$ and target outputs of cargo operation, $\boldsymbol{Y}$, and the predictive model is set to be $y_{i} = ax_{i} + b$, where $a, b$ are the parameters to be determined for this model.

\begin{align}
	\boldsymbol{X_{0}} = 
	\begin{bmatrix}
		x_1 \\
		x_2 \\
		\vdots \\
		x_{m}
	\end{bmatrix},
	~
	\boldsymbol{Y} = 
	\begin{bmatrix}
		y_1 \\
		y_2 \\
		\vdots \\
		y_{m}
	\end{bmatrix}
\end{align}

A simple manipulation over $\boldsymbol{X_{0}}$ for calculating constant parameter $b$ of the model is shown as follows:

\begin{align}
	\boldsymbol{X_{1}} = 
	\begin{bmatrix}
		1 & x_1 \\
		1 & x_2 \\
		\vdots & \vdots \\
		1 & x_{m}
	\end{bmatrix}
\end{align}

The parameters ($a, b$) of the model can be thus determined for the specific cargo type as follows, and predictive models of cargo operation are sub-models for the next step of berth stay prediction:

\begin{align}
	\begin{bmatrix}
		b \\
		a
	\end{bmatrix} = (\boldsymbol{X_{1}}^{T} \boldsymbol{X_{1}})^{-1} \boldsymbol{X_{1}}^{T} \boldsymbol{Y}
\end{align}

\subsection{Predictive models for Berth Stay}
Based on detailed operational data analyses in tanker terminals, other relevant components contributing to berth stay are also incorporated besides cargo operation as mentioned in the previous section. After analyzing the detailed operational data, the whole berth stay is divided into several blocks from ``All Fast'' to ``Cargo Arm Disconnected/Marpol Prewash Arm Disconnected'' for a specific berth stay.

There are 12 processing blocks identified in this study, namely (a) sampling block; (b) tank inspection block; (c) safety meeting block; (d) ullage before operation (UBO) block; (e) cargo arm connection block; (f) cargo operation block; (g) cargo arm disconnection block; (h) shifting block; (i) prewash arm connection block; (j) prewash block; (k) stripping block; (l) prewash arm disconnection block. It is important to note that it is not necessary to have all identified blocks present for each individual berth stay.

\subsubsection{Input}
In this section, various blocks and their corresponding inputs are defined based on the detailed operational data analyses.

\begin{itemize}
	\item[(a)] \underline{Sampling block}: This block models different types of sampling, such as sampling different cargo and shipment types at anchorage or at berth, from ``Commence Sampling'' to ``Sample Pass'' in the key port events.
	
	\item[(b)] \underline{Tank inspection block}: This block models different shipment types, such as loading or discharging, from ``Commence Tank Inspection'' to `` Complete Tank Inspection'' in the key port events.
	
	\item[(c)] \underline{Safety meeting block}: This block models different shipment types, such as loading or discharging, from ``Commence Safety Meeting'' to `` Complete Safety Meeting'' in the key port events.
	
	\item[(d)] \underline{Ullage before operation (UBO) block}: This block models different shipment types, such as loading or discharging, from ``Commence Ullage Before Operation'' to `` Complete Ullage Before Operation'' in the key port events.
	
	\item[(e)] \underline{Cargo arm connection block}: This block models different shipment types, such as loading or discharging, and different cargo arms, such as first arm or following arm, from ``Complete Ullage Before Operation'' to ``Cargo Arm Connected'' or from previous ``Cargo Arm Disconnected'' to next ``Cargo Arm Connected'' in the key port events.
	
	\item[(f)] \underline{Cargo operation block}: This block is basically deliberated in the previous section. However, it is slightly different from the previous one. In this block, multi-cargo operations can be also evaluated in different operation modes, such as sequential operations or concurrent operations. Whereas the previous section purely focuses on one cargo operation, which is an significant input for this model. This block models different shipment types, such as loading or discharging, and different cargo types, such as G1 and G2 in Appendix \ref{Appdx:Key_Cargo}, from ``Commence Operation'' to ``Complete Operation'' in the key port events.
	
	\item[(g)] \underline{Cargo arm disconnection block}: This block models different shipment types, such as loading or discharging, from very last ``Complete Operation'' to ``Cargo Arm Disconnected'' in the key port events.
	
	\item[(h)] \underline{Shifting block}: This block models different shipment types, such as loading or discharging, from ``Commence Shifting'' to ``Complete Shifting'' in the key port events. According to data analyses, shifting block may not be a necessary process for all. This block is required when cargo arms or prewash arms are not properly aligned for operations between vessel and shore.
	
	\item[(i)] \underline{Prewash arm connection block}: This block models certain cargo discharging operations from very last ``Cargo Arm Disconnected'' to ``Marpol Prewash Arm Connected'' in the key port events.
	
	\item[(j)] \underline{Prewash block}: This block models certain cargo discharging operations from ``Commence Prewash'' to ``Complete Prewash'' in the key port events.
	
	\item[(k)] \underline{Stripping block}: This block models certain cargo discharging operations from ``Commence Stripping to Shore'' to ``Complete Stripping to Shore'' in the key port events. For multi-cargo operations, it is highly likely to have prewash and stripping to shore on different arm hoses at the same moment. 
	
	\item[(l)] \underline{Prewash arm disconnection block}: This block models certain cargo discharging operations from ``Complete Stripping to Shore'' to ``Marpol Prewash Arm Disconnected'' in the key port events. 
	
\end{itemize}

\subsubsection{Approach}
In this section, approaches to model these blocks for berth stay prediction are illustrated.

\begin{itemize}
	\item[(a)] Linear Regression: The process block ``cargo operation block'' is equivalent to the previous section of predictive model for cargo operation, and it is trained by linear regression as mentioned in the previous section in details in unit of hour.
	
	\begin{align}
		T_{(f)} = F_{(f)}(a, b, x)
	\end{align}

	\begin{figure}[htbp]
		\centering
		\begin{minipage}{0.8\textwidth}
			\begin{algorithm}[H]
				\caption{Multi-Decomposed Gaussian Sampling (MDGS)}
				\label{Alg:Pseudocode_MDGS}
				\begin{algorithmic}
					\STATE Given a desired pdf with $n$ peaks, denoted as $f(x)$
					\STATE Initialization: $\mathbf{N} = \mathcal{N}_{i} \sim (\mu_{i}, \sigma_{i})$ for $i=1, 2, ..., n$
					
					\STATE Input: $f(x)$, $lb$, $ub$, $s$
					\STATE Output: $\mathbf{N} = \mathcal{N}_{i} \sim (\mu_{i}, \sigma_{i})$ for $i=1, 2, ..., n$
					\WHILE{TRUE}
					\STATE $S(s, f(x))$ $\leftarrow$ Generate $s$ samples based on $f(x)$ 
					\STATE $\hat{S}(s, \mathbf{N})$ $\leftarrow$ Generate $s$ samples based on $\mathbf{N}$
					\STATE $S^{'}, \hat{S}^{'}$ $\leftarrow$ Trim samples by boundaries $lb$, $ub$
					\STATE $D_c, D, p$ $\leftarrow$ KS-Test($S^{'}, \hat{S}^{'}$)
					
					\IF{$D_c \ge D$ and $p \ge 0.05$}
					\STATE Break while-loop

					\ELSE
					\STATE	$\mathcal{N}_{i}^{'} \sim (\mu_{i}^{'}, \sigma_{i}^{'})$  $\leftarrow$ Refine $\mathcal{N}_{i} \sim (\mu_{i}, \sigma_{i})$ for $i=1, 2, ..., n$
					
					\ENDIF
					\ENDWHILE
					
					\RETURN  $\mathbf{N}$
					
				\end{algorithmic}
			\end{algorithm}
		\end{minipage}
	\end{figure}
	
	\item[(b)] Multi-Decomposed Gaussian Sampling (MDGS): Due to variations of different operational backgrounds, this approach is applicable to many different blocks, which are sampling, tanker inspection, ullage before operation (UBO), cargo arm connection, cargo arm disconnection, prewash arm connection and prewash arm disconnection.
	
	A generic way of this approach is by decomposing a desired distribution into multiple Gaussian distributions, passing Kolmogorov–Smirnov test (KS-test) and conducting random samplings in the end. The pseudo-code of this approach is illustrated in Algorithm \ref{Alg:Pseudocode_MDGS}.

%
	
	Based on the approach illustrated, various blocks as described are formulated as follows, where $\boldsymbol{\mu}=\{\mu_{1}, \mu_{2}, ...... , \mu_{n} \}$, $\boldsymbol{\sigma}=\{\sigma_{1}, \sigma_{2}, ...... , \sigma_{n} \}$, $lb$ is lower bound and $ub$ is upper bound for a particular block in unit of hour:

	\begin{align}
		T_{(a)} = \mathcal{N}_{(a)}(\boldsymbol{\mu, \sigma}, lb, ub)_{a}
	\end{align}

	\begin{align}
		T_{(b)} = \mathcal{N}_{(b)}(\boldsymbol{\mu, \sigma}, lb, ub)_{b}
	\end{align}

	\begin{align}
		T_{(d)} = \mathcal{N}_{(d)}(\boldsymbol{\mu, \sigma}, lb, ub)_{d}
	\end{align}

	\begin{align}
		T_{(e)} = \mathcal{N}_{(e)}(\boldsymbol{\mu, \sigma}, lb, ub)_{e}
	\end{align}

	\begin{align}
		T_{(g)} = \mathcal{N}_{(g)}(\boldsymbol{\mu, \sigma}, lb, ub)_{g}
	\end{align}

	\begin{align}
		T_{(i)} = \mathcal{N}_{(i)}(\boldsymbol{\mu, \sigma}, lb, ub)_{i}
	\end{align}

	\begin{align}
		T_{(l)} = \mathcal{N}_{(l)}(\boldsymbol{\mu, \sigma}, lb, ub)_{l}
	\end{align}

	\item[(c)] Proportional Constant: This approach is applicable to blocks that are shifting, prewash and stripping. Based on historical data, these blocks are closely related to the number of cargoes that are operated, and the relationship is proportional to the number of cargoes to be handled, as $C_{h}$, $C_{j}$ and $C_{k}$ are respective unitary timings in unit of hour.
	
	\begin{align}
		T_{(h)} = n \times C_{h}
	\end{align}

	\begin{align}
		T_{(j)} = n \times C_{j}
	\end{align}

	\begin{align}
		T_{(k)} = n \times C_{k}
	\end{align}

	\item[(d)] Fixed Constant: Based on historical data, safety meeting is a fixed period of time around 1 hour in general.
	
	\begin{align}
		T_{(c)} = 1
	\end{align}
	
\end{itemize}

\subsubsection{Berth Stay Scenario}
In this study, four scenarios are covered for berth stay prediction. The scenarios are designed to accommodate the uncertainty under tanker cargo operations. Based on the most concerning turning-points of tanker cargo operations from domain experts, the scenarios are selected and defined as follows: (1) All Fast, (2) Sample Pass, (3) Commence Operation and (4) Complete Operation. Based on these four scenarios, it is also clear to indicate real-time prediction capability in some extents if live and dynamic information could be fed into the system. The results and discussion are illustrated in the following sections.

\section{Result and Discussion}
Firstly, preliminary data statistics are demonstrated, and corresponding results of predictive models and aforementioned scenarios are presented and discussed in this section.

\subsection{Terminal Operation Data Analysis}
\begin{figure}[htbp]
	\centering
	\includegraphics[width=\textwidth]{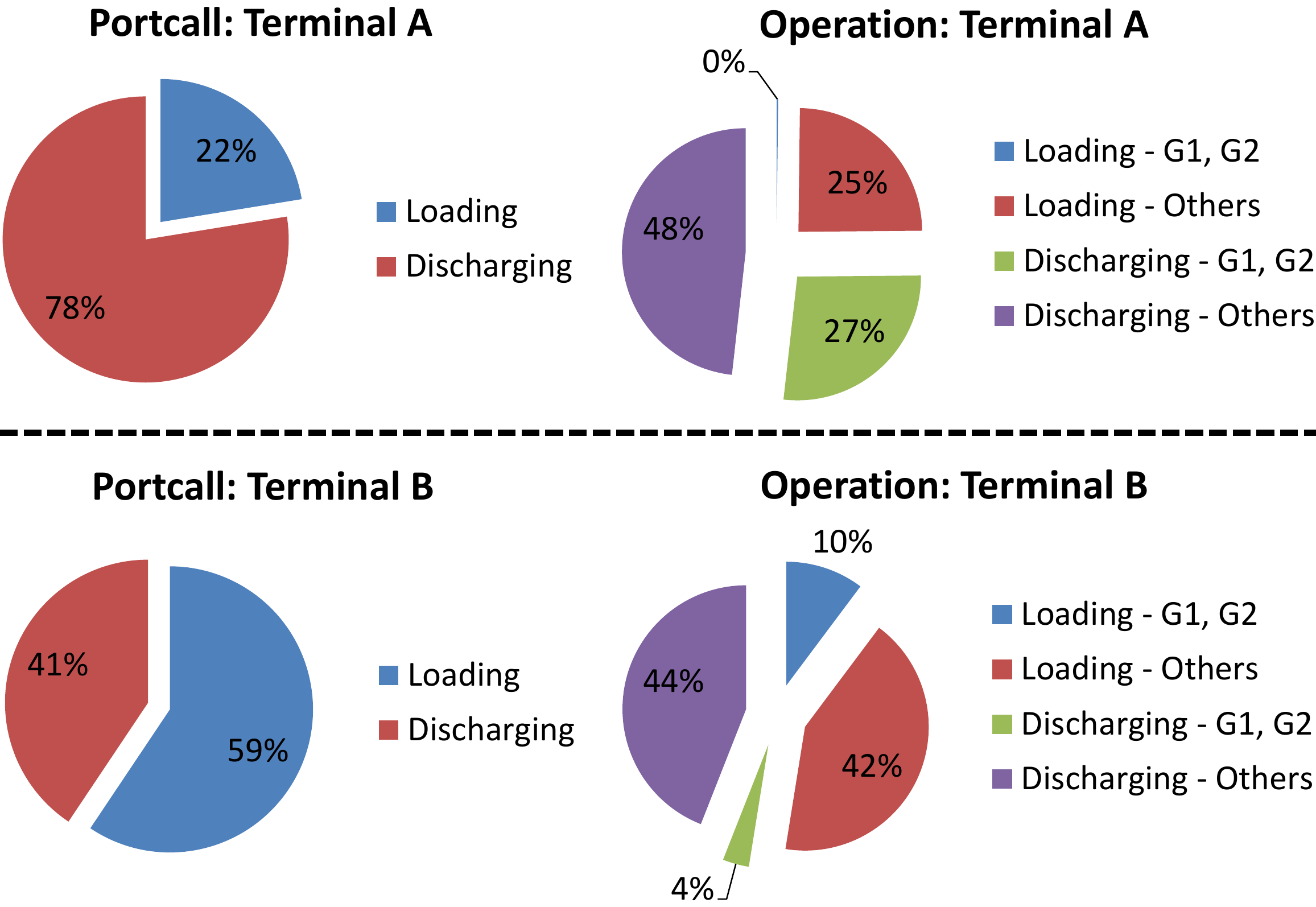} 
	\caption{Data Statistics}
	\label{Fig:Data_Statistics}
\end{figure}

In this section, preliminary data statistics are illustrated. The data comes from two different terminals, for simplicity, namely Terminal A and B. Since the data is from different parties, their formats and coverage are also quite different from each other, therefore, certain parts of information may not be known for these terminals. The statistics are shown in Table \ref{Tab:Data_Statistics} and Figure \ref{Fig:Data_Statistics} as follows:

\begin{table}[htbp]
	\begin{adjustbox}{width=0.65\textwidth, center}
	\centering
	\begin{tabular}{l|cc}
		\hline
		\textbf{Item}        & \textbf{Terminal A}         & \textbf{Terminal B}         \\
		\hline
		\textbf{Period}      & 2018-01 to 2020-08 & 2017-01 to 2018-12 \\
		\hdashline
		\textbf{Portcall}   & 254                & 559                \\
		- Loading (All)     & 57                 & 332              \\
		- Discharging (All) & 197                & 227              \\
		
		- Loading (G1, G2)  & 1 & 62 \\
		\quad - Single  & 1 & 36 \\
		\quad - Multiple  & 0 & 26 \\
		\quad\quad - Sequential  & 0 & 2 \\
		\quad\quad - Concurrent  & 0 & 24 \\
		
		- Discharging (G1, G2)  & 77 &  28 \\
		\quad - Single  & 28 & 26 \\
		\quad - Multiple  & 49 & 2 \\
		\quad\quad - Sequential  & 49 & 0 \\
		\quad\quad - Concurrent  & 0 & 2 \\
		
		\hdashline
		\textbf{Operation}  & 535                & 988                \\
		- Loading (All)     & 133 (1/132)$^{\dagger}$      & 519 (101/418)$^{\dagger}$  \\
		- Discharging (All) & 402 (144/258)$^{\dagger}$     & 469 (34/435)$^{\dagger}$  \\
		
		- Loading (G1) & 0 & 52 \\
		\quad - Single  & 0 & 21 \\
		\quad - Multiple  & 0 & 31 \\
		\quad\quad - Sequential  & 0 & 0 \\
		\quad\quad - Concurrent  & 0 & 31 \\
		
		- Loading (G2) & 1 & 49 \\
		\quad - Single  & 1 & 15 \\
		\quad - Multiple  & 0 & 34 \\
		\quad\quad - Sequential  & 0 & 4 \\
		\quad\quad - Concurrent  & 0 & 30 \\
		
		- Discharging (G1)  & 99  & 19 \\
		\quad - Single  & 25 & 14 \\
		\quad - Multiple  & 74 & 5 \\
		\quad\quad - Sequential  & 74 & 0 \\
		\quad\quad - Concurrent  & 0 & 5 \\
		
		- Discharging (G2)  & 41   & 15 \\
		\quad - Single  & 3 & 12 \\
		\quad - Multiple  & 38 & 3 \\
		\quad\quad - Sequential  & 38 & 0 \\
		\quad\quad - Concurrent  & 0 & 3 \\
		
		\hline
	\end{tabular}
	\end{adjustbox}
	\newline
	\centering $\dagger$: Bracket numbers are focused cargoes and the other cargoes.
	
	\caption{Detailed Data Statistics}
	\label{Tab:Data_Statistics}
	
\end{table}

Based on the table and figure of data statistics, the operation modes of two terminals are distinctively different. Terminal A is much less frequently dealing with portcalls and operations than Terminal B. In other words, Terminal B is much busier than Terminal A for vessel berthing. Besides this difference, Terminal A mostly takes discharging operations, while Terminal B behaves oppositely.

Moreover, by zooming into different cargoes, the interested cargoes that belongs to G1 and G2 are 27\% occurring in discharging operations at Terminal A, while Terminal B only has 14\% in loading and discharging combined operations. Their distinctive behaviors are well pictured and differentiated through their data statistics. In addition, Terminal A captures key port events as detailed as possible, while Terminal B only covers four key port events that are All Fast, Commence Operation, Complete Operation and Cargo Arm Disconnected as shown in Table \ref{Table:Key_Port_Event}.

\subsection{Tanker Cargo Operation Prediction Modelling}
As the data analysis demonstrates above, the modelling results of cargo operation prediction are validated and verified by different terminals, including Terminal A and Terminal B, among the focused cargoes in G1 and G2 as listed in Table \ref{Table:Key_Cargo}. Due to different procedures, personnel, equipment, berth condition and etc. of two different terminals,  the results can effectively demonstrate that the proposed approach is able to be applied to heterogeneous characteristics of terminals. It also presents the universal capability of the approach.

\subsubsection{Terminal A}
As the data statistics show above, Terminal A mostly have discharging operations, while only one operation happens in loading operation on the focused cargoes. Therefore, the cargo operation prediction is merely on discharging operations for this terminal.

\begin{figure}[htbp]
	\centering
	\subfloat[G1]{
		\label{SubFig:Distribution_Terminal_A_G1_Discharging}
		\includegraphics[width=0.4\textwidth]{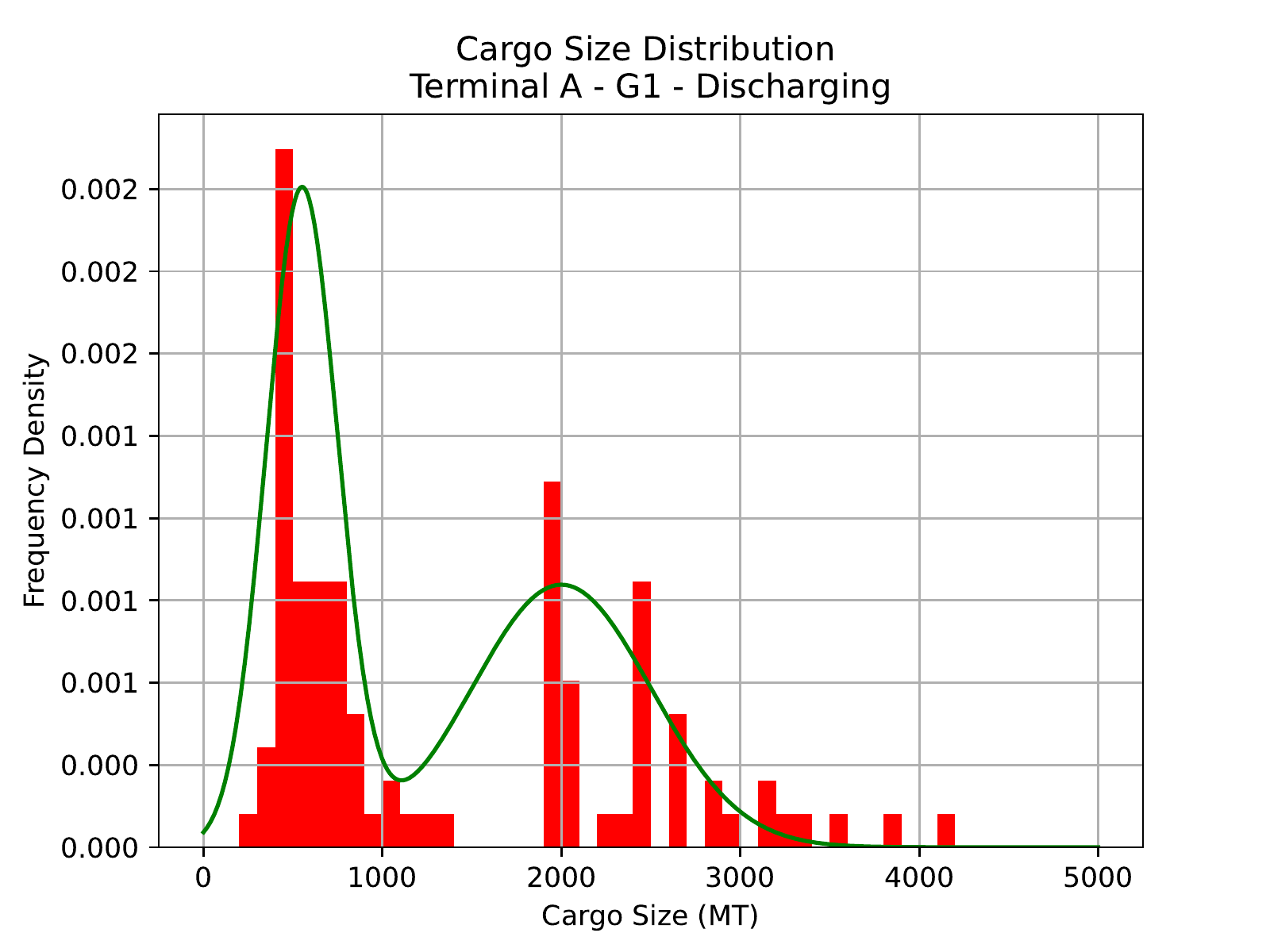} }~
	\subfloat[G2]{
		\label{SubFig:Distribution_Terminal_A_G2_Discharging}
		\includegraphics[width=0.4\textwidth]{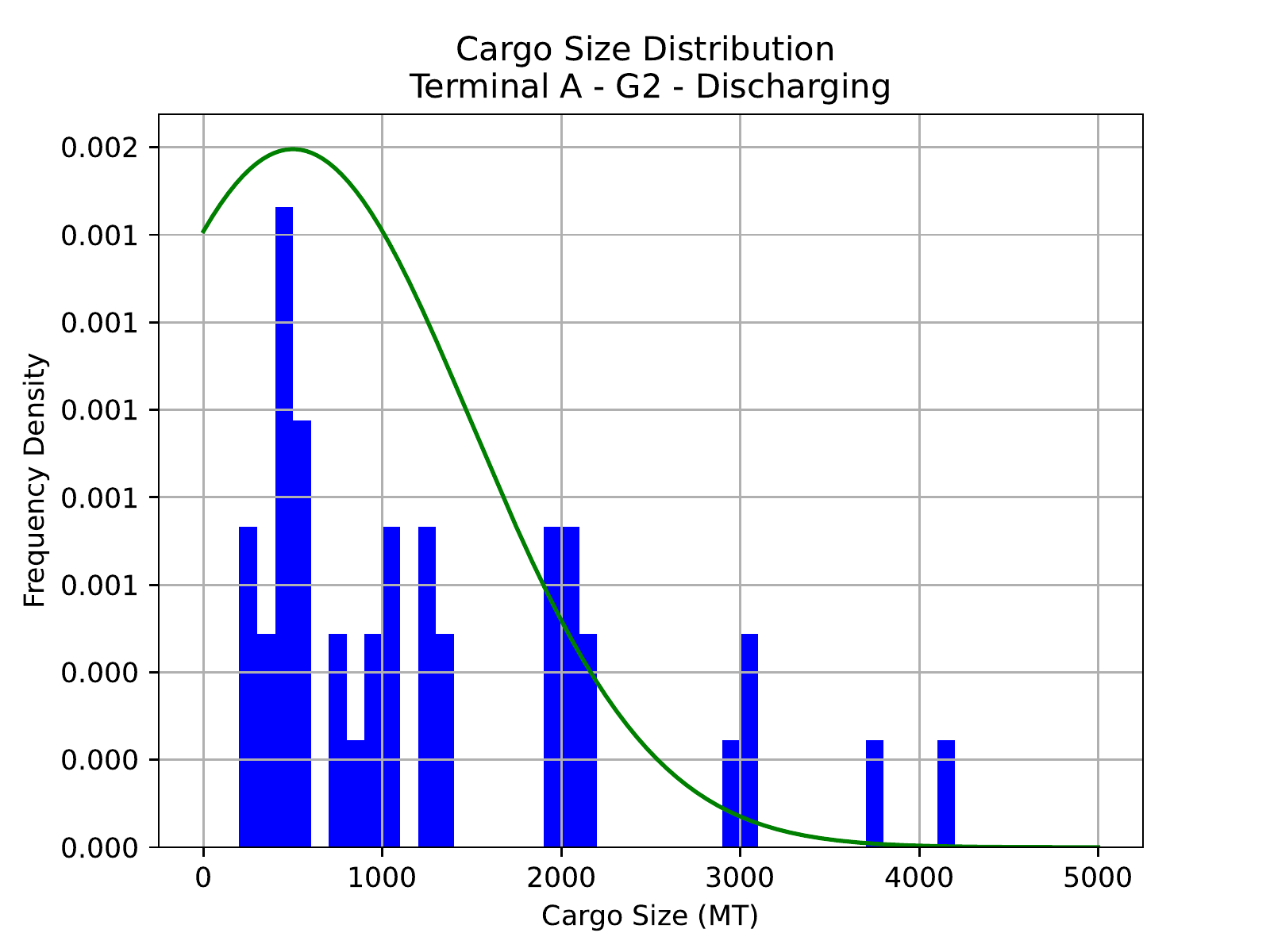} } 
	
	\caption{Size Distributions of Focused Cargoes in Terminal A - Discharging}
	\label{Fig:Distribution_Terminal_A_Focused_Discharging}
\end{figure}

\begin{figure}[htbp]
	\centering
	\subfloat[G1]{
		\label{SubFig:Operation_Stay_Terminal_A_G1_Discharging}
		\includegraphics[width=0.4\textwidth]{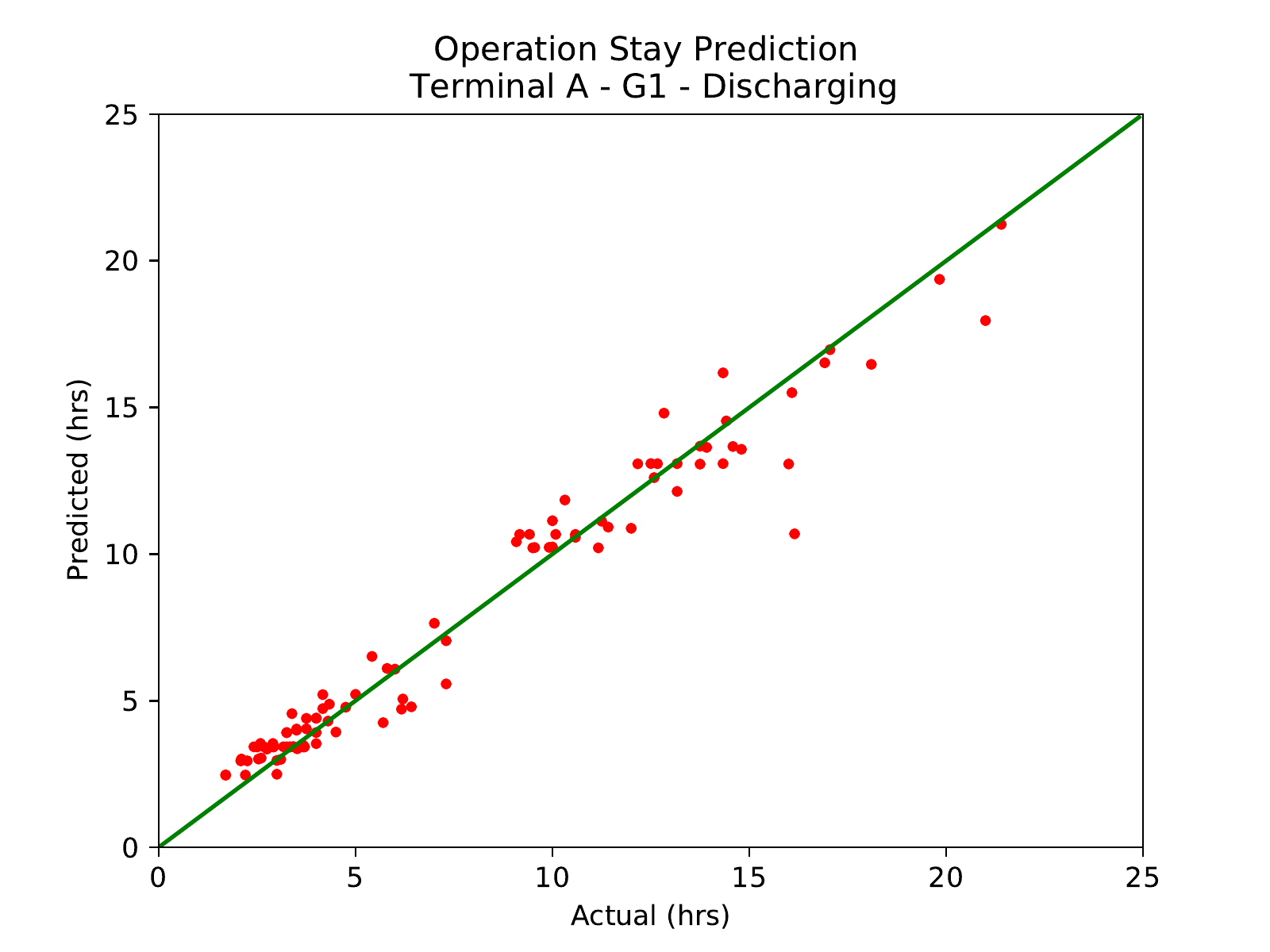} }~
	\subfloat[G2]{
		\label{SubFig:Operation_Stay_Terminal_A_G2_Discharging}
		\includegraphics[width=0.4\textwidth]{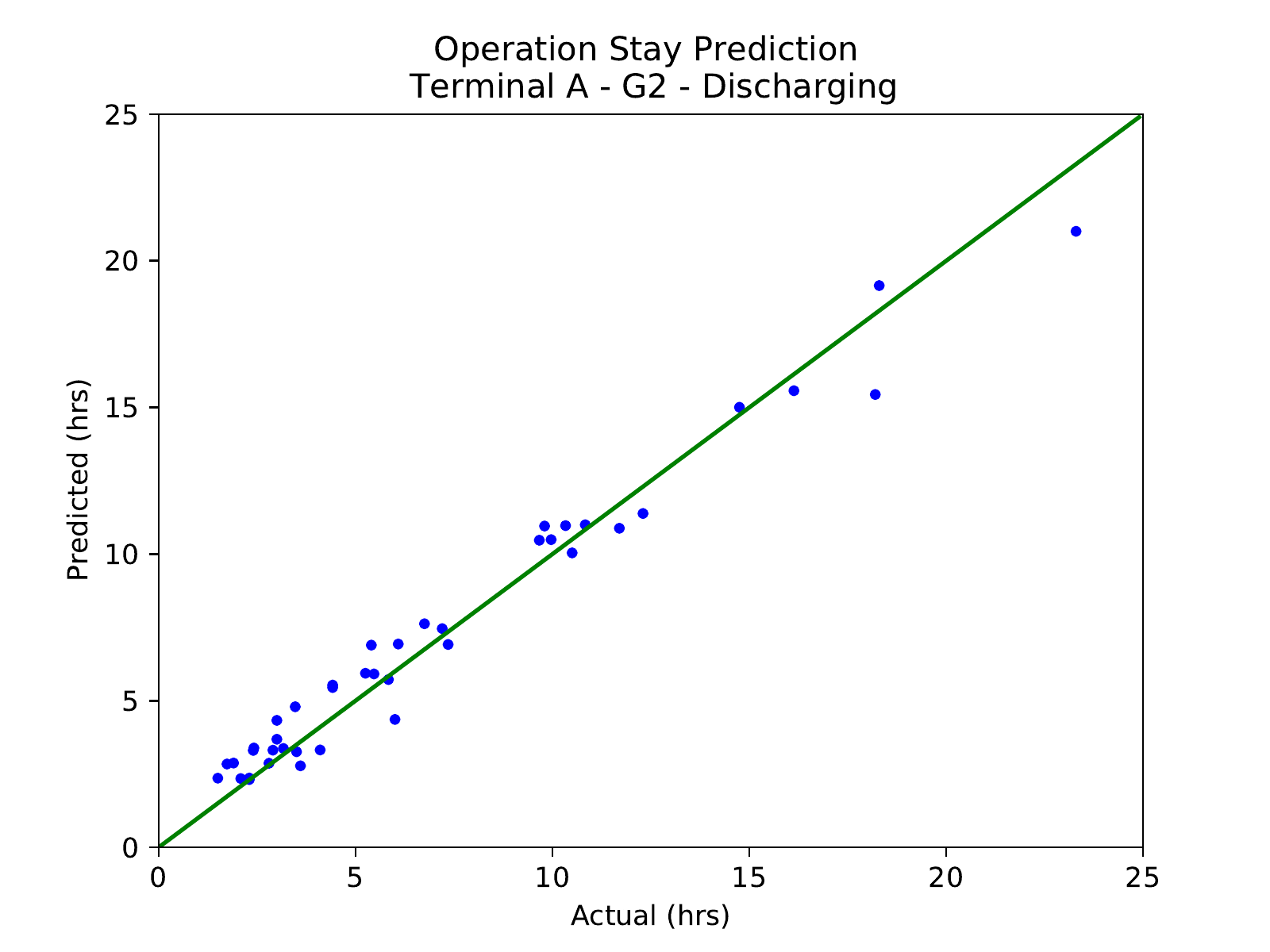} } 
	
	\caption{Cargo Operation Predictions of Focused Cargoes in Terminal A - Discharging}
	\label{Fig:Operation_Stay_Prediction_Terminal_A_Focused_Discharging}
\end{figure}

The distributions and cargo operation predictions of focused cargoes are presented in Figure \ref{Fig:Distribution_Terminal_A_Focused_Discharging} and Figure \ref{Fig:Operation_Stay_Prediction_Terminal_A_Focused_Discharging}. Based on the prediction results, there are two peaks in G1 distribution, which are around 500 MT and 2000 MT. Similarly, the peak of G2 distribution is only around 500 MT for Terminal A. The predictive models can well predict cargo operation for the focused cargoes in Terminal A. The predictions are more accurate as the points are more aligned with the reference green line. 

\begin{figure}[htbp]
	\centering
	\subfloat[G1]{
		\label{SubFig:Error_Distribution_Terminal_A_G1_Discharging}
		\includegraphics[width=0.4\textwidth]{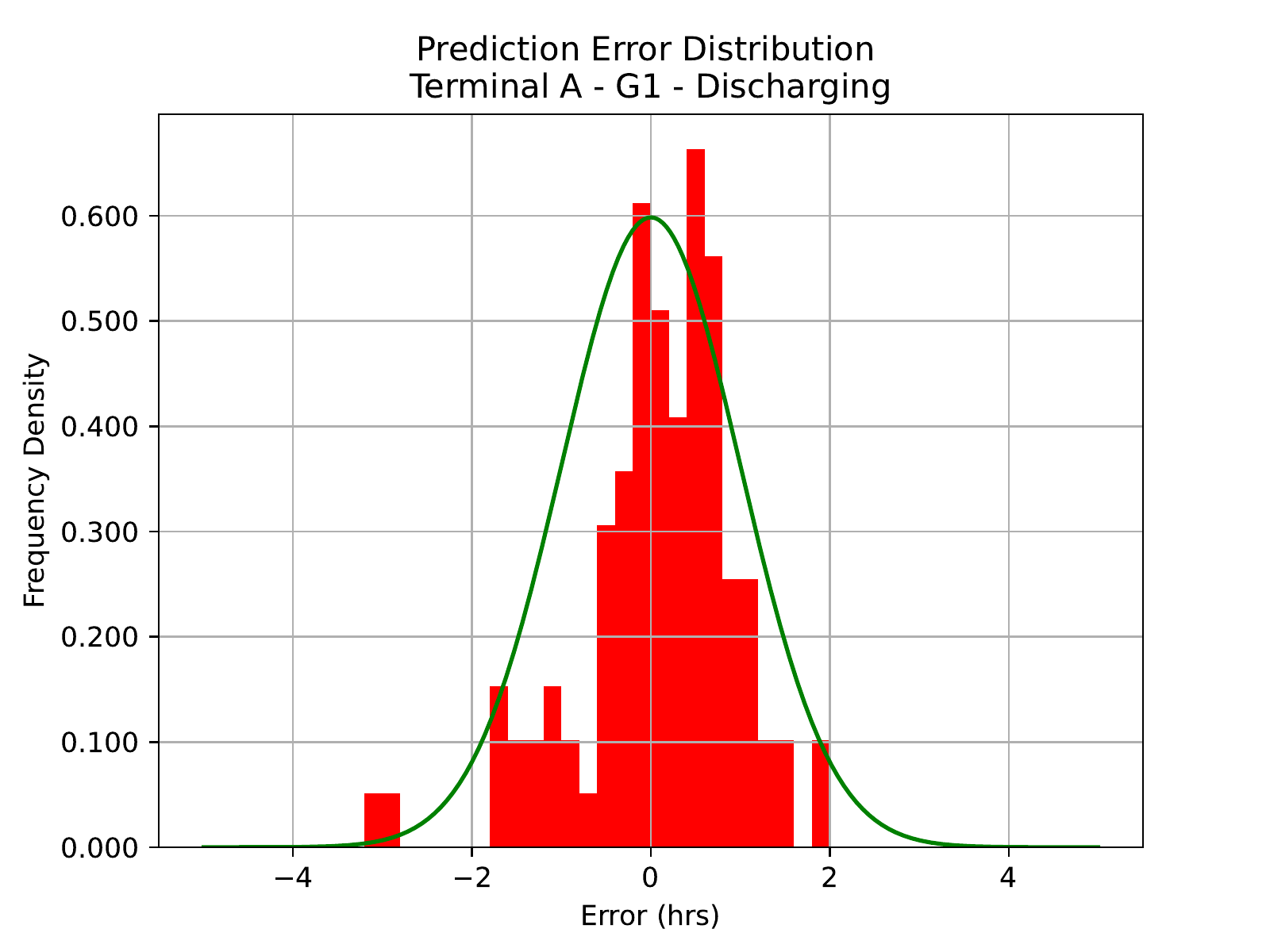} }~
	\subfloat[G2]{
		\label{SubFig:Error_Distribution_Terminal_A_G2_Discharging}
		\includegraphics[width=0.4\textwidth]{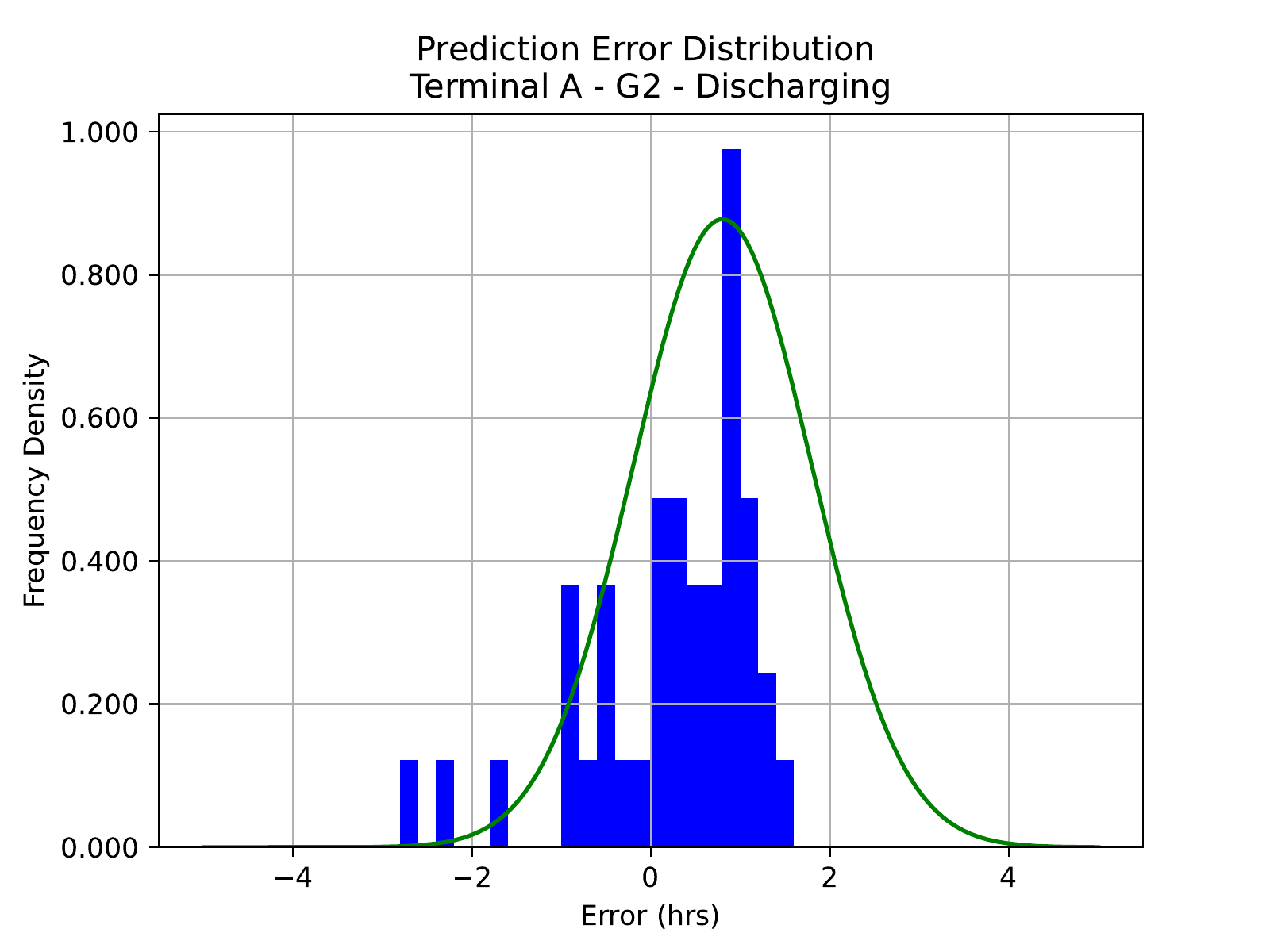} } 
	
	\caption{Prediction Error Distribution of Focused Cargoes in Terminal A - Discharging}
	\label{Fig:Error_Distribution_Terminal_A_Focused_Discharging}
\end{figure}

According to the prediction results of Terminal A, some statistics metrics are introduced and evaluated to quantify the performances of predictive models. The metrics are tabulated in Table \ref{Tab:Metrics_Prediction_Performance_Terminal_A_Discharging}. The prediction error distributions are presented in Figure  \ref{Fig:Error_Distribution_Terminal_A_Focused_Discharging}.

\begin{table}[htbp]
	\centering
	\begin{tabular}{c|ll}
		\hline
		Metric & G1          & G2          \\
		\hline
		$\tilde{\mu}$ & 0.12267333  & 0.4120028   \\
		$\mu$   & 0.028330514 & 0.206535282 \\
		$\sigma$  & 1.040935438 & 0.956142992 \\
		MSE    & 1.073404288 & 0.934568453 \\
		MAE    & 0.713927824 & 0.784176543 \\
		\hline
	\end{tabular}
	\newline
	Note: $\tilde{\mu}$ denotes median values.
	\caption{Metrics of Cargo Operation Prediction Performance in Terminal A - Discharging}
	\label{Tab:Metrics_Prediction_Performance_Terminal_A_Discharging}
\end{table}

\subsubsection{Terminal B}
\begin{figure}[htbp]
	\centering
	\subfloat[G1]{
		\label{SubFig:Distribution_Terminal_B_G1_Loading}
		\includegraphics[width=0.4\textwidth]{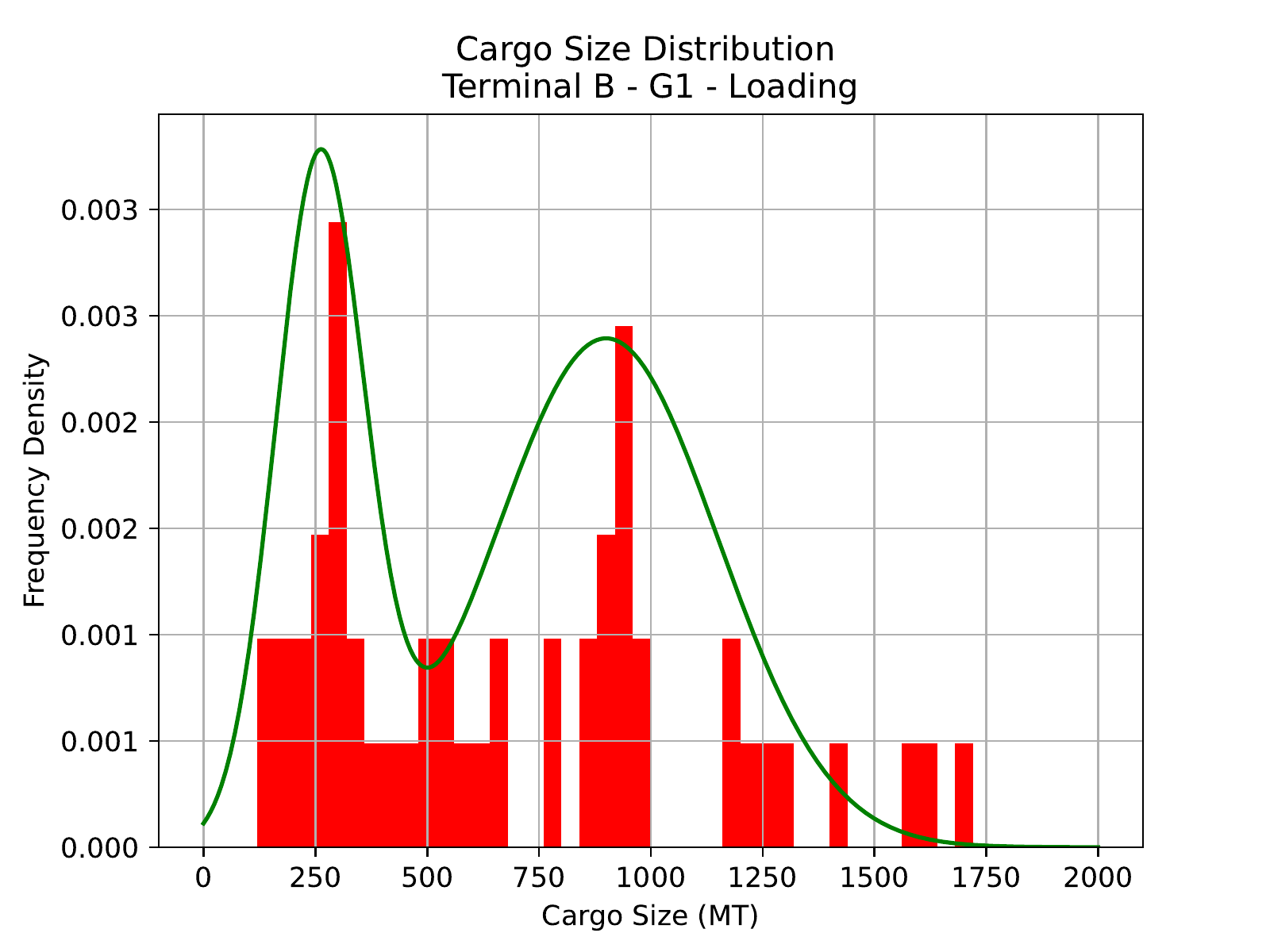} }~
	\subfloat[G2]{
		\label{SubFig:Distribution_Terminal_B_G2_Loading}
		\includegraphics[width=0.4\textwidth]{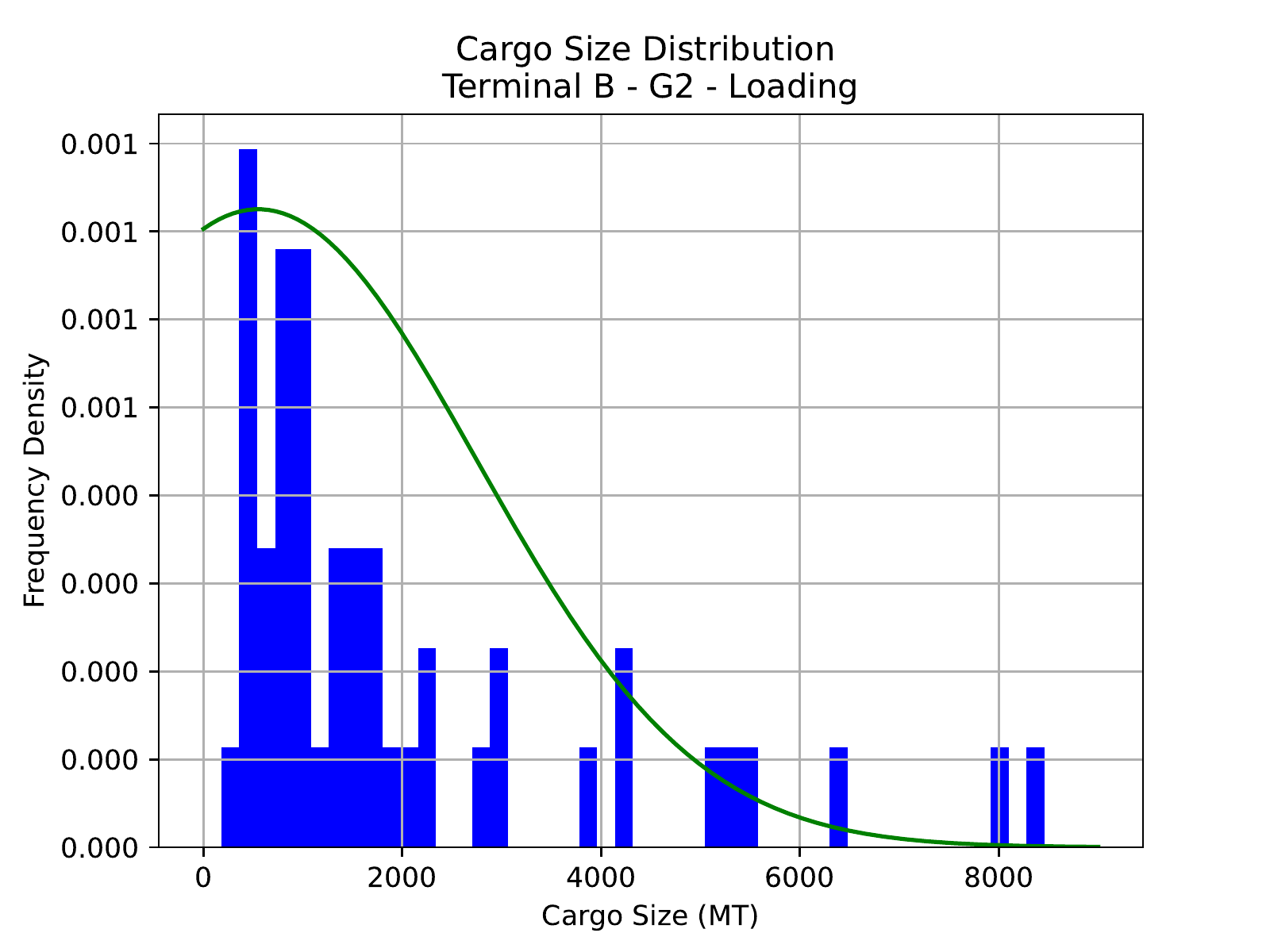} } 
	
	\caption{Size Distributions of Focused Cargoes in Terminal B - Loading}
	\label{Fig:Distribution_Terminal_B_Focused_Loading}
\end{figure}

\begin{figure}[htbp]
	\centering
	\subfloat[G1]{
		\label{SubFig:Operation_Stay_Terminal_B_G1_Loading}
		\includegraphics[width=0.4\textwidth]{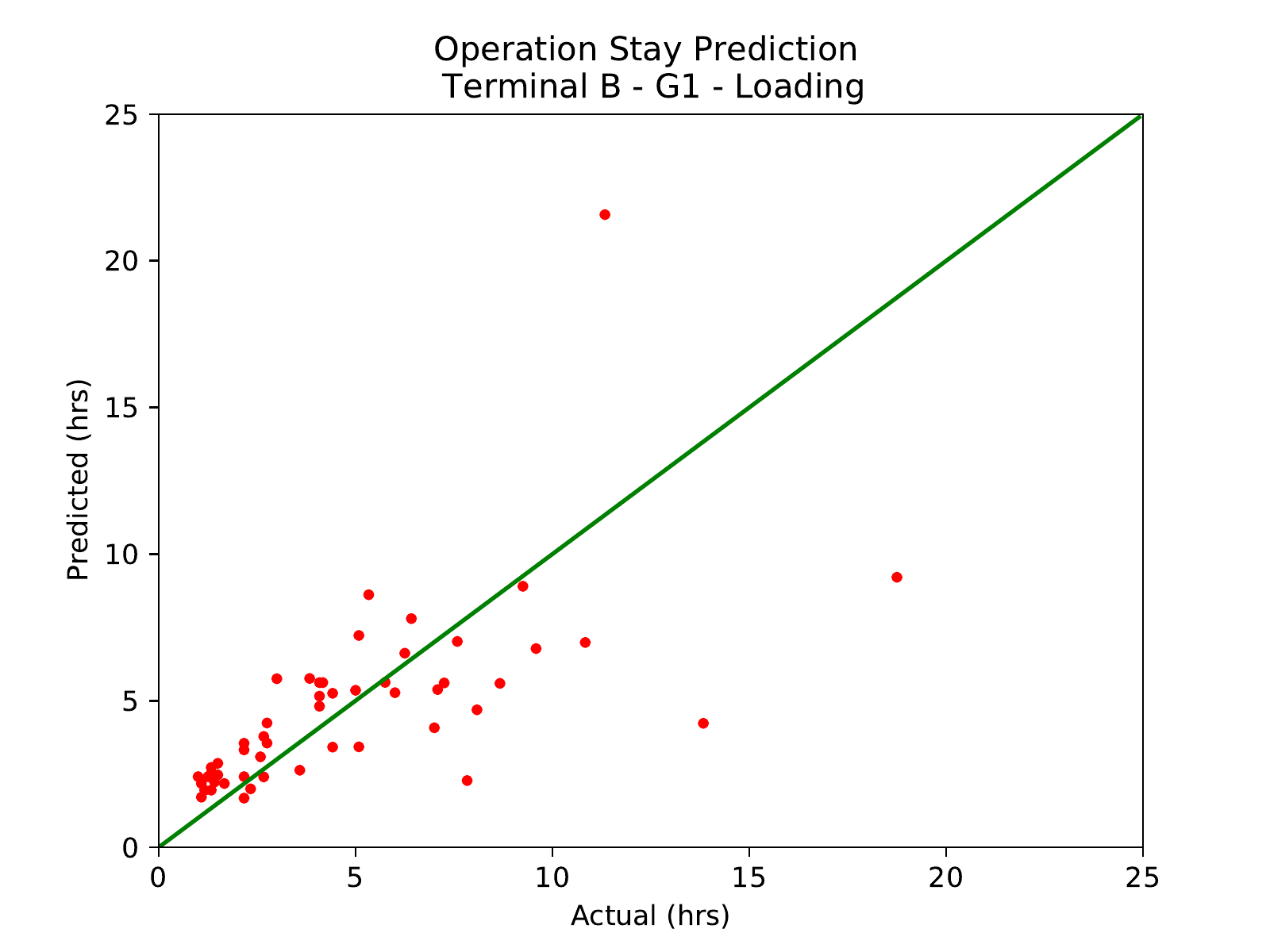} }~
	\subfloat[G2]{
		\label{SubFig:Operation_Stay_Terminal_B_G2_Loading}
		\includegraphics[width=0.4\textwidth]{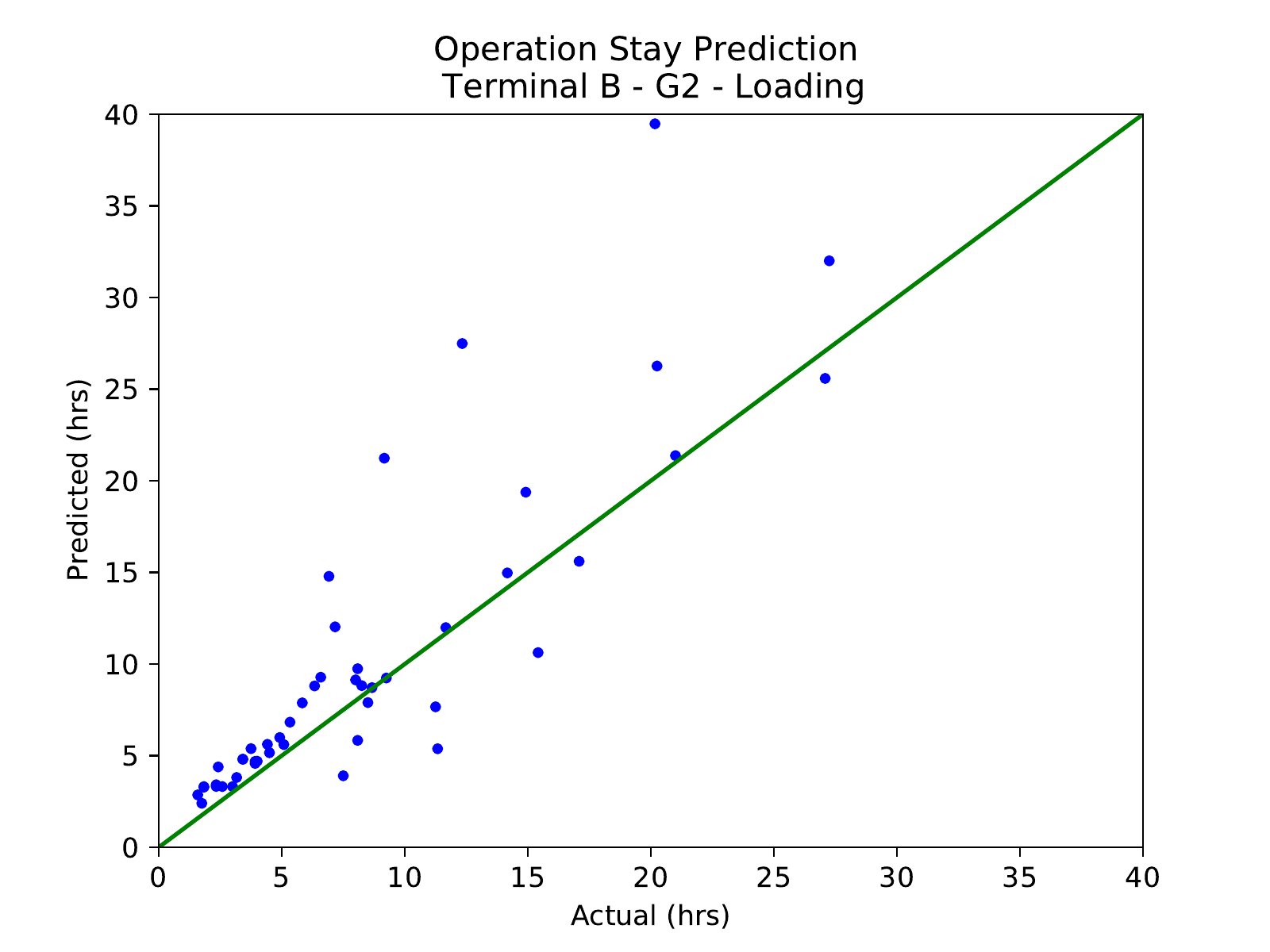} } 
	
	\caption{Cargo Operation Predictions of Focused Cargoes in Terminal B - Loading}
	\label{Fig:Operation_Stay_Prediction_Terminal_B_Focused_Loading}
\end{figure}

\begin{figure}[htbp]
	\centering
	\subfloat[G1]{
		\label{SubFig:Error_Distribution_Terminal_B_G1_Loading}
		\includegraphics[width=0.4\textwidth]{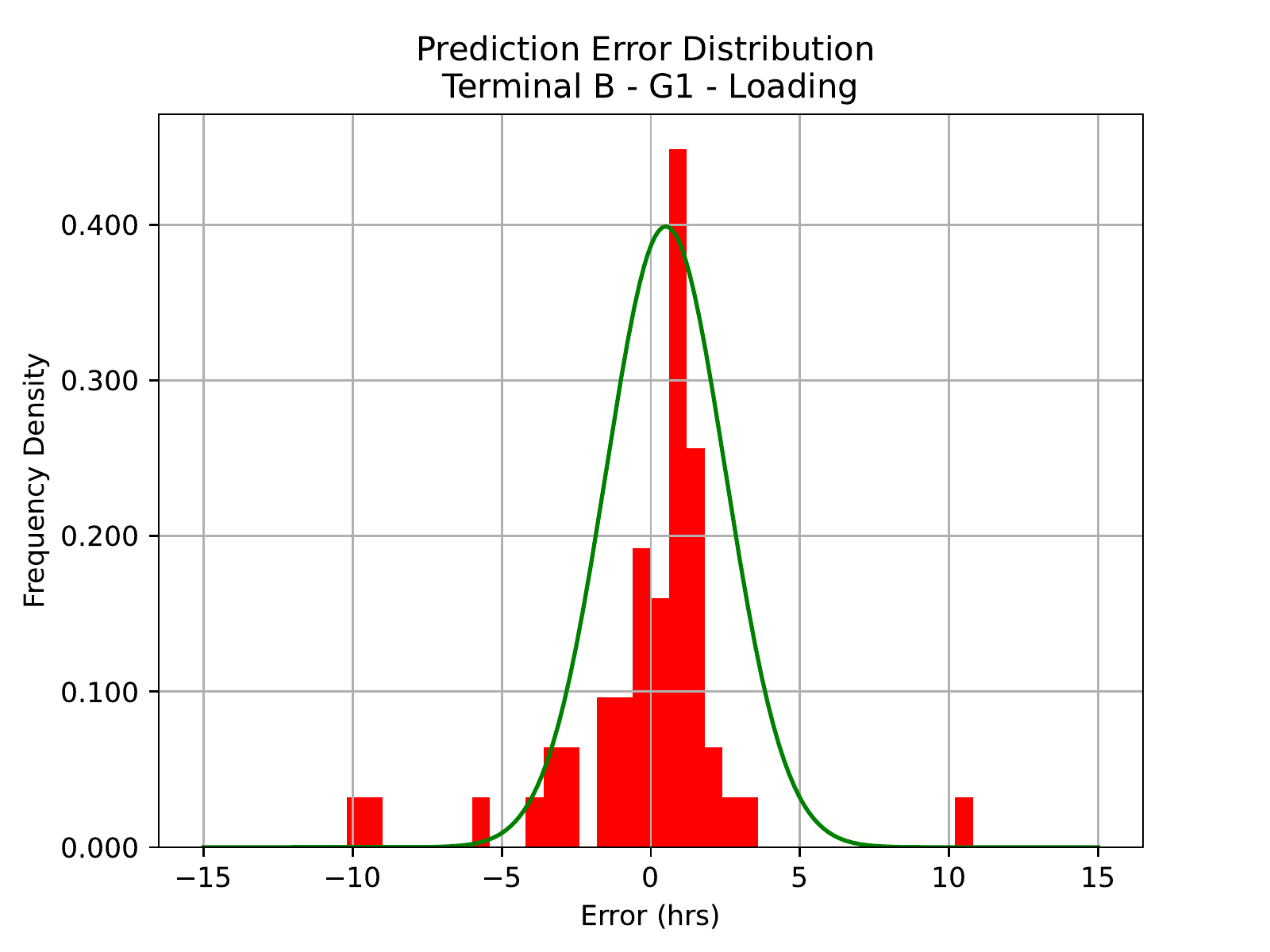} }~
	\subfloat[G2]{
		\label{SubFig:Error_Distribution_Terminal_B_G2_Loading}
		\includegraphics[width=0.4\textwidth]{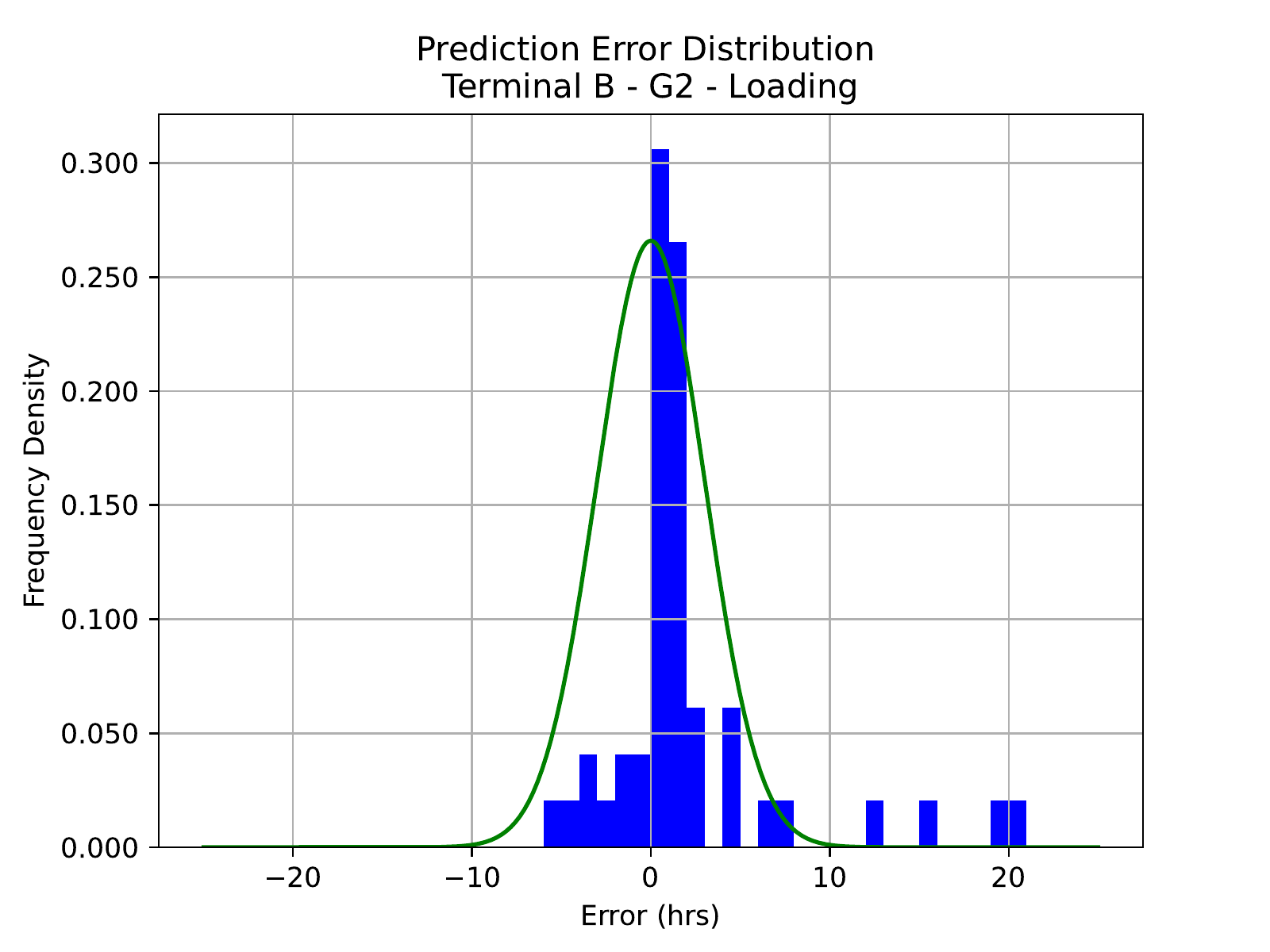} } 
	
	\caption{Prediction Error Distribution of Focused Cargoes in Terminal B - Loading}
	\label{Fig:Error_Distribution_Terminal_B_Focused_Loading}
\end{figure}

Different from Terminal A, both loading and discharging take places on focused cargoes in Terminal B. For loading operations, the distribution and cargo operation prediction of focused cargoes are presented in Figure \ref{Fig:Distribution_Terminal_B_Focused_Loading} and Figure \ref{Fig:Operation_Stay_Prediction_Terminal_B_Focused_Loading}. Based on the prediction results, Two peaks in G1 distribution are around 250 MT and 1000 MT. One peak in G2 distribution is found around 500 MT. The predictions are more convergent when the operations are less than 10 hrs. The prediction performances are shown in Table \ref{Tab:Metrics_Prediction_Performance_Terminal_B_Loading} and Figure \ref{Fig:Error_Distribution_Terminal_B_Focused_Loading}.

\begin{table}[htbp]
	\centering
	\begin{tabular}{c|ll}
		\hline
		Metric & G1          & G2          \\
		\hline
		$\tilde{\mu}$ & 0.620644317   & 1.073380133    \\
		$\mu$         & -0.075593138  & 2.148568085    \\
		$\sigma$      & 2.933683211   & 5.095291035    \\
		MSE           & 8.446701941   & 30.048499      \\
		MAE           & 1.869863185   & 3.119073901    \\
		\hline
	\end{tabular}
	\newline
	Note: $\tilde{\mu}$ denotes median values.
	\caption{Metrics of Cargo Operation Prediction Performance in Terminal B - Loading}
	\label{Tab:Metrics_Prediction_Performance_Terminal_B_Loading}
\end{table}

\begin{figure}[htbp]
	\centering
	\subfloat[G1]{
		\label{SubFig:Distribution_Terminal_B_G1_Discharging}
		\includegraphics[width=0.4\textwidth]{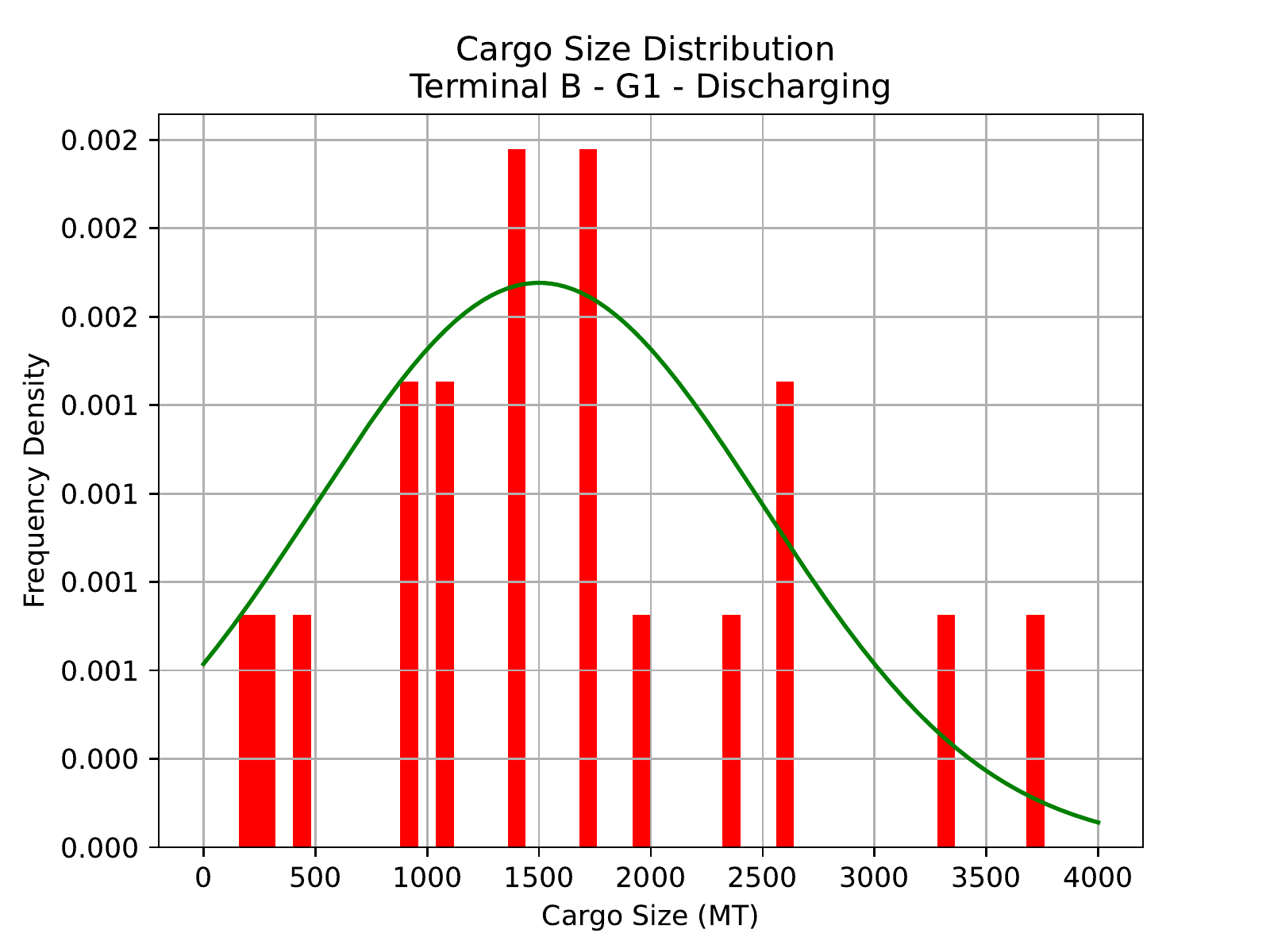} }~
	\subfloat[G2]{
		\label{SubFig:Distribution_Terminal_B_G2_Discharging}
		\includegraphics[width=0.4\textwidth]{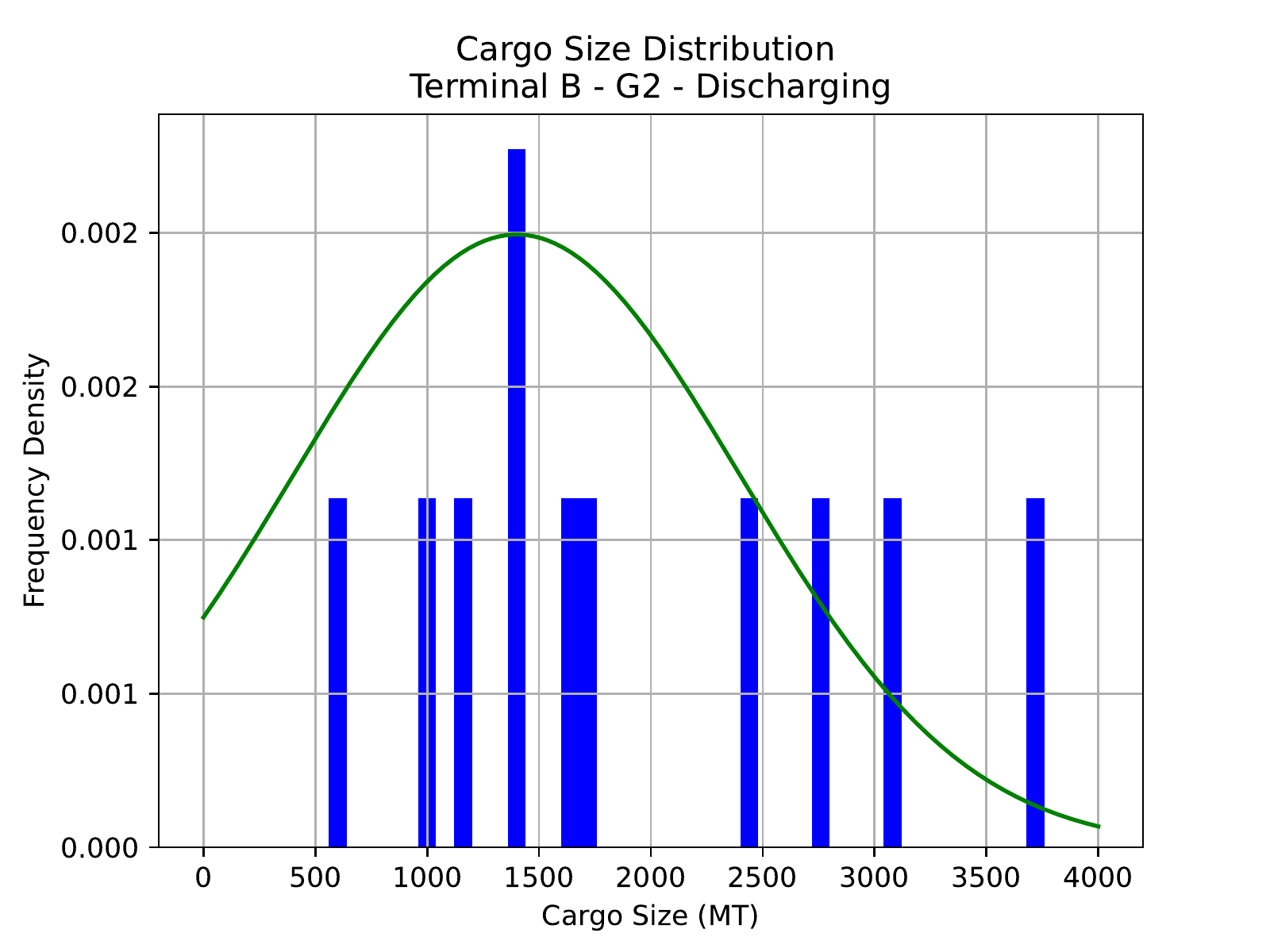} } 
	
	\caption{Size Distributions of Focused Cargoes in Terminal B - Discharging}
	\label{Fig:Distribution_Terminal_B_Focused_Discharging}
\end{figure}

\begin{figure}[htbp]
	\centering
	\subfloat[G1]{
		\label{SubFig:Operation_Stay_Terminal_B_G1_Discharging}
		\includegraphics[width=0.4\textwidth]{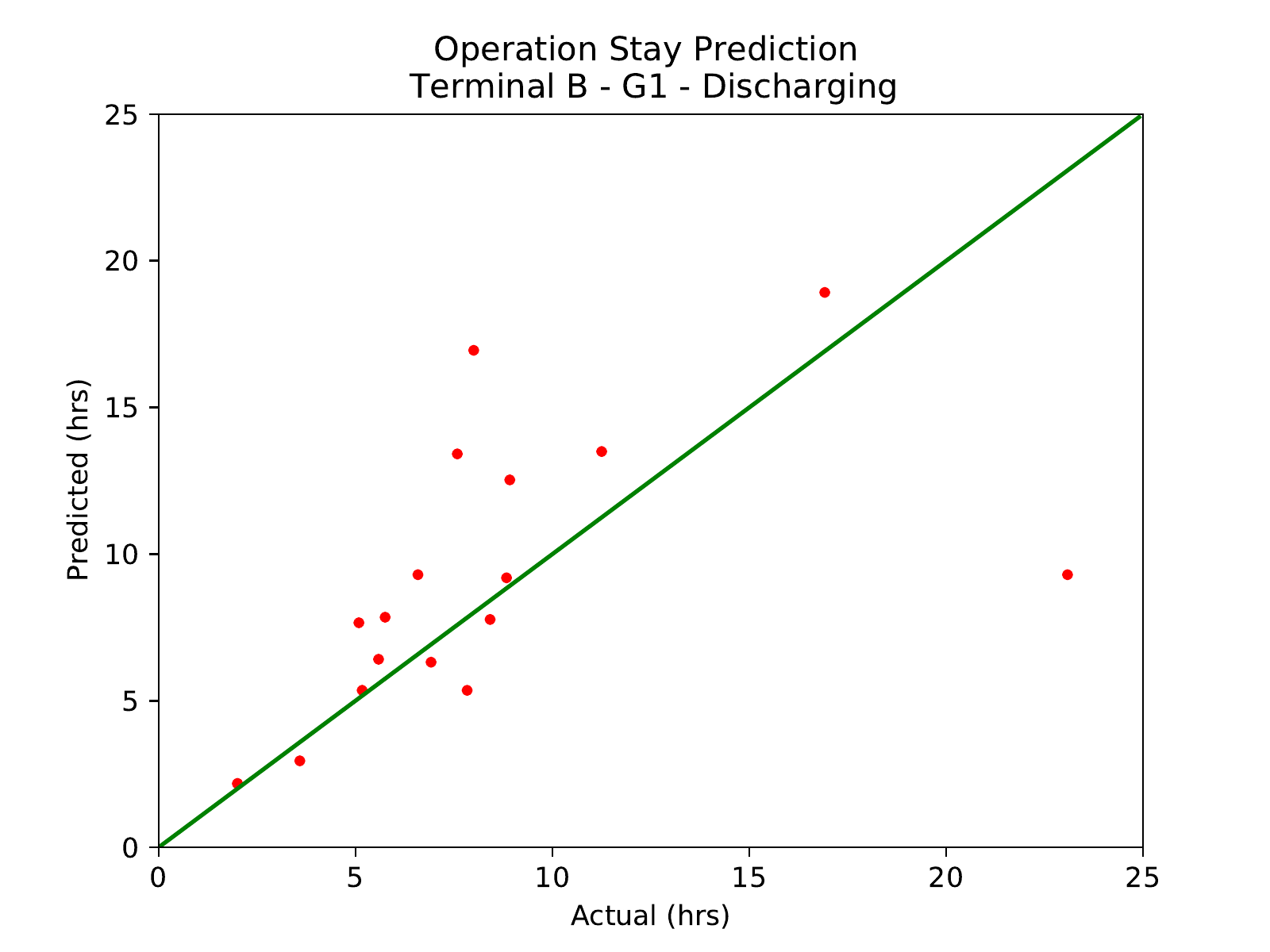} }~
	\subfloat[G2]{
		\label{SubFig:Operation_Stay_Terminal_B_G2_Discharging}
		\includegraphics[width=0.4\textwidth]{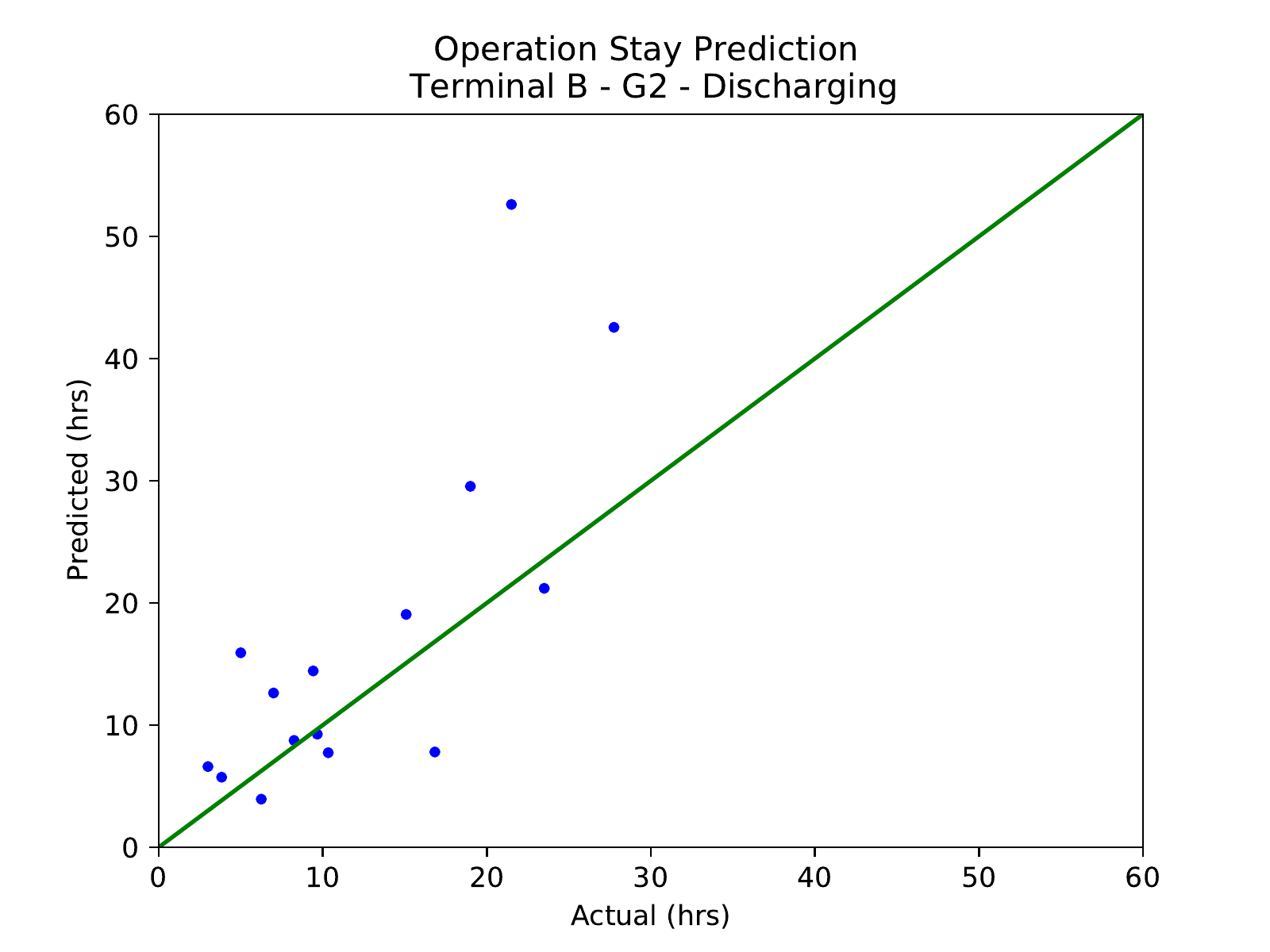} } 
	
	\caption{Cargo Operation Predictions of Focused Cargoes in Terminal B - Discharging}
	\label{Fig:Operation_Stay_Prediction_Terminal_B_Focused_Discharging}
\end{figure}

\begin{figure}[htbp]
	\centering
	\subfloat[G1]{
		\label{SubFig:Error_Distribution_Terminal_B_G1_Discharging}
		\includegraphics[width=0.4\textwidth]{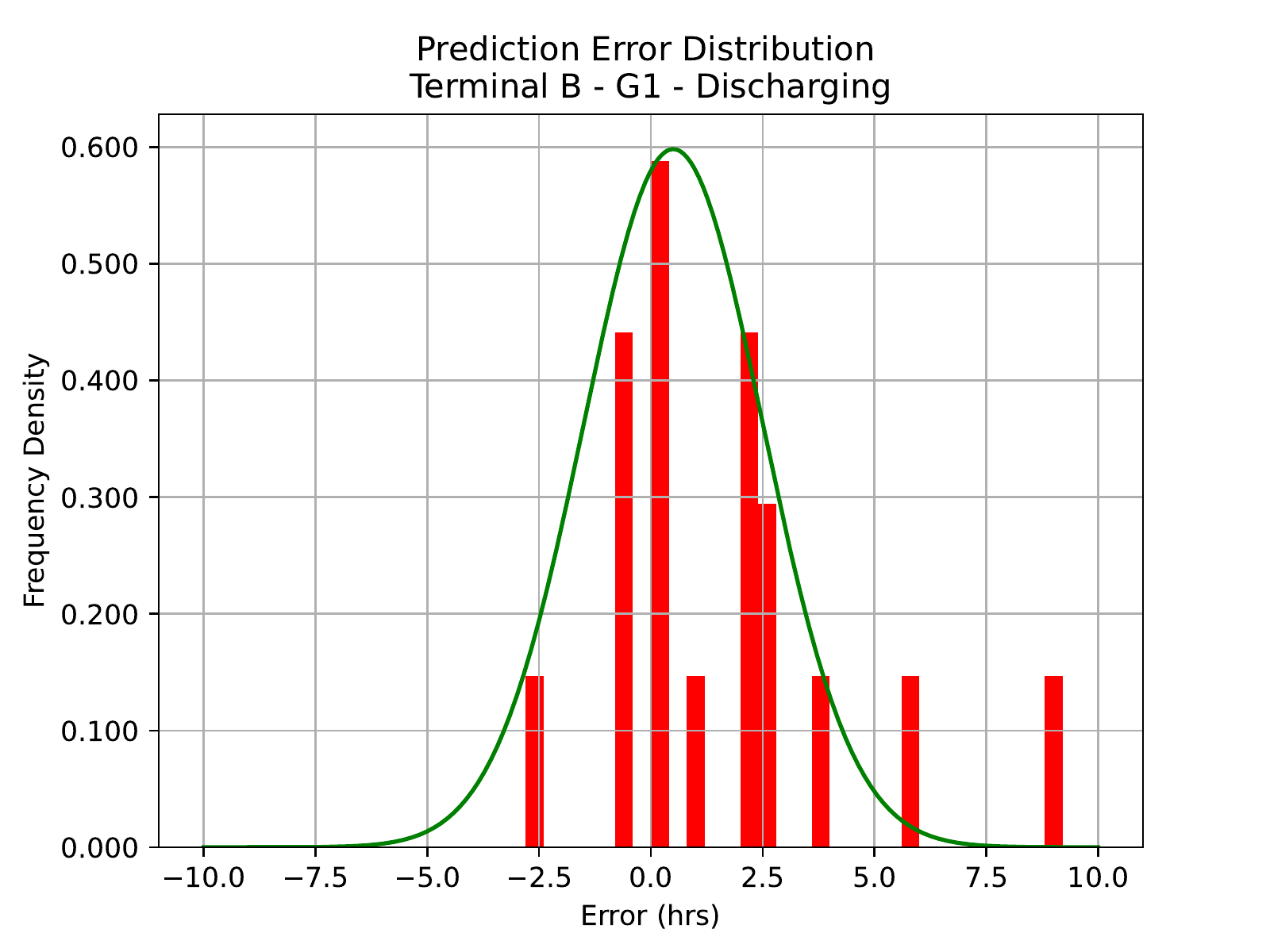} }~
	\subfloat[G2]{
		\label{SubFig:Error_Distribution_Terminal_B_G2_Discharging}
		\includegraphics[width=0.4\textwidth]{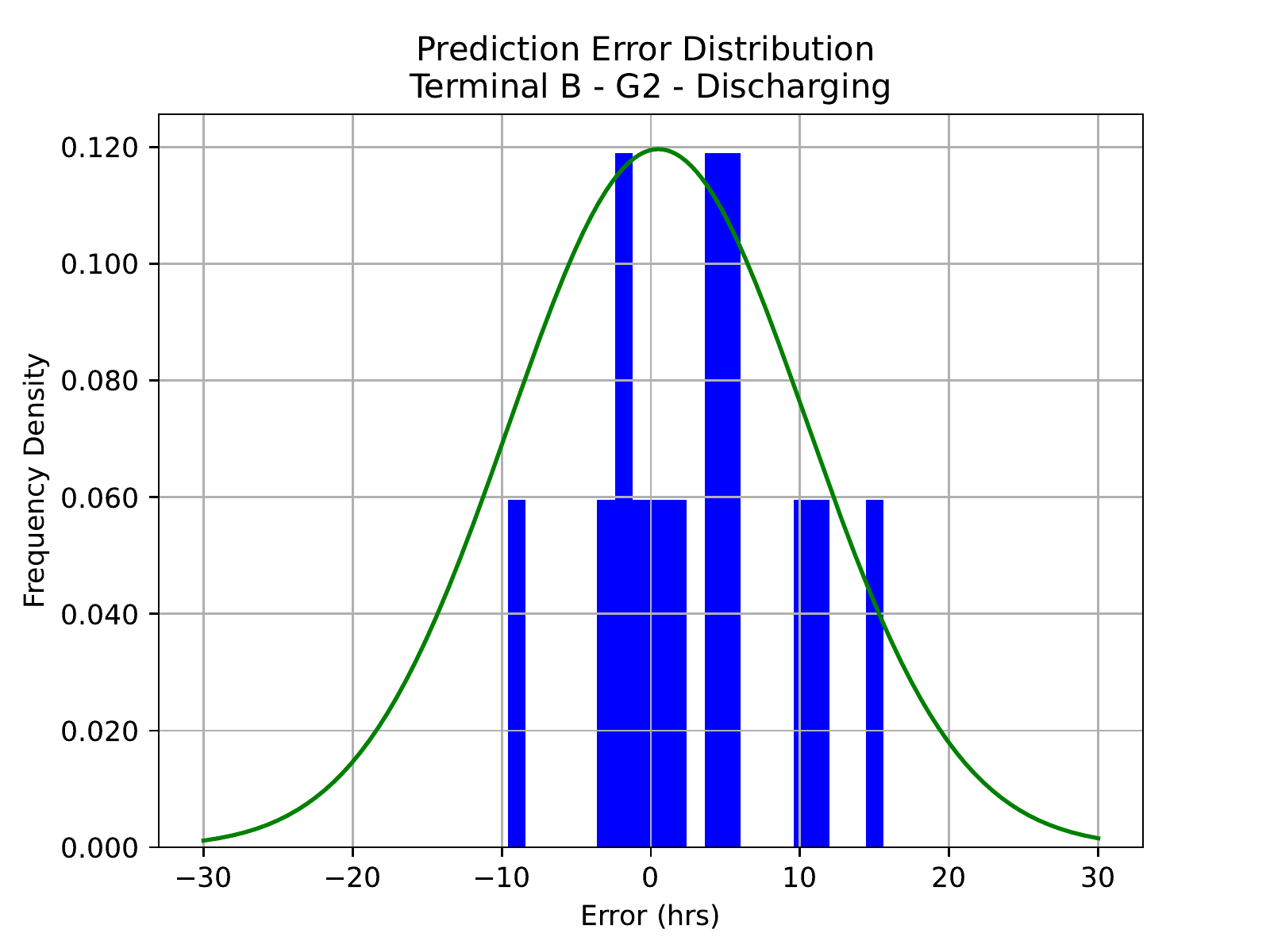} } 
	
	\caption{Prediction Error Distribution of Focused Cargoes in Terminal B - Discharging}
	\label{Fig:Error_Distribution_Terminal_B_Focused_Discharging}
\end{figure}

For discharging operations, the distribution and cargo operation prediction of focused cargoes are presented in Figure \ref{Fig:Distribution_Terminal_B_Focused_Discharging} and Figure \ref{Fig:Operation_Stay_Prediction_Terminal_B_Focused_Discharging}. Based on the prediction results, the peaks in G1 and G2 distributions are around 1500 MT and 2000 MT, respectively. There are large variances for both G1 and G2, compared with loading operations. The metrics of prediction performance are demonstrated in Table \ref{Tab:Metrics_Prediction_Performance_Terminal_B_Discharging} and Figure \ref{Fig:Error_Distribution_Terminal_B_Focused_Discharging}.

\begin{table}[htbp]
	\centering
	\begin{tabular}{c|ll}
		\hline
		Metric & G1          & G2          \\
		\hline
		$\tilde{\mu}$ & 0.356685067   & 3.6040847    \\
		$\mu$         & -0.259207812  & 4.757091683    \\
		$\sigma$      & 6.21158139    & 9.513482189    \\
		MSE           & 36.62020872   & 107.1025084      \\
		MAE           & 3.6009041     & 6.977218966    \\
		\hline
	\end{tabular}
	\newline
	Note: $\tilde{\mu}$ denotes median values.
	\caption{Metrics of Cargo Operation Prediction Performance in Terminal B - Discharging}
	\label{Tab:Metrics_Prediction_Performance_Terminal_B_Discharging}
\end{table}

Based on the prediction results of Terminal A and Terminal B, it is clear to note that Terminal A is more consistent in handling both G1 and G2 cargoes than Terminal B in this case. The predictive models perform more accurately in Terminal A than Terminal B. According to the prediction error distributions of Terminal A - Discharging, 75.76\% and 75.61\% prediction errors falls within $\pm$ 1 hr for G1 and G2, respectively. Whereas, 40.38\% and 34.69\% prediction errors satisfy $\pm$ 1 hr for G1 and G2 in Terminal B - Loading, while only 42.11\% and 13.33\% for G1 and G2 in Terminal B - Discharging, respectively. Considering this large prediction variance in Terminal B for cargo operation, we anticipate that there could be large uncertainty for the following berth stay prediction in Terminal B too. This is due to the propagated and accumulative effects of prediction errors from different blocks. Above all, the proposed approaches perform reasonably well in terms of accuracy and precision after multiple cross-validations among two terminals.

\subsection{Berth Stay Prediction}
As aforementioned, berth stay prediction is more broader and wider than previous tanker cargo operation prediction. The tanker cargo operation is a key and the most contributing part of berth stay. Besides, berth stay is defined from all fast to all clear. Therefore, the berth stay predictions over four scenarios are conducted and the subsequent results and discussion are presented in this section. Based on data analysis, a generic workflow for handling the focused cargoes is identified, and its corresponding chain with four scenarios as circled numbers indicated in Figure \ref{Fig:Generic_Block_Chain} below:

\begin{figure}[htbp]
	\centering
	\includegraphics[width=0.8\textwidth]{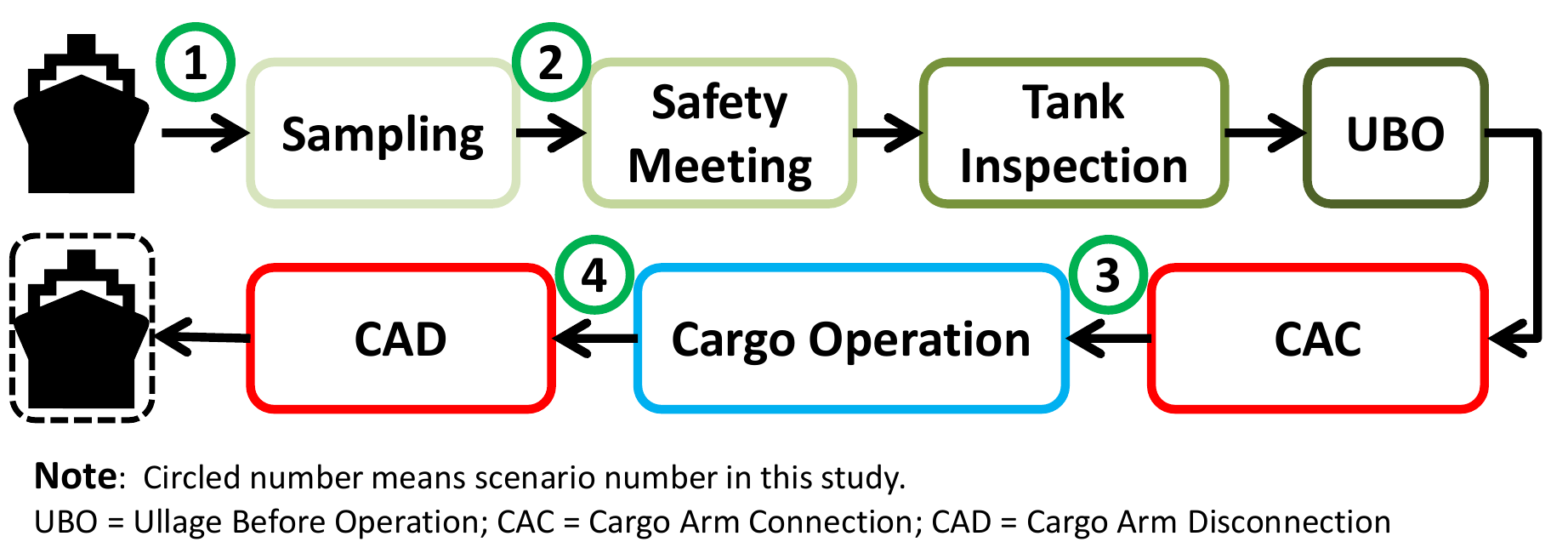} 
	\caption{Generic Chain of Focused Cargoes}
	\label{Fig:Generic_Block_Chain}
\end{figure}

Based on the historical data, the distributions of specific blocks as aforementioned are investigated and presented in the following Figure \ref{Fig:Distribution_Blocks_Berth_Stay}. It is also important to note that it is not necessary for all blocks to be existing. Therefore, a general weighted benchmark berth stay duration is about 18.68 hrs in this study. The individuals are 20.13 hrs for Terminal A - Discharging, 15.39 hrs and 22.52 hrs for Terminal B - Loading and Discharging, respectively. Due to data limitations in this study, a general weighted benchmark duration of Terminal A is treated as 20.13 hrs, and a general weighted benchmark duration of Terminal B is calculated as 17.18 hrs.

\begin{figure}[htbp]
	\centering
	\subfloat[]{
		\label{SubFig:Distribution_Sampling}
		\includegraphics[width=0.2\textwidth]{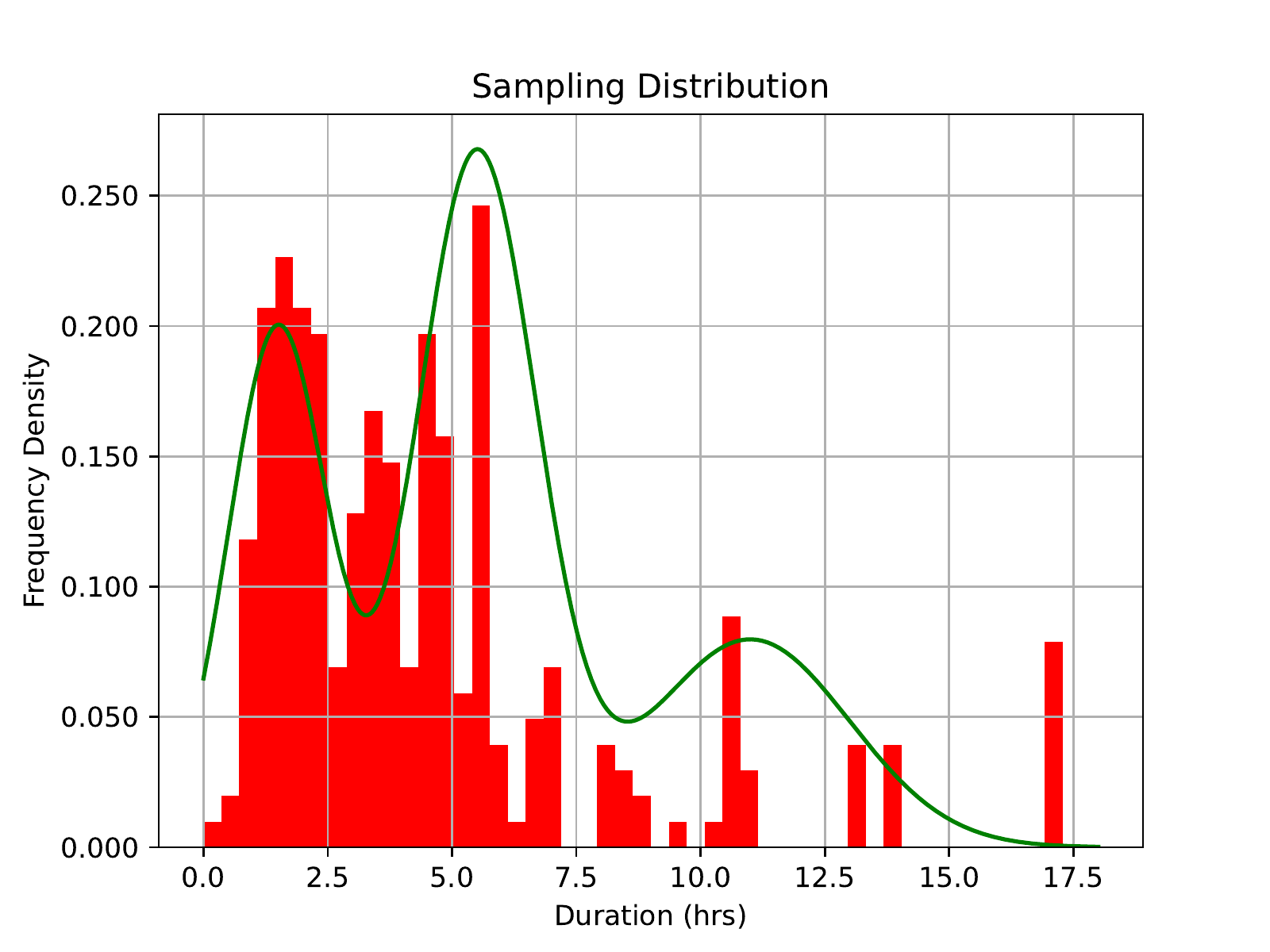} }~
	\subfloat[]{
		\label{SubFig:Distribution_SafetyMeeting}
		\includegraphics[width=0.2\textwidth]{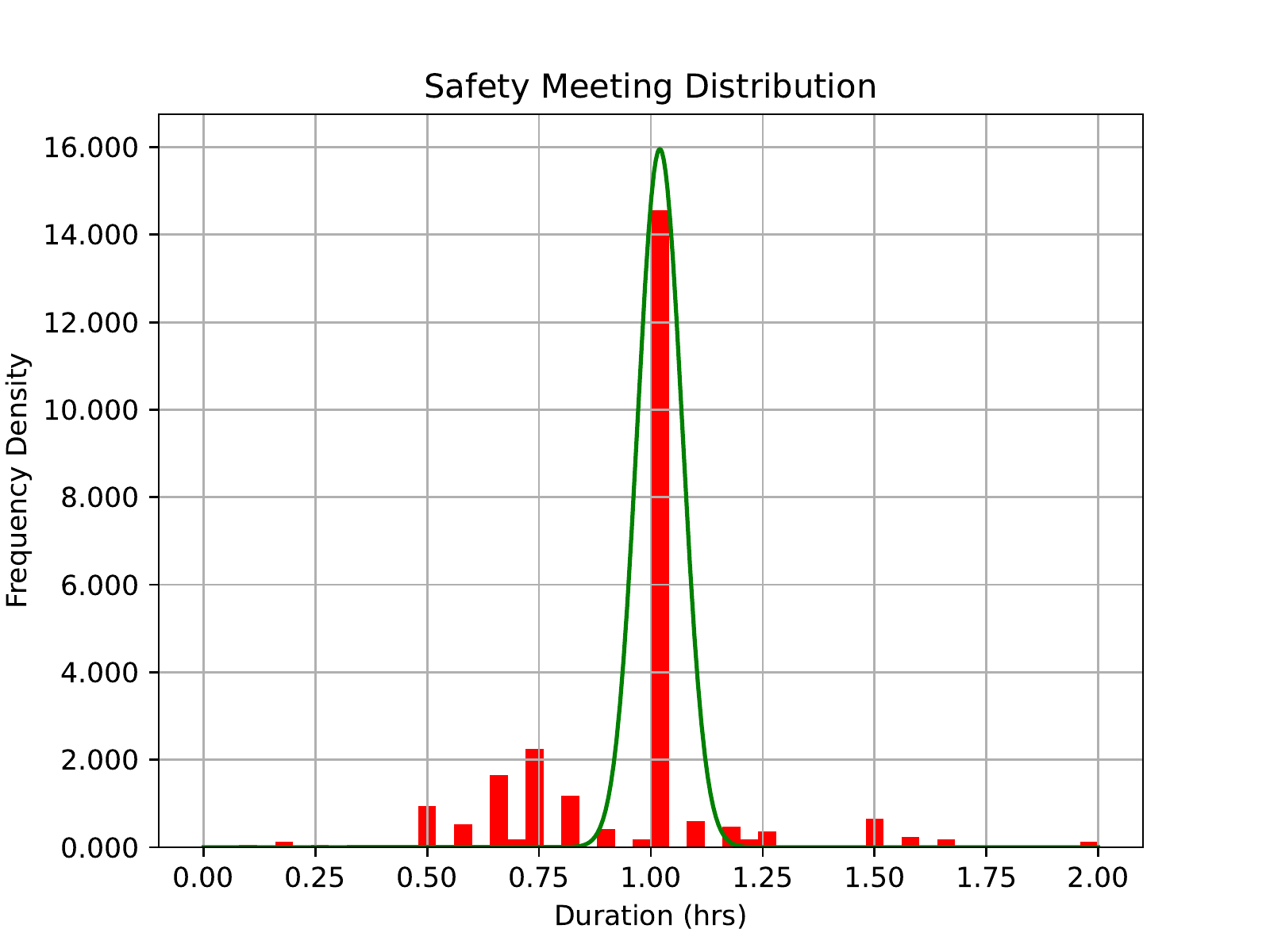} }~ 
	\subfloat[]{
		\label{SubFig:Distribution_TankInspection}
		\includegraphics[width=0.2\textwidth]{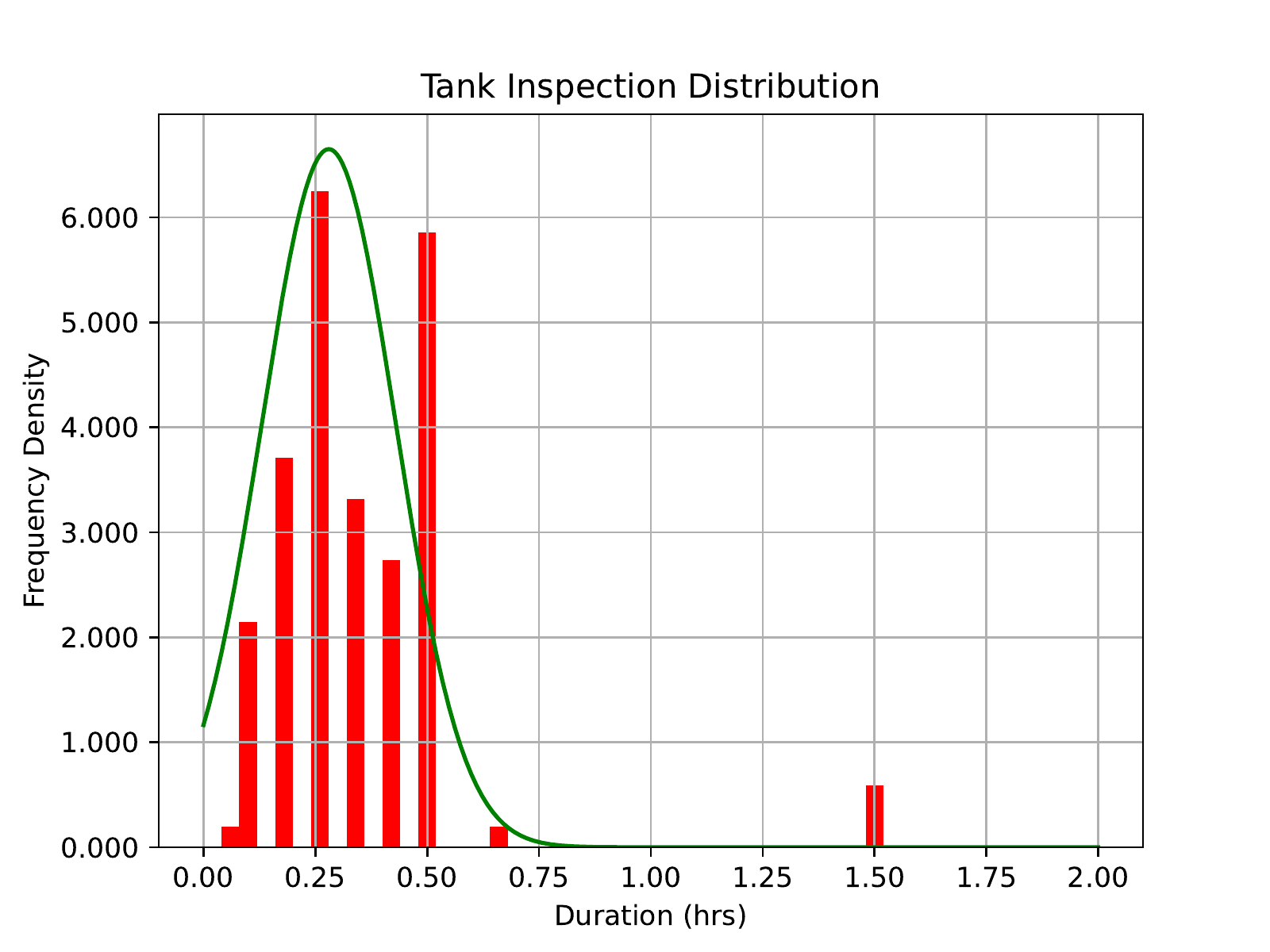} }~ 
	\subfloat[]{
		\label{SubFig:Distribution_UBO}
		\includegraphics[width=0.2\textwidth]{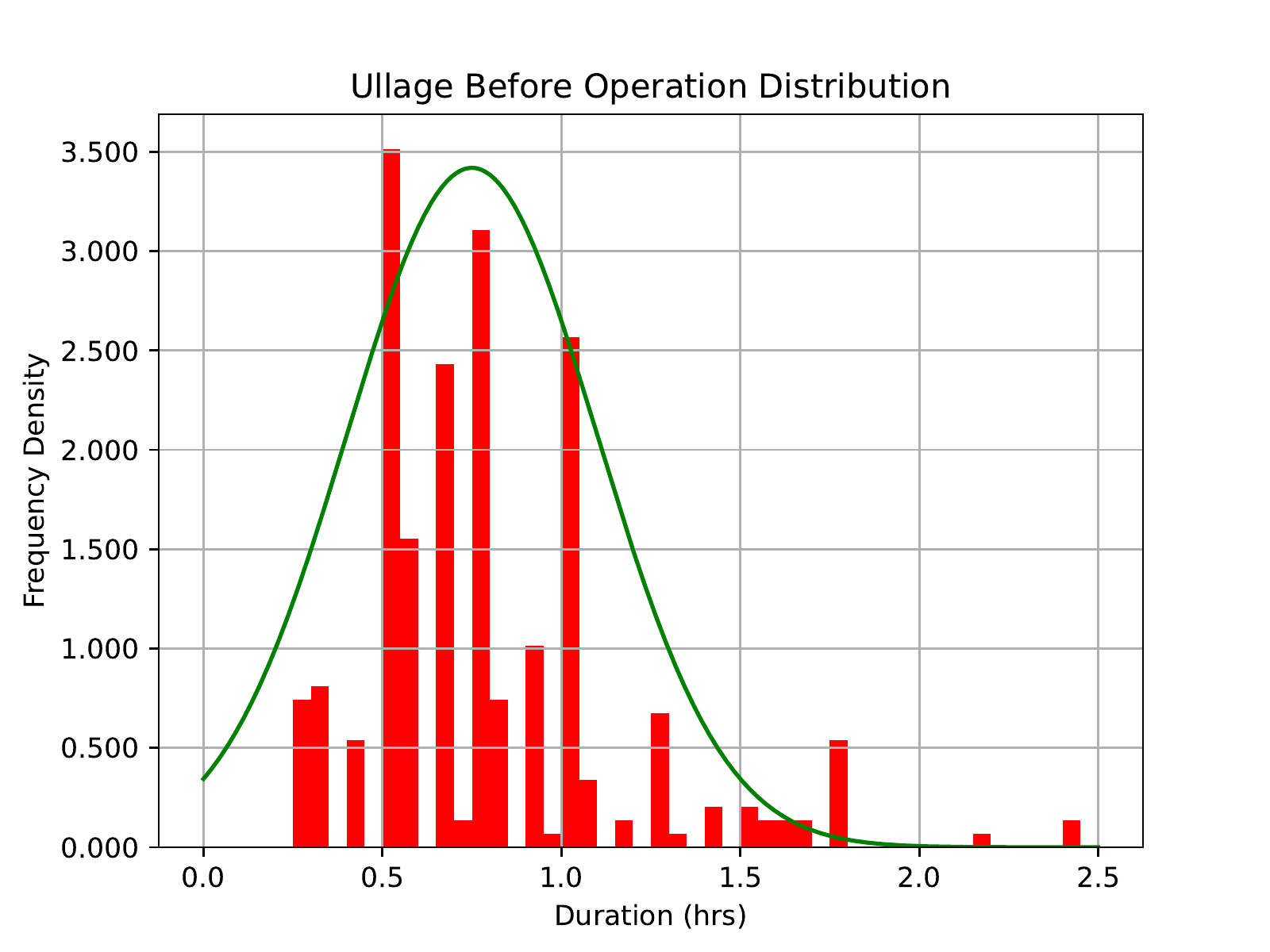} }

	\subfloat[]{
		\label{SubFig:Distribution_CAC}
		\includegraphics[width=0.2\textwidth]{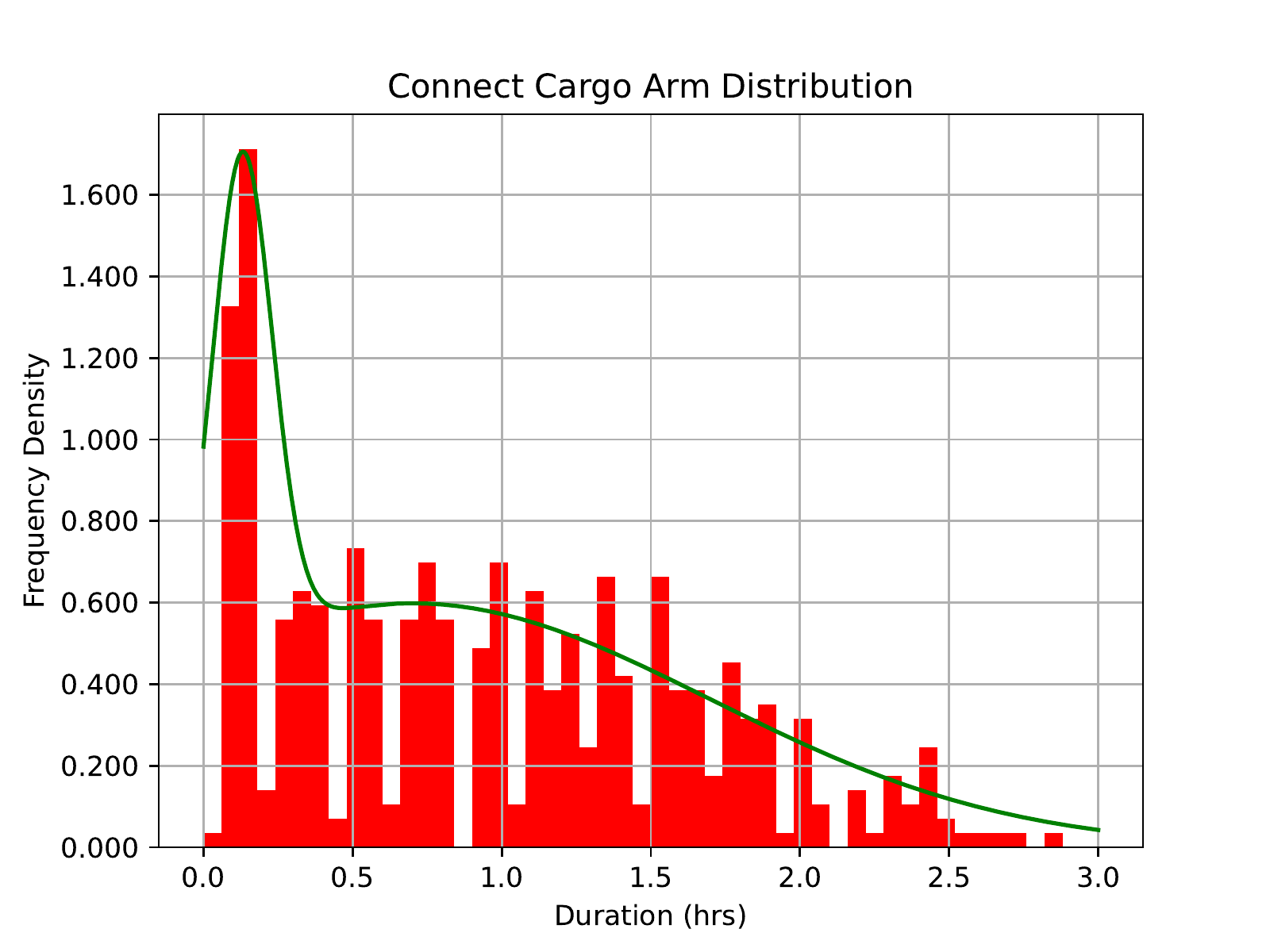} }~
	\subfloat[]{
		\label{SubFig:Distribution_CAD}
		\includegraphics[width=0.2\textwidth]{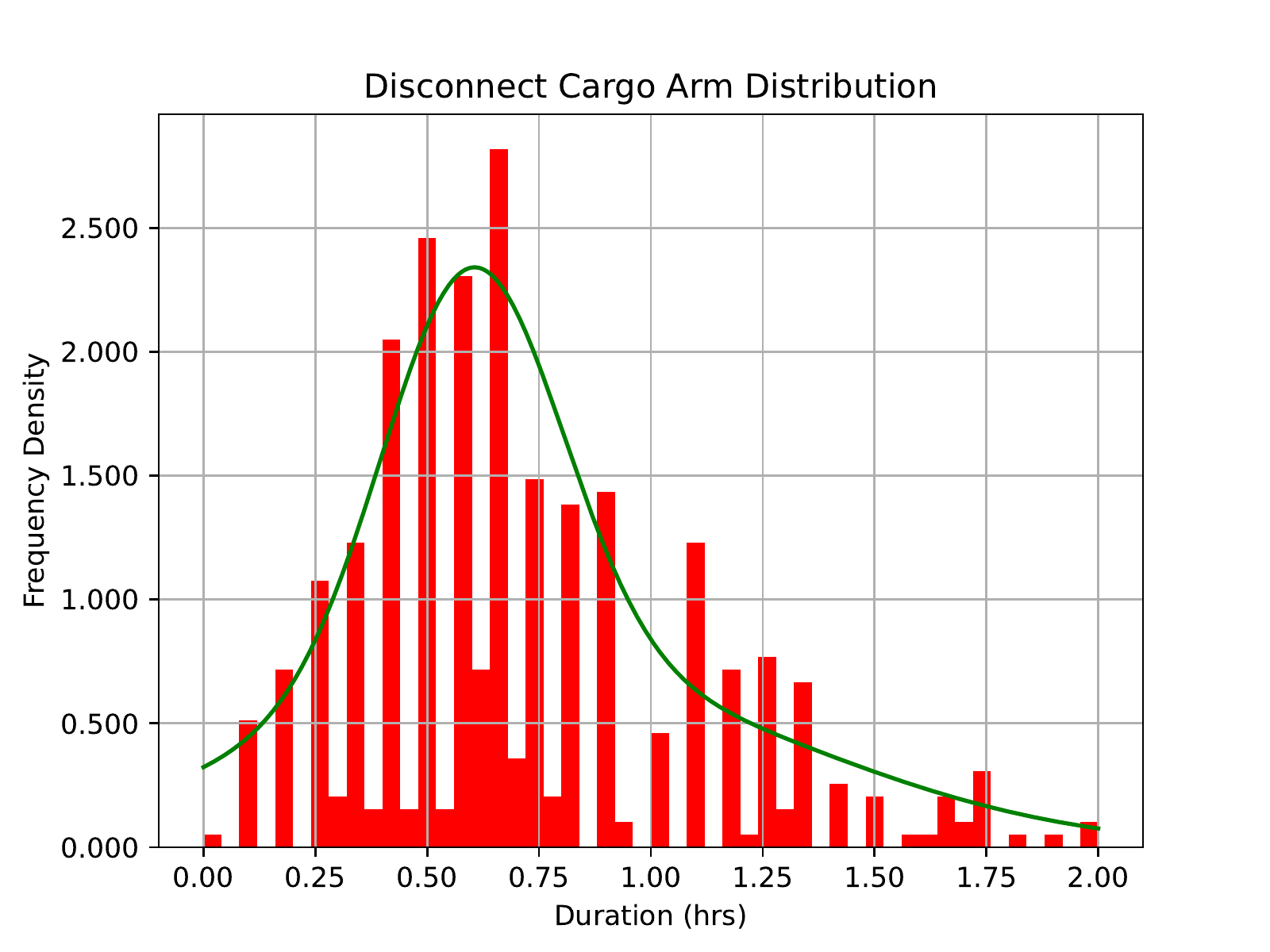} }~ 
	\subfloat[]{
		\label{SubFig:Distribution_PAC}
		\includegraphics[width=0.2\textwidth]{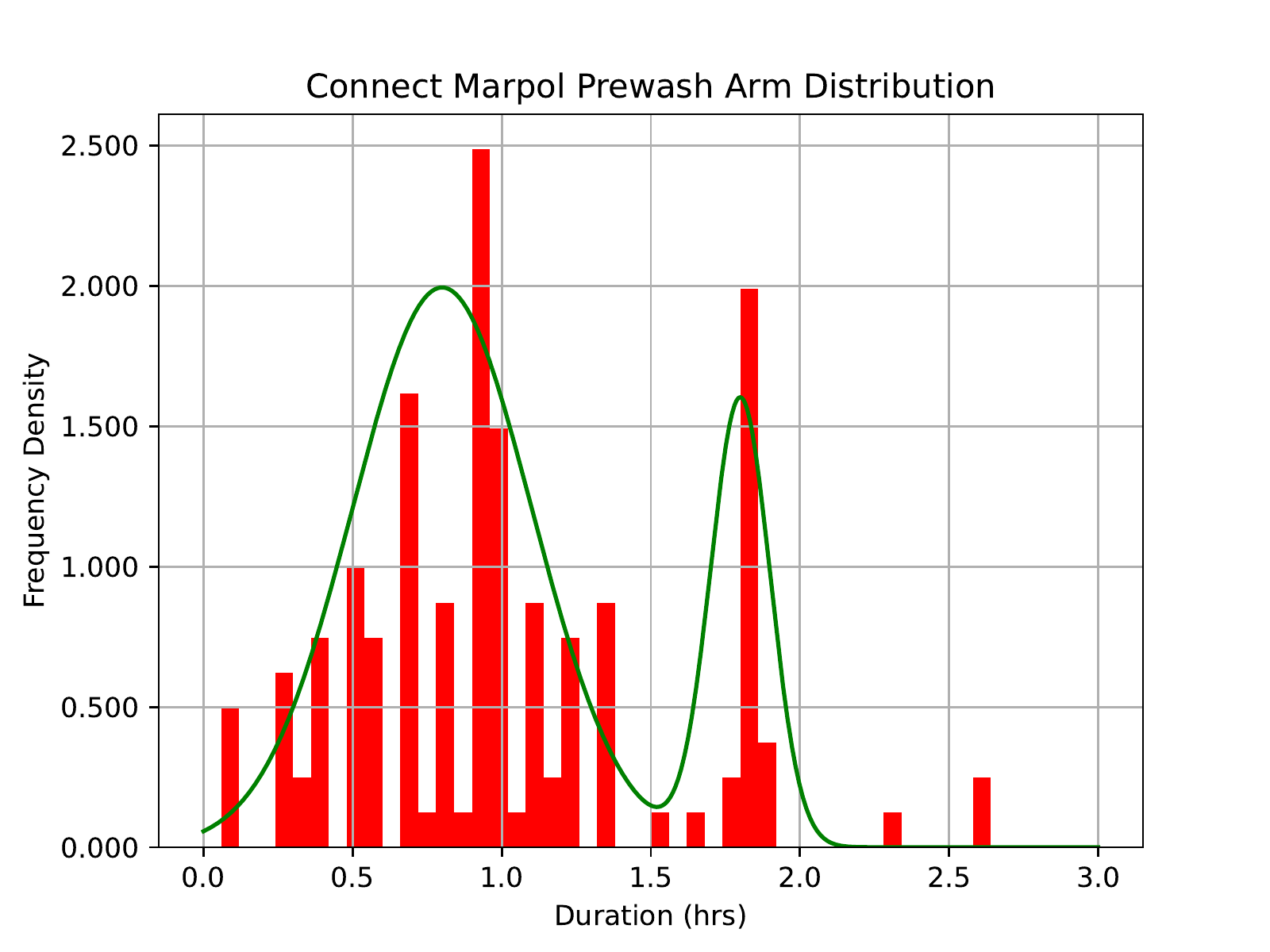} }~ 
	\subfloat[]{
		\label{SubFig:Distribution_PAD}
		\includegraphics[width=0.2\textwidth]{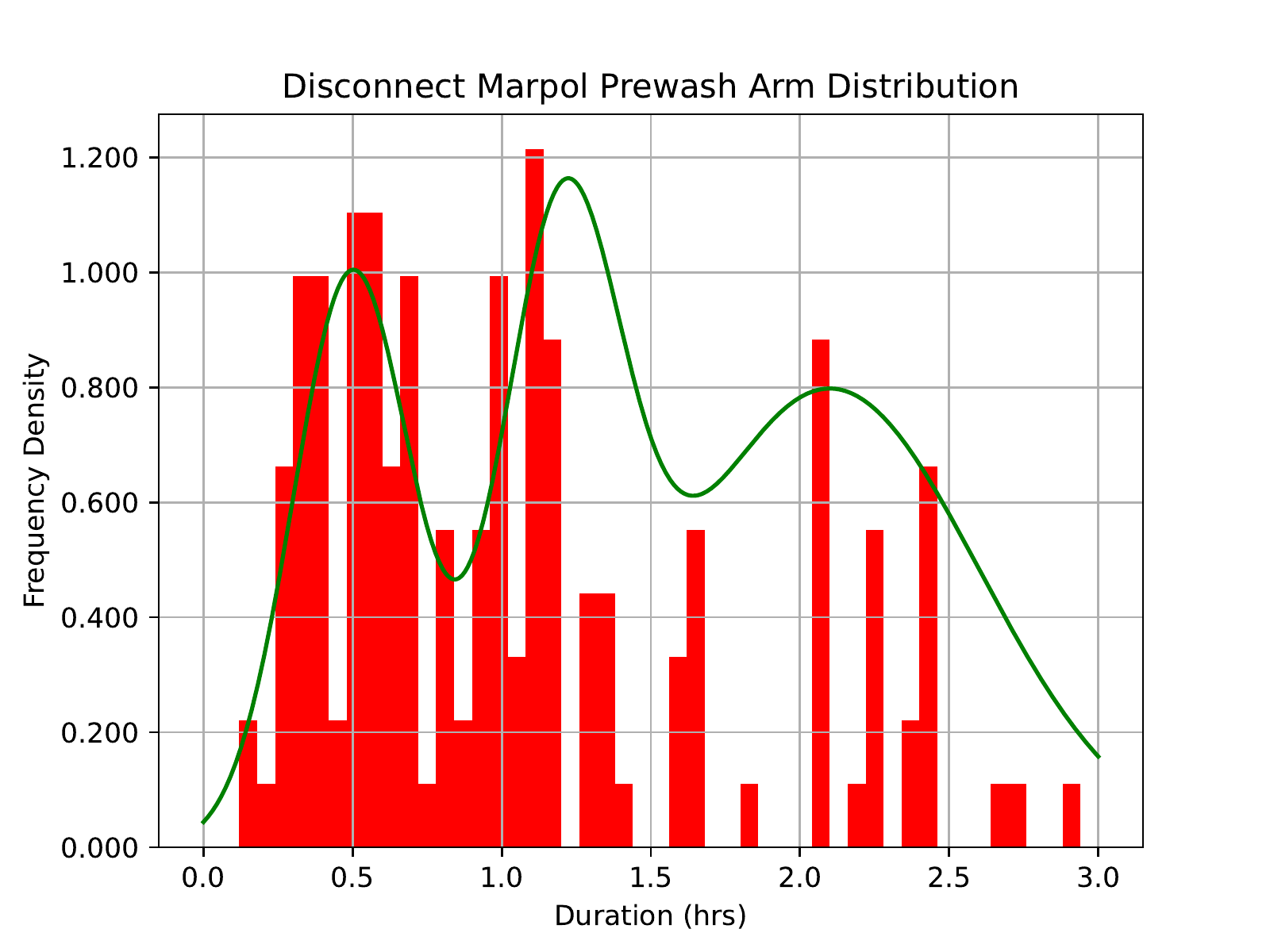} }
	
	\caption{Distributions of Blocks for Historical Berth Stay}
	\label{Fig:Distribution_Blocks_Berth_Stay}
\end{figure}

According to MDGS algorithm and distribution of historical data, the blocks are trained and the corresponding prediction distributions are tabulated in the following Table \ref{Tab:Distribution_Block_Berth_Stay_Prediction}, where UBO = Ullage Before Operation, CAC = Cargo Arm Connection, CAD = Cargo Arm Disconnection, PAC = Prewash Arm Connection and PAD = Prewash Arm Disconnection.

\begin{table}[htbp]
	\centering
	\begin{tabular}{lll}
		\hline
		Block           & Distribution                                                                                                               & Boundary        \\
		\hline
		Sampling        & \begin{tabular}[c]{@{}l@{}}$\mathcal{N}$$\sim$(1.5, 2)\\ $\mathcal{N}$$\sim$(1.8, 0.5)\\ $\mathcal{N}$$\sim$(3.4, 1.8)\end{tabular}                            & {[}0.25, 5{]}   \\ \hdashline
		Tank Inspection & \begin{tabular}[c]{@{}l@{}}$\mathcal{N}$$\sim$(0.08, 0.25)\\ $\mathcal{N}$$\sim$(0.25, 0.07)\\ $\mathcal{N}$$\sim$(0.5, 0.08)\end{tabular}                     & {[}0.05, 0.6{]} \\ \hdashline
		UBO             & \begin{tabular}[c]{@{}l@{}}$\mathcal{N}$$\sim$(0.5, 0.1)\\ $\mathcal{N}$$\sim$(0.75, 0.1)\\ $\mathcal{N}$$\sim$(1, 0.13)\end{tabular}                          & {[}0.25, 2{]}   \\ \hdashline
		CAC             & \begin{tabular}[c]{@{}l@{}}$\mathcal{N}$$\sim$(0.15, 0.07)\\ $\mathcal{N}$$\sim$(0.58, 0.2)\end{tabular}                                           & {[}0.08, 2{]}   \\ \hdashline
		CAD             & \begin{tabular}[c]{@{}l@{}}$\mathcal{N}$$\sim$(0.6, 0.06)\\ $\mathcal{N}$$\sim$(0.9, 0.125)\\ $\mathcal{N}$$\sim$(1.2, 0.15)\end{tabular}                      & {[}0.03, 2{]}   \\ \hdashline
		PAC             & \begin{tabular}[c]{@{}l@{}}$\mathcal{N}$$\sim$(0.6, 0.05)\\ $\mathcal{N}$$\sim$(0.9, 0.025)\\ $\mathcal{N}$$\sim$(1.3, 0.05)\\ $\mathcal{N}$$\sim$(1.8, 0.04)\end{tabular} & {[}0.08, 3{]}   \\ \hdashline
		PAD             & $\mathcal{N}$$\sim$(0.81, 0.25)                                                                                                        & {[}0.17, 3{]}  \\
		\hline
	\end{tabular}

	\caption{Distributions of Blocks for Berth Stay Prediction}
	\label{Tab:Distribution_Block_Berth_Stay_Prediction}
\end{table}

\subsubsection{Scenario 1 - All Fast}
In this scenario, the starting point of berth stay prediction is ``All Fast'', which means that it tries to predict how long vessel will stay at berth since mooring completed and all fast. The ending point of berth stay should theoretically be ``Cargo Arm Disconnected'' without the requirement of prewash, and ``Marpol Prewash Arm Disconnected'' with the requirement of prewash afterwards.

\begin{figure}[htbp]
	\centering
	\includegraphics[width=0.5\textwidth]{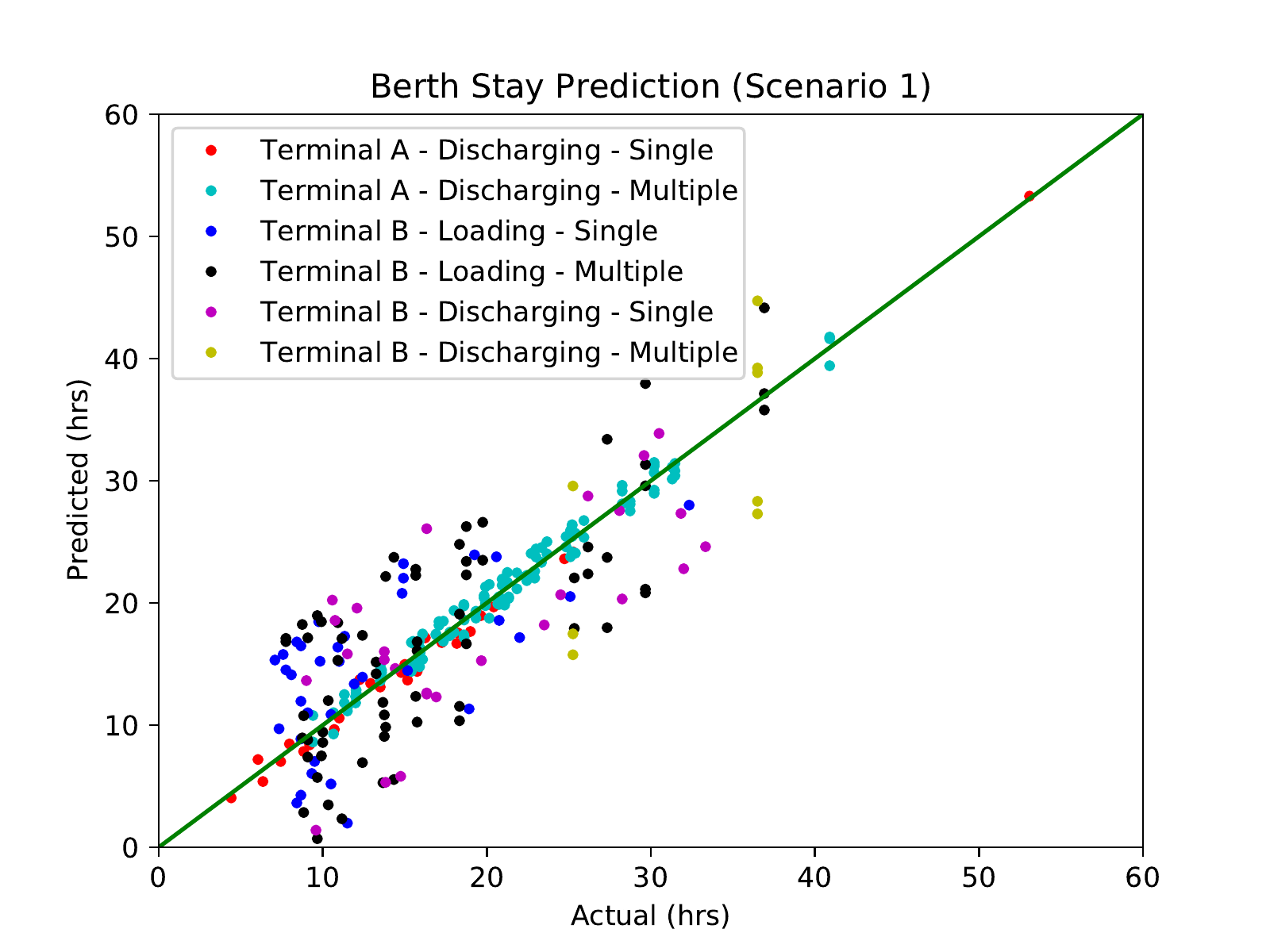} 
	\caption{Berth Stay Prediction - Scenario 1}
	\label{Fig:Berth_Stay_Prediction_S1}
\end{figure}

\begin{figure}[htbp]
	\centering
	\subfloat[Terminal A]{
		\label{SubFig:Error_Distribution_S1_Terminal_A}
		\includegraphics[width=0.4\textwidth]{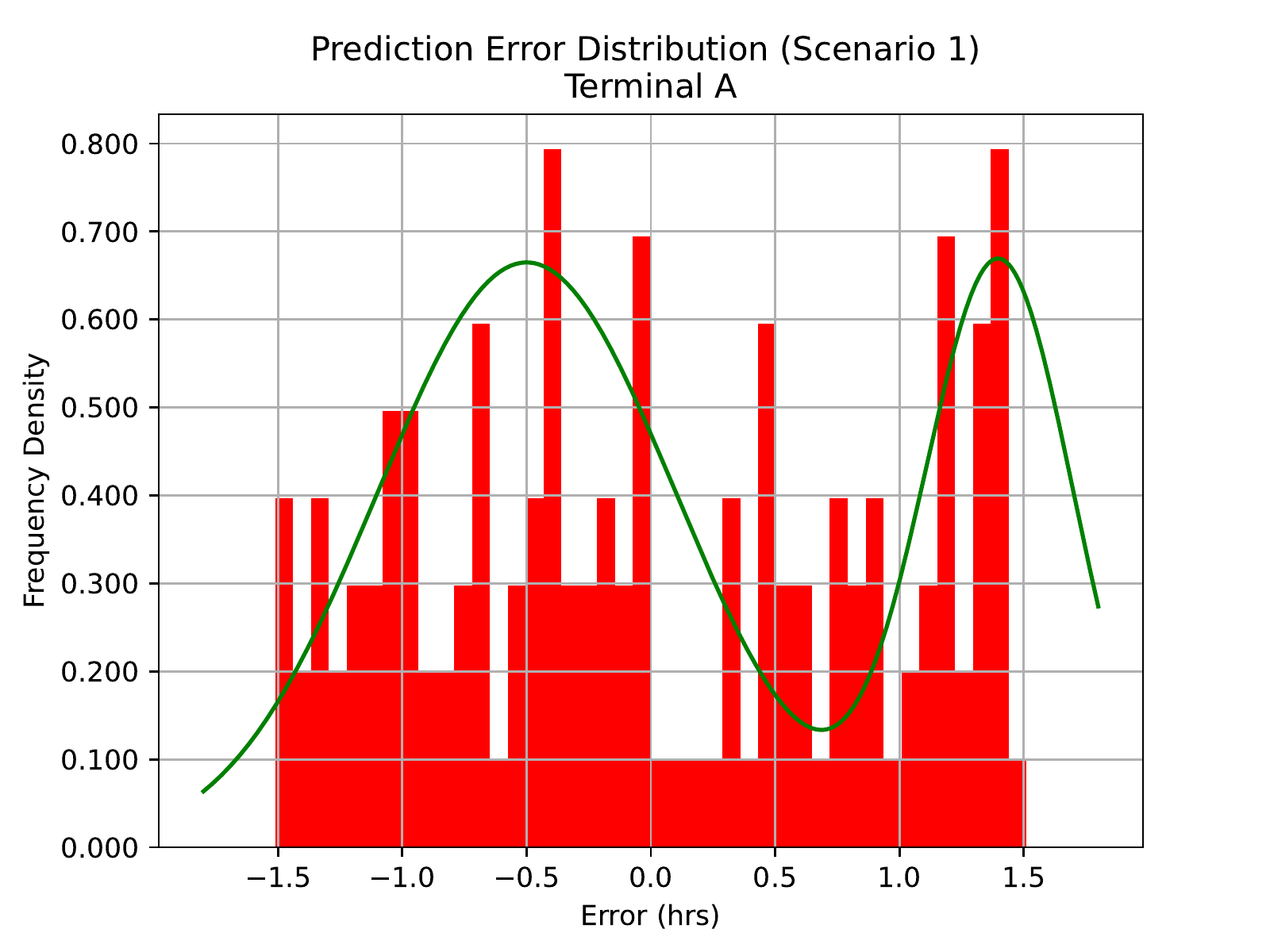} }~
	\subfloat[Terminal B]{
		\label{SubFig:Error_Distribution_S1_Terminal_B}
		\includegraphics[width=0.4\textwidth]{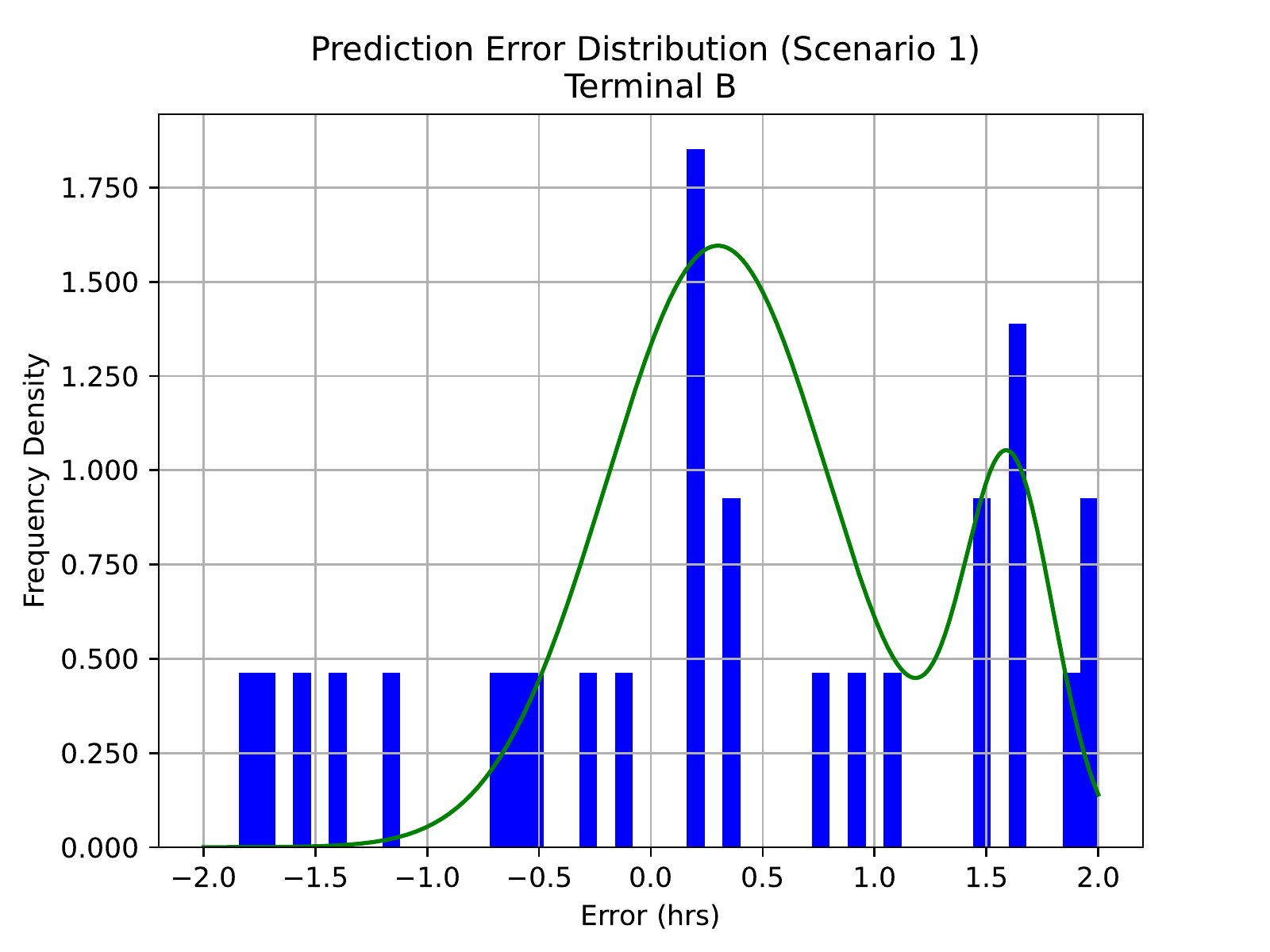} } 
	
	\caption{Prediction Error Distributions of Focused Cargoes - Scenario 1}
	\label{Fig:Error_Distribution_S1_Terminal_A_B}
\end{figure}

Based on the generic chain of focused cargoes in this study, the results of berth stay prediction for scenario 1 are shown in Figure \ref{Fig:Berth_Stay_Prediction_S1}, and the corresponding distributions of prediction errors are presented in Figure \ref{Fig:Error_Distribution_S1_Terminal_A_B} for both terminals.

\subsubsection{Scenario 2 - Sample Pass}
In this scenario, the starting point of berth stay prediction is ``Sample Pass'', which means that it tries to predict how long vessel will stay at berth since the samples of cargoes pass the relevant tests before conducting operations. Similarly, the ending point of berth stay should theoretically be ``Cargo Arm Disconnected'' without the requirement of prewash, and ``Marpol Prewash Arm Disconnected'' with the requirement of prewash afterwards.

\begin{figure}[htbp]
	\centering
	\includegraphics[width=0.5\textwidth]{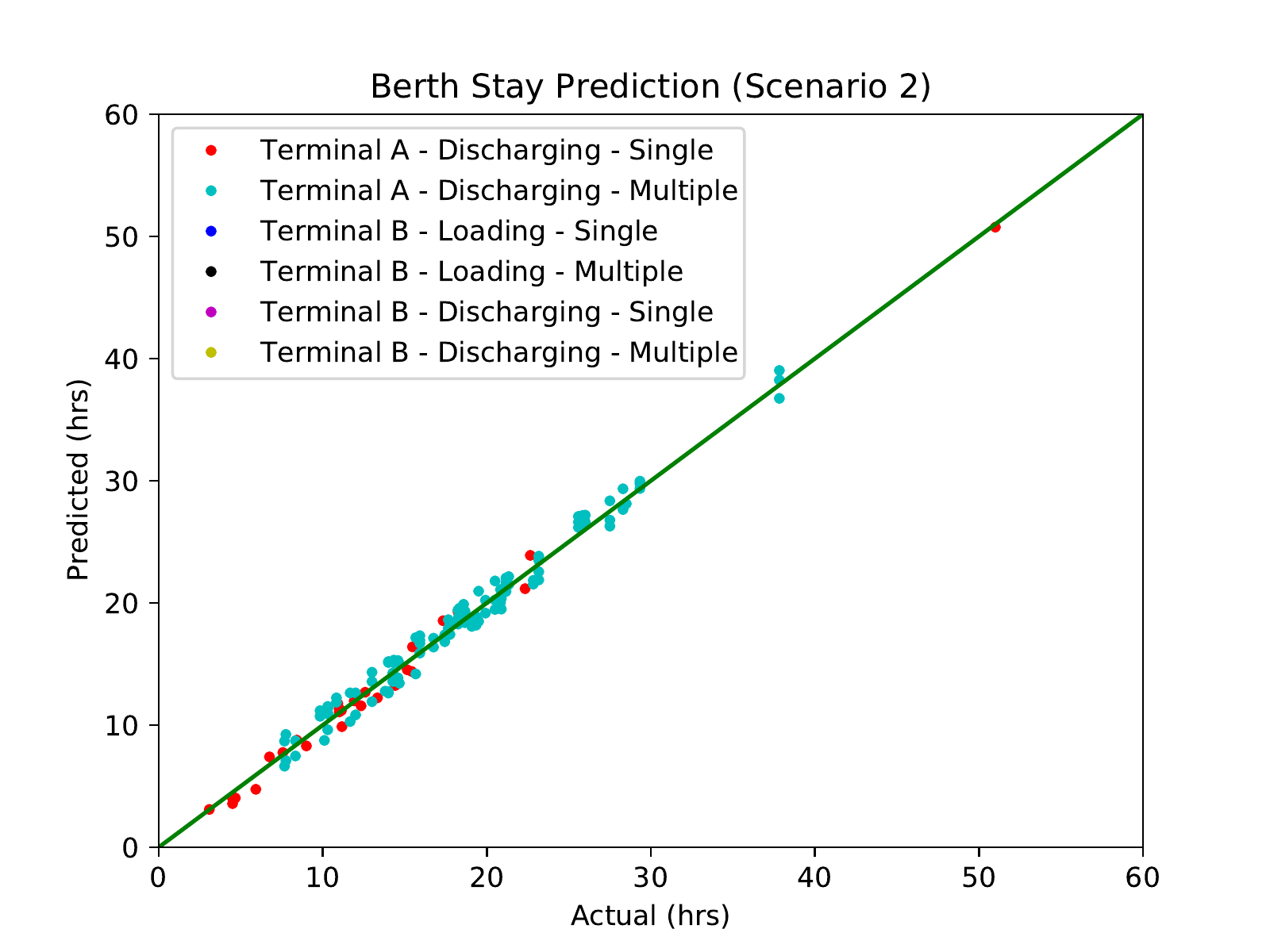} 
	\caption{Berth Stay Prediction - Scenario 2}
	\label{Fig:Berth_Stay_Prediction_S2}
\end{figure}

\begin{figure}[htbp]
	\centering
	\subfloat[Terminal A]{
		\label{SubFig:Error_Distribution_S2_Terminal_A}
		\includegraphics[width=0.4\textwidth]{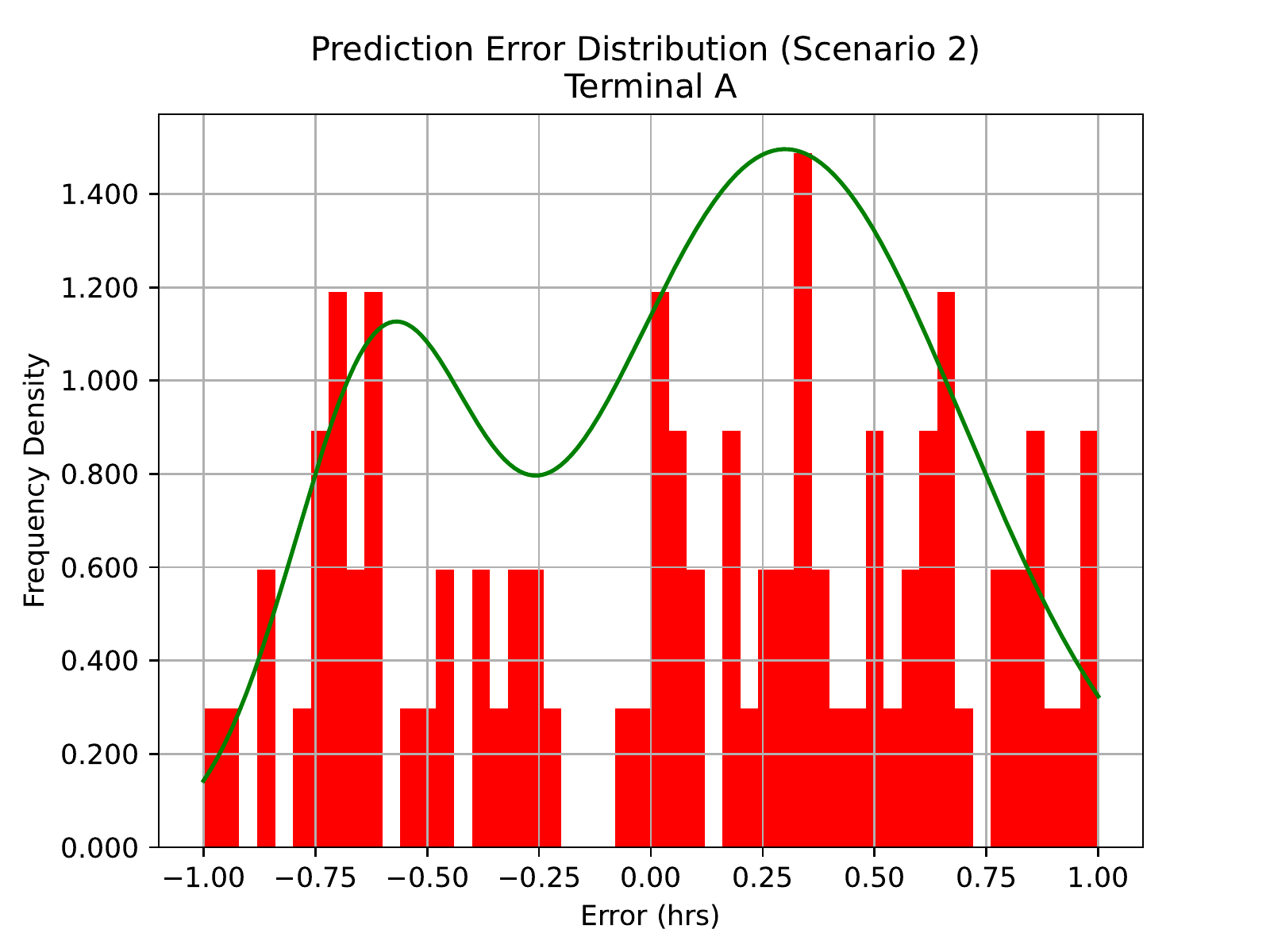} }~
	\subfloat[Terminal B]{
		\label{SubFig:Error_Distribution_S2_Terminal_B}
		\includegraphics[width=0.4\textwidth]{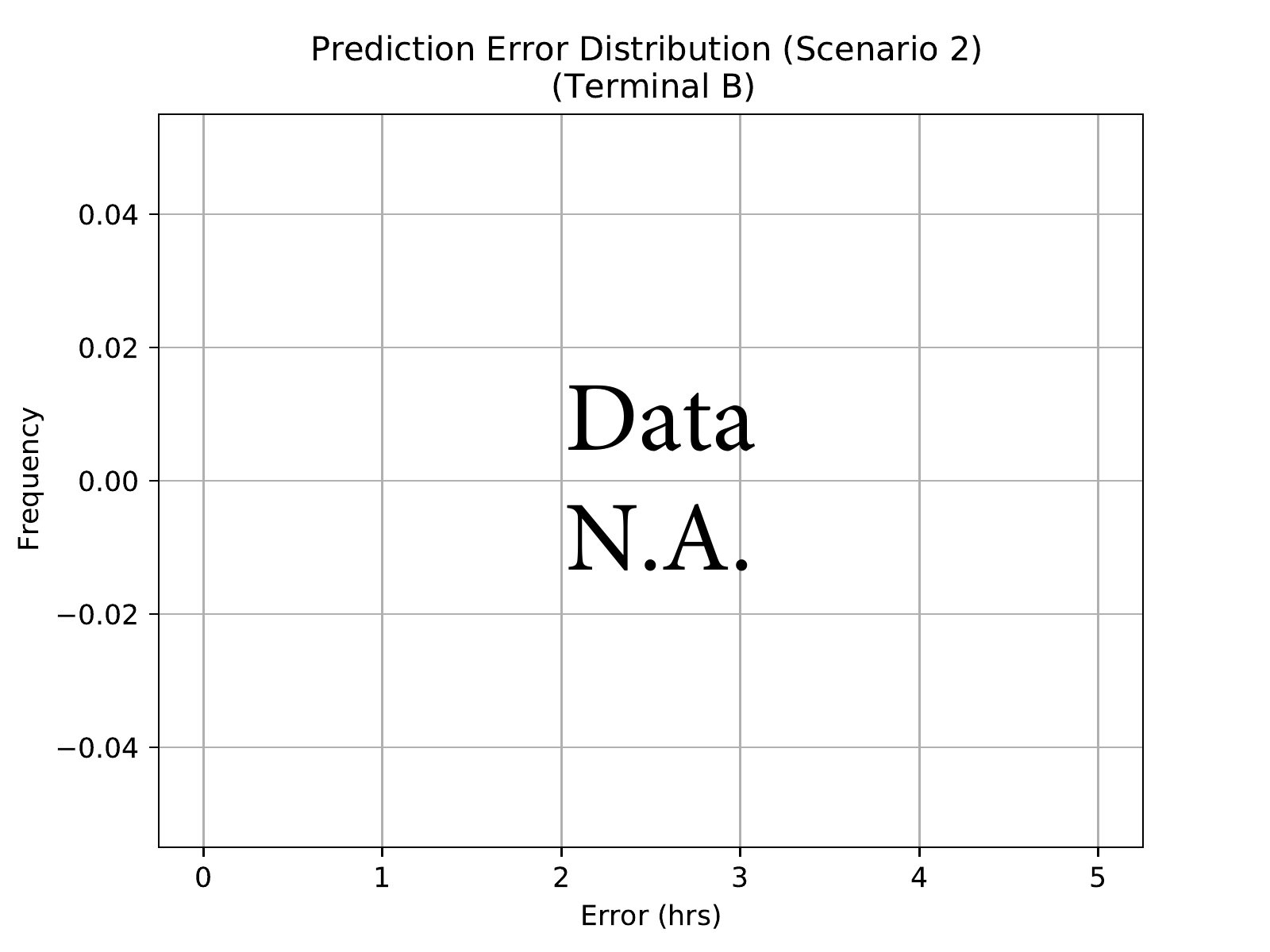} } 
	
	\caption{Prediction Error Distributions of Focused Cargoes - Scenario 2}
	\label{Fig:Error_Distribution_S2_Terminal_A_B}
\end{figure}

Based on the generic chain of focused cargoes in this study, the results of berth stay prediction for scenario 2 are shown in Figure \ref{Fig:Berth_Stay_Prediction_S2}, and the corresponding distributions of prediction errors are presented in Figure \ref{Fig:Error_Distribution_S2_Terminal_A_B} for both terminals.

\subsubsection{Scenario 3 - Commence Operation}
In this scenario, the starting point of berth stay prediction is ``Commence Operation'', which means that it tries to predict how long vessel will stay at berth since the cargo operations get started. Similarly, the ending point of berth stay should theoretically be ``Cargo Arm Disconnected'' without the requirement of prewash, and ``Marpol Prewash Arm Disconnected'' with the requirement of prewash afterwards.

\begin{figure}[htbp]
	\centering
	\includegraphics[width=0.5\textwidth]{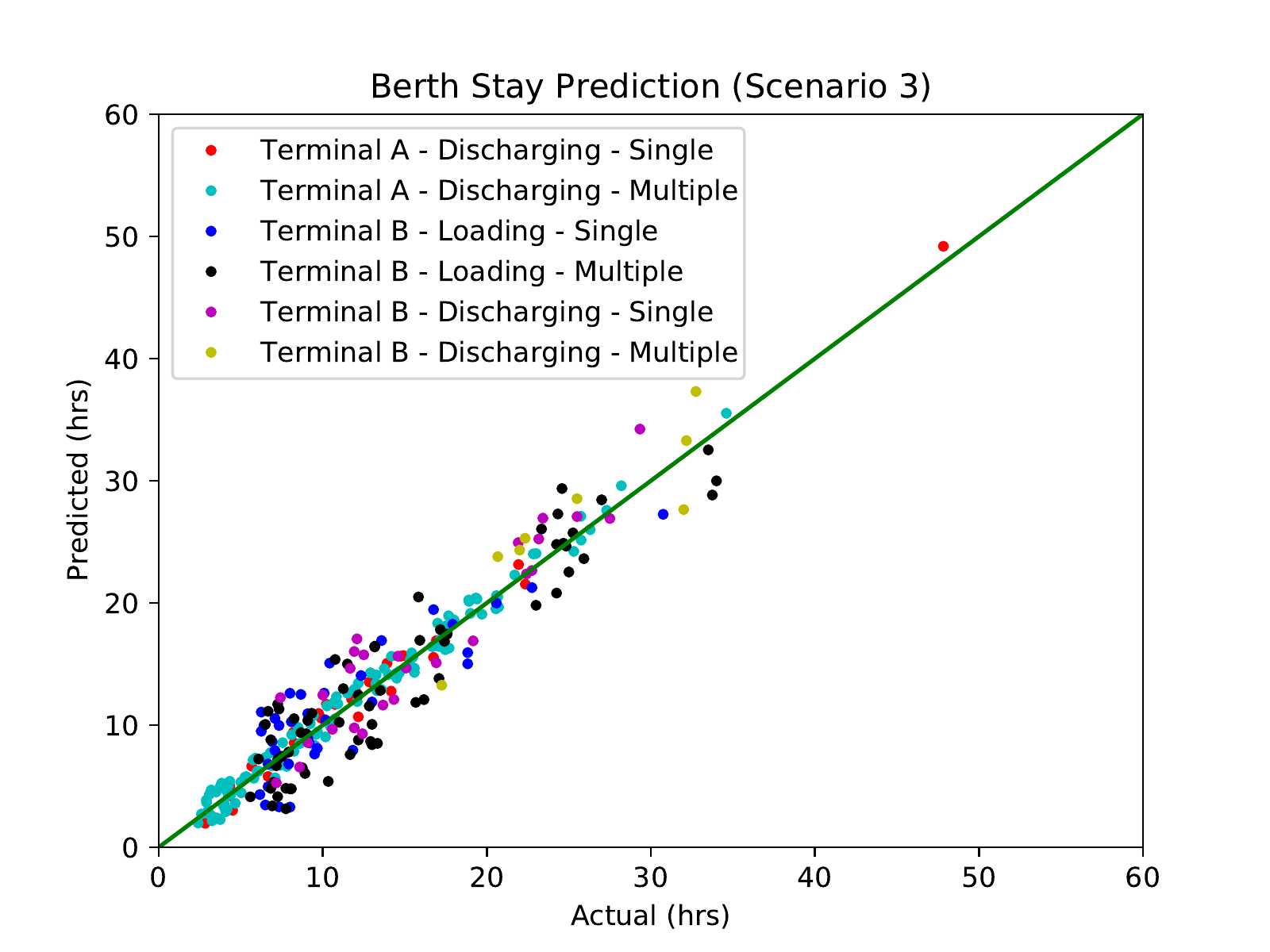} 
	\caption{Berth Stay Prediction - Scenario 3}
	\label{Fig:Berth_Stay_Prediction_S3}
\end{figure}

\begin{figure}[htbp]
	\centering
	\subfloat[Terminal A]{
		\label{SubFig:Error_Distribution_S3_Terminal_A}
		\includegraphics[width=0.4\textwidth]{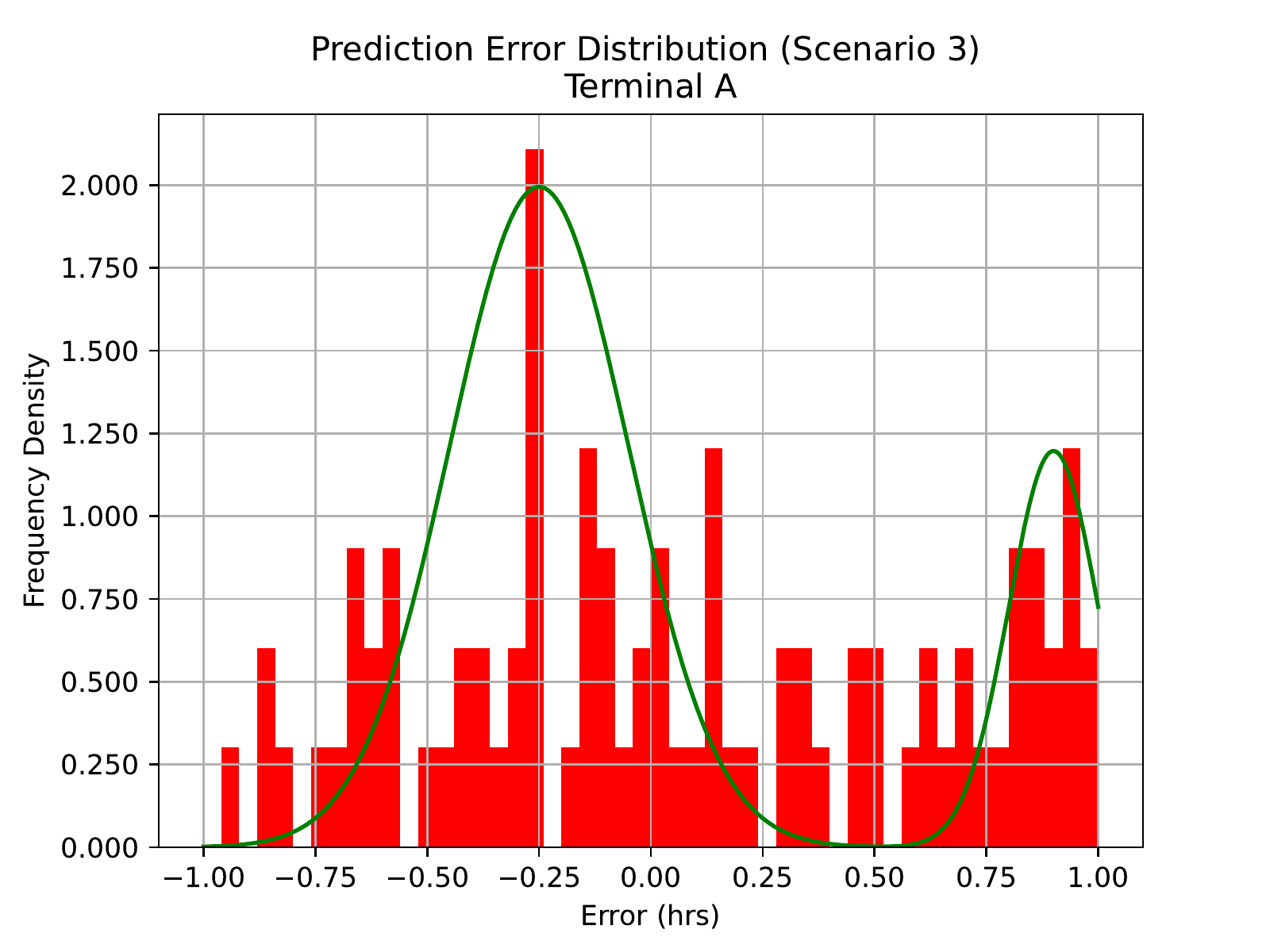} }~
	\subfloat[Terminal B]{
		\label{SubFig:Error_Distribution_S3_Terminal_B}
		\includegraphics[width=0.4\textwidth]{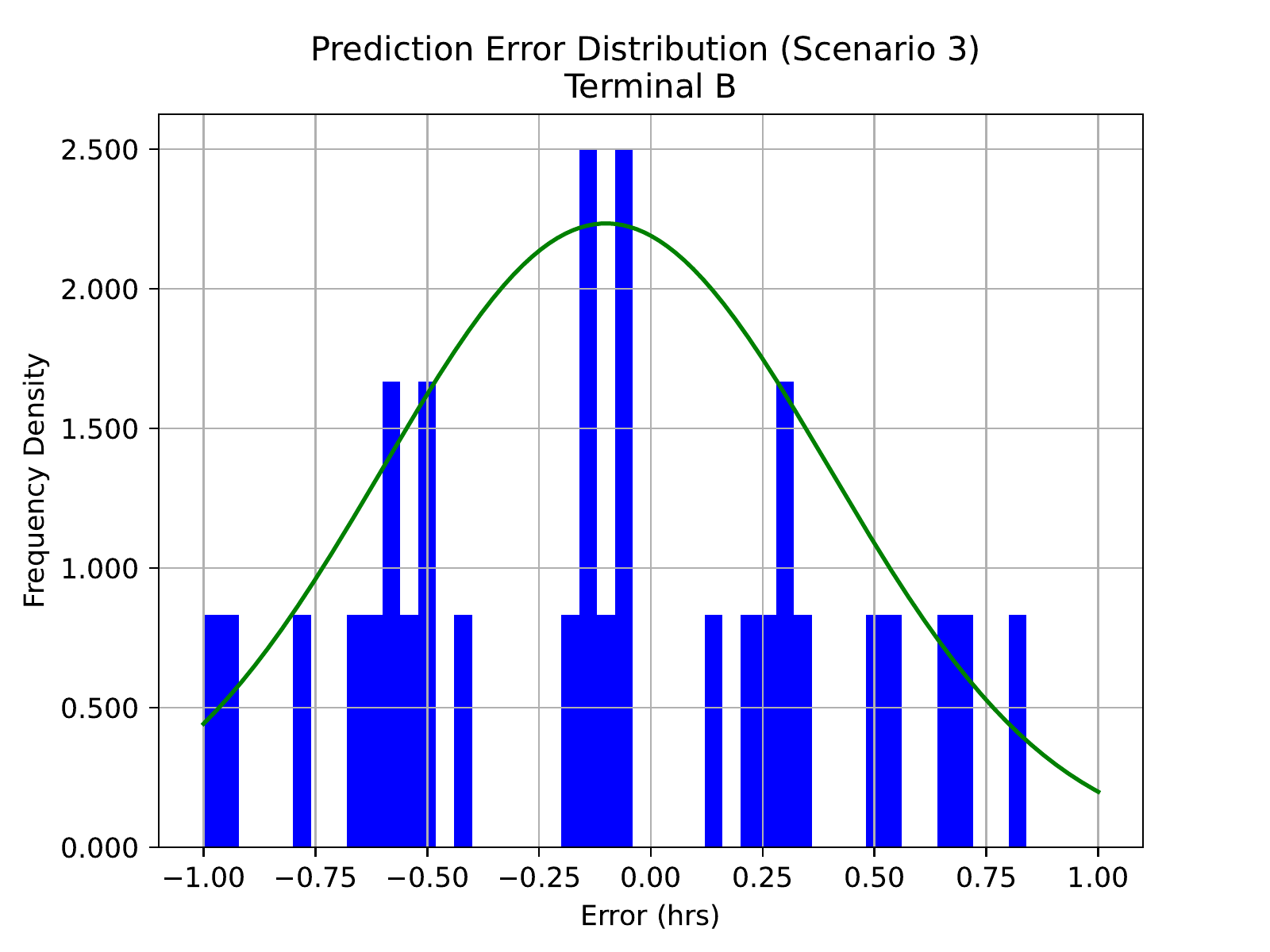} } 
	
	\caption{Prediction Error Distributions of Focused Cargoes - Scenario 3}
	\label{Fig:Error_Distribution_S3_Terminal_A_B}
\end{figure}

Based on the generic chain of focused cargoes in this study, the results of berth stay prediction for scenario 3 are shown in Figure \ref{Fig:Berth_Stay_Prediction_S3}, and the corresponding distributions of prediction errors are presented in Figure \ref{Fig:Error_Distribution_S3_Terminal_A_B} for both terminals.

\subsubsection{Scenario 4 - Complete Operation}
In this scenario, the starting point of berth stay prediction is ``Complete Operation'', which means that it tries to predict how long vessel will stay at berth since the cargo operations get finished. Similarly, the ending point of berth stay should theoretically be ``Cargo Arm Disconnected'' without the requirement of prewash, and ``Marpol Prewash Arm Disconnected'' with the requirement of prewash afterwards.

\begin{figure}[htbp]
	\centering
	\includegraphics[width=0.5\textwidth]{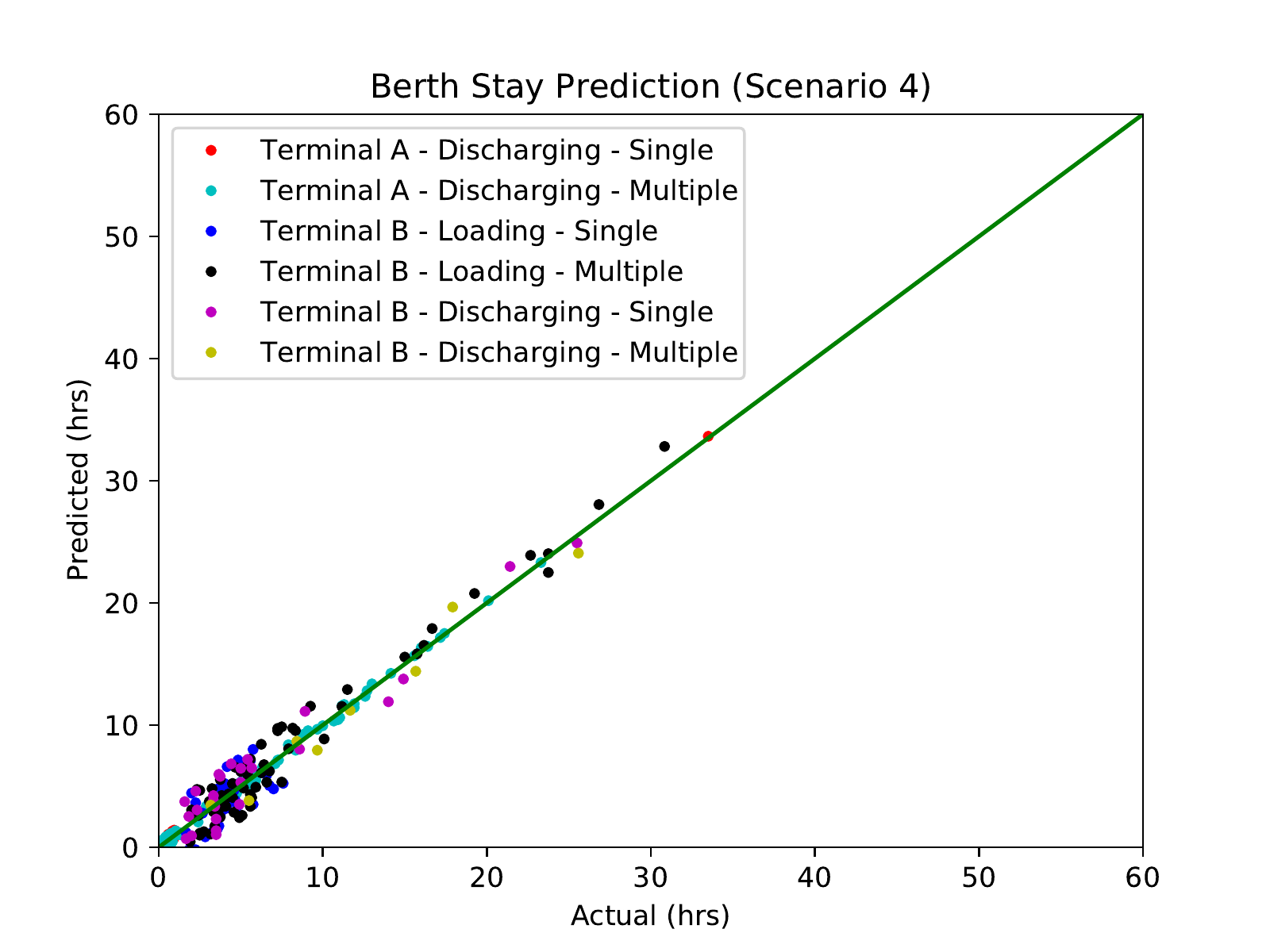} 
	\caption{Berth Stay Prediction - Scenario 4}
	\label{Fig:Berth_Stay_Prediction_S4}
\end{figure}

\begin{figure}[htbp]
	\centering
	\subfloat[Terminal A]{
		\label{SubFig:Error_Distribution_S4_Terminal_A}
		\includegraphics[width=0.4\textwidth]{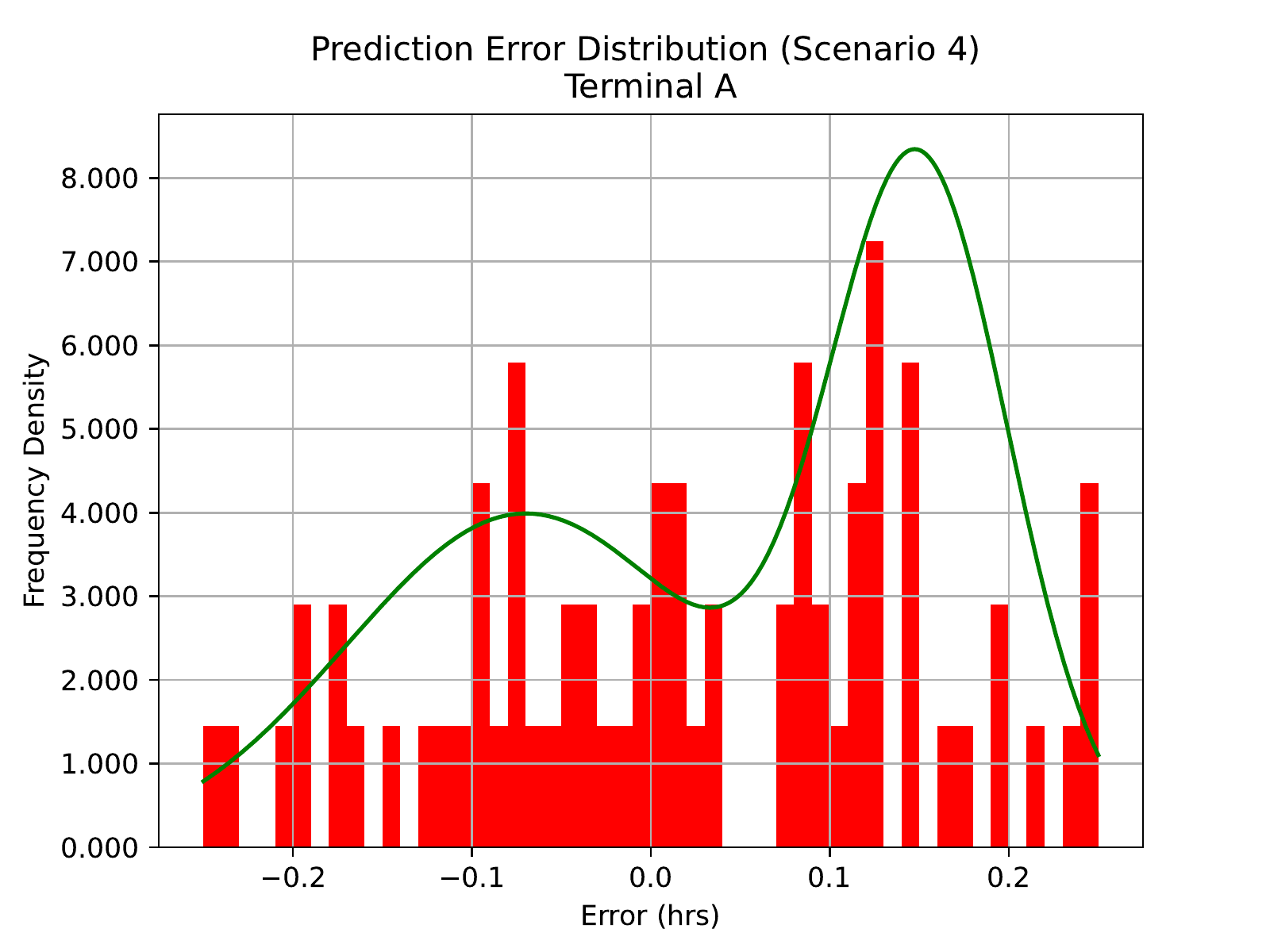} }~
	\subfloat[Terminal B]{
		\label{SubFig:Error_Distribution_S4_Terminal_B}
		\includegraphics[width=0.4\textwidth]{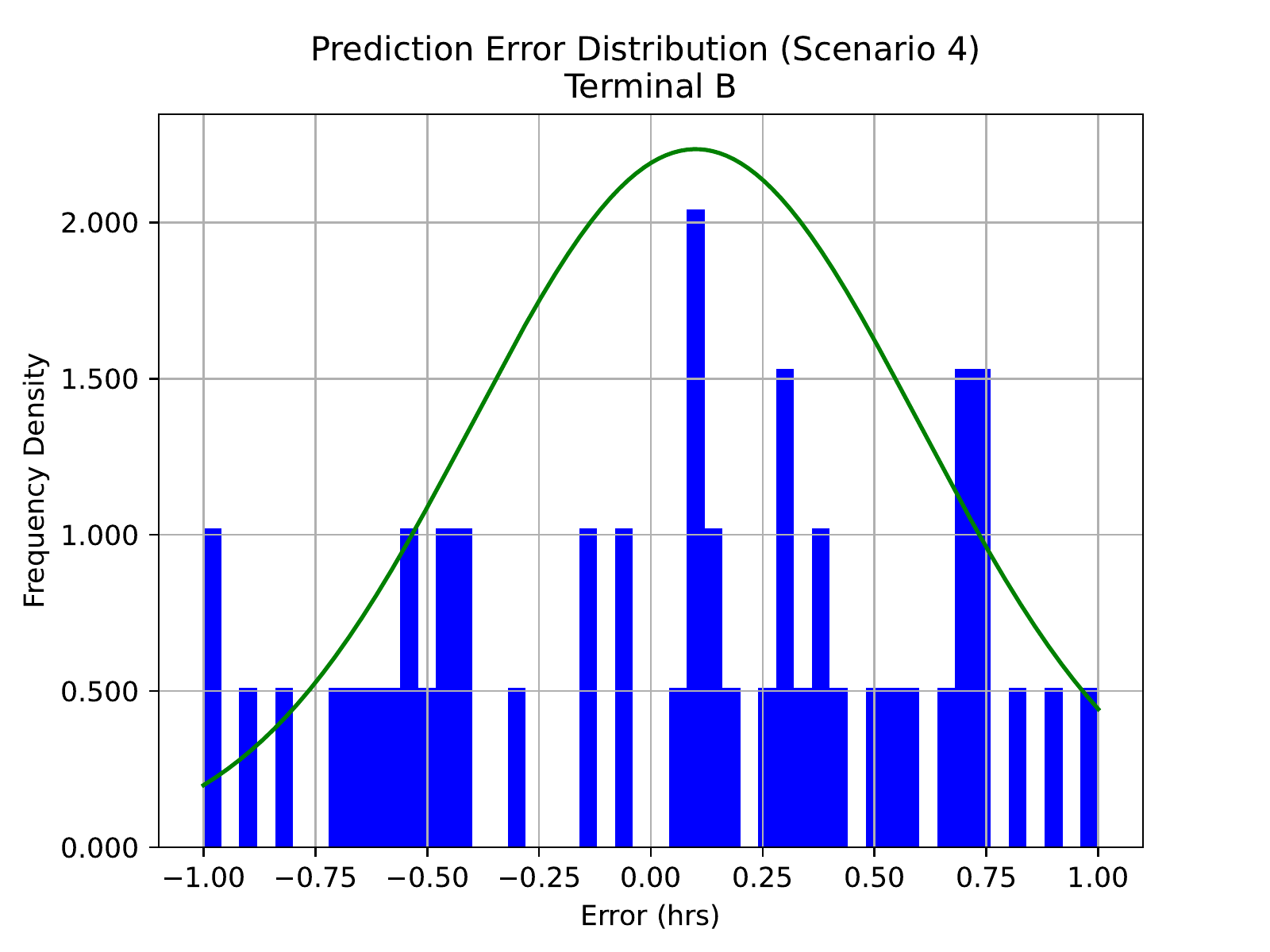} } 
	
	\caption{Prediction Error Distributions of Focused Cargoes - Scenario 4}
	\label{Fig:Error_Distribution_S4_Terminal_A_B}
\end{figure}

Based on the generic chain of focused cargoes in this study, the results of berth stay prediction for scenario 4 are shown in Figure \ref{Fig:Berth_Stay_Prediction_S4}, and the corresponding distributions of prediction errors are presented in Figure \ref{Fig:Error_Distribution_S4_Terminal_A_B} for both terminals.

Based on the prediction results of four scenarios for different terminals and operations, the all-in-one statistical metrics of investigations are presented as follows in Table \ref{Tab:Metrics_Berth_Stay_Prediction_Performance_Scenarios}, and some findings are illustrated in the following section.

\begin{table}[htbp]
	\begin{adjustbox}{max width=\textwidth}
		\begin{tabular}{ccllllll}
			\hline
			\multicolumn{3}{c}{Scheme}                                                       & $\tilde{\mu}$   & $\mu$     & $\sigma$    & MSE      & MAE      \\
			\hline
			\multirow{6}{*}{Scenario 1} & \multirow{2}{*}{Terminal A} & Discharging Single   & -0.58399 & -0.4273  & 0.819119 & 0.829576 & 0.818936 \\
			&                             & Discharging Multiple & 0.019267 & 0.127319 & 0.890336 & 0.801832 & 0.778537 \\
			
			& \multirow{4}{*}{Terminal B} & Loading Single       & 2.145928625	& 1.471799722	& 5.469884871	& 31.2547338	& 4.945469262 \\
			&                             & Loading Multiple     & 0.20193017	& 0.347555804	& 5.883884398	& 34.20827378	& 4.975122609 \\
			&                             & Discharging Single   & -2.112751585	& -0.998379449	& 6.089212291	& 36.64917145	& 5.322180732 \\
			&                             & Discharging Multiple & -2.715470635	& -2.129360239	& 7.224871676	& 50.20809942	& 6.536110629 \\
			
			\hdashline
			
			\multirow{6}{*}{Scenario 2} & \multirow{2}{*}{Terminal A} & Discharging Single   & 0.034482 & -0.10909 & 0.825367 & 0.668802 & 0.698766 \\
			&                             & Discharging Multiple & 0.329962 & 0.143473 & 0.897999 & 0.819787 & 0.806109 \\
			& \multirow{4}{*}{Terminal B} & Loading Single       & N.A.     & N.A.     & N.A.     & N.A.     & N.A.     \\
			&                             & Loading Multiple     & N.A.     & N.A.     & N.A.     & N.A.     & N.A.     \\
			&                             & Discharging Single   & N.A.     & N.A.     & N.A.     & N.A.     & N.A.     \\
			&                             & Discharging Multiple & N.A.     & N.A.     & N.A.     & N.A.     & N.A.     \\
			
			\hdashline
			
			\multirow{6}{*}{Scenario 3} & \multirow{2}{*}{Terminal A} & Discharging Single   & 0.510896 & 0.206053 & 0.945436 & 0.904383 & 0.842775 \\
			&                             & Discharging Multiple & 0.135593 & 0.199043 & 0.877694 & 0.803086 & 0.770508 \\
			& \multirow{4}{*}{Terminal B} & Loading Single       & 0.202168229 & 0.293102984 & 2.806112318 & 7.74144608 & 2.400476293 \\
			&                             & Loading Multiple     & -0.49169084	& -0.536168676	& 2.867391334	& 8.382918634	& 2.440462353 \\
			&                             & Discharging Single   & -0.071487695	& 0.716103246	& 2.621488679	& 7.120691258	& 2.264205844 \\
			&                             & Discharging Multiple & 2.64377256	& 1.097839691	& 3.388669667	& 11.25294884	& 3.183254294 \\
			
			\hdashline
			
			\multirow{6}{*}{Scenario 4} & \multirow{2}{*}{Terminal A} & Discharging Single   & 0.130712 & 0.0579   & 0.279427 & 0.078643 & 0.251066 \\
			&                             & Discharging Multiple & 0.030558 & 0.031283 & 0.283017 & 0.080362 & 0.238986 \\
			& \multirow{4}{*}{Terminal B} & Loading Single       & -0.545943774 & -0.303345495	& 1.596874382	& 2.571192733	& 1.418807102 \\
			&                             & Loading Multiple     & 0.347459943	& 0.206347551	& 1.408183425	& 1.995052476	& 1.210069987 \\
			&                             & Discharging Single   & 0.593047753	& 0.329992849	& 1.556057556	& 2.437082893	& 1.380503727 \\
			&                             & Discharging Multiple & -0.857570145	& -0.540861979	& 1.23600163	& 1.629269205	& 1.114090448 \\
			
			\hline
			
		\end{tabular}
	\end{adjustbox}
	\newline
	Note: $\tilde{\mu}$ denotes median values.
	\caption{Prediction Performance (Errors) of Proposed Approaches}
	\label{Tab:Metrics_Berth_Stay_Prediction_Performance_Scenarios}
\end{table}

\noindent \textbf{Findings}: \\
\noindent(1) It is clear to note that Terminal A operates the focused cargoes more consistently and predictably than Terminal B. However, the focus of this study is not to judge which terminal performs better. This is just a pure fact from the study. 

\noindent(2) For both terminals, it is generally natural that the prediction results trend to be more accurate and precise when the scenarios become more approaching to the end of berth stay. Cross-validations between terminals show that the proposed approach is flexible and acceptable enough to predict reasonable berth stays with limited information provided.

\noindent(3) For Terminal A, except Scenario 4, other scenarios are almost under a similar level of prediction errors. For Terminal B, except Scenario 4, other scenarios show that predictions on discharging operations are more divergent than those of loading operations, and multiple operations have larger errors than those of single operation.

\section{Conclusion}
In this section, conclusion is drawn, and limitations faced in the study are named out. Finally, the research direction is pointed out for the future.

\subsection{Research Conclusion}
In this study, proposed approach for predicting berth stay of vessels is validated by the historical datasets with respect to focused cargoes among two heterogeneous terminals. After investigating four scenarios of berth stay prediction, the proposed approach can predict berth stay accurately and precisely in reasonable manners under highly uncertain environment with limited information provided. 

\noindent \underline{Scenario 1}: Terminal A reaches 0.78$\sim$0.82 hrs (95.93\%$\sim$96.13\%) in MAE (Accuracy), while Terminal B has 4.95$\sim$6.54 hrs (61.93\%$\sim$71.19\%) in MAE (Accuracy).

\noindent \underline{Scenario 2}: It is not applicable to investigate for Terminal B due to data limitation. While Terminal A results in 0.70$\sim$0.81 hrs (95.98\%$\sim$96.52\%) in MAE (Accuracy).

\noindent \underline{Scenario 3}: The results approach to 0.77$\sim$0.84 hrs (95.83\%$\sim$96.17\%) and 2.26$\sim$3.18 hrs (81.49\%$\sim$86.85\%) for Terminal A and Terminal B in MAE (Accuracy), respectively.

\noindent \underline{Scenario 4}: Terminal A obtains 0.24$\sim$0.25 hrs (98.76\%$\sim$98.81\%), while Terminal B gets 1.11$\sim$1.42 hrs (91.73\%$\sim$93.54\%) in MAE (Accuracy). 

With the experimental and evaluation results shown above, the main focus of this study as stated above is not to judge the performance between terminals. However, this is just a result of fact from the study to validate the flexibility and effectiveness of proposed approach among heterogeneous characteristics of terminals. This also demonstrates the proposed approach has dynamic capability of predicting berth stay among different dynamic scenarios. The approach may be potentially applied for short-term pilot-booking or scheduling optimizations within a reasonable time frame for advancement of port intelligence and logistics efficiency.

\subsection{Research Limitation}
Even though many aspects in berth stay prediction are considered and investigated in this study, there are still limitations to be noted. 1) Only focused cargoes (i.e. G1 and G2) of two designated terminals are investigated in this study. Other cargoes may have different profiles over berth stay. 2) Due to limited operational data in terminals, only two designated terminals are studied in this work. Different terminals may also have other different standards of procedures in handling cargoes. This work may not be comprehensive for all terminals, but the proposed approaches could be similarly adopted to other terminals. 3) Berth stay prediction is merely based on the information of terminals and cargoes in this study, while tanker vessel information could also have potential impacts on berth stay duration. For instance, discharging may depend on vessel sizes and pump capability, which could affect the discharging rates along operations.

\subsection{Future Direction}
To follow up this study, we would like to direct our future study to investigate more cargoes and terminals together with vessels for more advanced in berth stay prediction. In that case, a more comprehensive results of berth stay prediction would be reached, and insightful and potential applications of this berth stay prediction could also be explored and deployed, so as to benefit stakeholders from different perspectives, such as facilitate pilot-booking, schedule-optimizing, etc.


%

\section*{Acknowledgment}
The authors would like to thank Mr. Chua Chye Poh and many others from ShipsFocus Group (https://www.shipsfocus.com) for providing helpful discussion and domain knowledge in the field of maritime. The authors would also gratefully thank the anonymous reviewers for the kind reviews, constructive comments and suggestions.

\bibliography{mybibfile}

\end{document}